\documentclass[aps,pra,reprint,superscriptaddress,floatfix,nofootinbib]{revtex4-2}

\usepackage[english]{babel}
\usepackage[utf8]{inputenc}
\usepackage{amsfonts,amsmath,amssymb,amsthm,mathtools,bm}
\usepackage{centernot}
\usepackage{enumerate}
\usepackage{hyperref}
\usepackage{graphicx}
\usepackage[dvipsnames]{xcolor} 

\usepackage{physics}
\usepackage{braket} 

\usepackage{natbib}
\usepackage{circuitikz}
\usepackage{booktabs}
\usepackage{siunitx}

\usepackage[normalem]{ulem} 
\newcommand{\middlebar}[1]{\text{\sout{\ensuremath{#1}}}}

\newcommand{\ii}{\mathrm{i}}
\newcommand{\diff}[1]{\mathrm{d}#1\,}
\renewcommand*\d[2]{
	\mathrm{d}
	\ifx\relax#1\relax\else
	\rule{-0.02em}{1.5ex}^{#1}\rule{0.08em}{0ex}\!
	\fi
	#2\,
}

\DeclareMathOperator{\sign}{Sign}
\DeclareMathOperator{\Heaviside}{\Theta}

\renewcommand{\a}[1]{\hat{a}_{#1}}
\newcommand{\ad}[1]{\hat{a}_{#1}^\dagger}

\DeclareMathOperator{\cutoff}{C}

\DeclareMathOperator{\DawsonF}{D_+}
\DeclareMathOperator{\JGaussian}{Q_G}
\DeclareMathOperator{\JIsoTrap}{Q_T}
\DeclareMathOperator{\erfi}{erfi}
\DeclareMathOperator{\erfc}{erfc}
\DeclareMathOperator{\sinc}{sinc}
\DeclareMathOperator{\expint}{E_1}
\DeclareMathOperator{\Jchi}{J_\chi}
\DeclareMathOperator{\JIsoTrapAux}{Y}

\newcommand{\Jmodes}{\mathcal J}
\newcommand{\Imodes}{\mathcal I}

\newcommand{\Mclean}{\text{M}}

\let\Re\relax
\let\Im\relax
\DeclareMathOperator{\Re}{Re}
\DeclareMathOperator{\Im}{Im}

\newcommand{\cut}{\text{cut}} 
\newcommand{\cpx}{\varphi} 
\newcommand{\unc}{0} 
\newcommand{\fll}{\text{f}} 

\newcommand{\Omcut}{\Omega_\cut}

\newcommand{\Omu}[1]{\Omega^{\unc}_{#1}}
\newcommand{\Omv}[1]{\Delta\Omega_{#1}}
\newcommand{\Ti}{T} 

\newcommand{\atc}{s} 
\newcommand{\aTf}{S_\fll} 

\newcommand{\aswitch}{X}

\makeatletter
\pgfcircdeclarebipolescaled{instruments}
{
    \anchor{text}{\pgfextracty{\pgf@circ@res@up}{\northeast}
        \pgfpoint{-.5\wd\pgfnodeparttextbox}{
            \dimexpr.5\dp\pgfnodeparttextbox+.5\ht\pgfnodeparttextbox+\pgf@circ@res@up\relax
        }
    }
}
{\ctikzvalof{bipoles/oscope/height}}
{josephson}
{\ctikzvalof{bipoles/oscope/height}}
{\ctikzvalof{bipoles/oscope/width}}
{
    \pgf@circ@setlinewidth{bipoles}{\pgfstartlinewidth}
    \pgfextracty{\pgf@circ@res@up}{\northeast}
    \pgfextractx{\pgf@circ@res@right}{\northeast}
    \pgfextractx{\pgf@circ@res@left}{\southwest}
    \pgfextracty{\pgf@circ@res@down}{\southwest}
    \pgfmathsetlength{\pgf@circ@res@step}{0.25*\pgf@circ@res@up}
    \pgfscope
        \pgfpathrectanglecorners{\pgfpoint{\pgf@circ@res@left}{\pgf@circ@res@down}}{\pgfpoint{\pgf@circ@res@right}{\pgf@circ@res@up}}
        \pgf@circ@draworfill
    \endpgfscope
    \pgfscope
      \pgfpathmoveto{\pgfpoint{\pgf@circ@res@left}{\pgf@circ@res@up}}%
      \pgfpathlineto{\pgfpoint{\pgf@circ@res@right}{\pgf@circ@res@down}}%
      \pgfpathmoveto{\pgfpoint{\pgf@circ@res@right}{\pgf@circ@res@up}}%
      \pgfpathlineto{\pgfpoint{\pgf@circ@res@left}{\pgf@circ@res@down}}%
      \pgfusepath{draw}
    \endpgfscope
}
\def\pgf@circ@josephson@path#1{\pgf@circ@bipole@path{josephson}{#1}}
\tikzset{josephson/.style = {\circuitikzbasekey, /tikz/to path=\pgf@circ@josephson@path, l=#1}}

\begin{document}

\title{Towards an experimental implementation of entanglement harvesting in superconducting circuits: effect of detector gap variation on entanglement harvesting}

\author{Adam Teixid\'{o}-Bonfill}
\email{adam.teixido-bonfill@uwaterloo.ca}
\affiliation{Department of Applied Mathematics, University of Waterloo, Waterloo, Ontario, N2L 3G1, Canada}
\affiliation{Institute for Quantum Computing, University of Waterloo, Waterloo, Ontario, N2L 3G1, Canada}
\affiliation{Perimeter Institute for Theoretical Physics, Waterloo, Ontario, N2L 2Y5, Canada}

\author{Xi Dai}
\email{xi.dai@phys.ethz.ch. Current address: Department of Physics, ETH Zurich, CH-8093 Zurich, Switzerland. }
\affiliation{Institute for Quantum Computing, University of Waterloo, Waterloo, Ontario, N2L 3G1, Canada}

\author{Adrian Lupascu}
\email{adrian.lupascu@uwaterloo.ca}
\affiliation{Institute for Quantum Computing, University of Waterloo, Waterloo, Ontario, N2L 3G1, Canada}
\affiliation{Department of Physics and Astronomy, University of Waterloo, Waterloo, Ontario, Canada N2L 3G1}
\affiliation{Waterloo Institute for Nanotechnology, University of Waterloo, Waterloo, Ontario, Canada N2L 3G1}

\author{Eduardo Mart\'{i}n-Mart\'{i}nez}
\email{emartinmartinez@uwaterloo.ca}

\affiliation{Department of Applied Mathematics, University of Waterloo, Waterloo, Ontario, N2L 3G1, Canada}
\affiliation{Institute for Quantum Computing, University of Waterloo, Waterloo, Ontario, N2L 3G1, Canada}
\affiliation{Perimeter Institute for Theoretical Physics, Waterloo, Ontario, N2L 2Y5, Canada}


\begin{abstract}
Motivated by the prospect of experimental implementations of entanglement harvesting in superconducting circuits, we propose a model of variable-gap particle detector that aims to bridge some of the gaps between Unruh-DeWitt (UDW) models and  realistic implementations. Using parameters tailored to potential experimental setups, we investigate entanglement harvesting in both spacelike-separated and causally connected scenarios. Our findings reveal that while variations in the energy gap reduce the ability to harvest entanglement for spacelike-separated detectors, detectors in causal contact can still become entangled through their interaction with the field. Notably, our analysis shows that (due to the derivative coupling nature of the model) even for causally connected detectors, the entanglement primarily originates from the field's correlations. This demonstrates the potential for genuine entanglement harvesting in the lab and opens the door to near-future entanglement harvesting experiments in superconducting circuits.

\end{abstract}

\maketitle


\section{Introduction}

It is known that localized probes can extract entanglement from the vacuum state of quantum fields. The entanglement extraction is possible even when the probes are spacelike separated and therefore cannot communicate with each other~\cite{Valentini1991,Reznik2003,Reznik2005BellInequalities,MenicucciEntanglingPower}. This is the Relativistic Quantum Information (RQI) protocol that has become known as \textit{entanglement harvesting} (see e.g.,~\cite{Pozas2015,Rick2024Fully}). 

In entanglement harvesting, the entanglement acquired by the probes must come from the correlations inherently present in the vacuum. Theoretical analyses long established that the vacuum of quantum fields contains entanglement~\cite{Redhead1995MoreAdo,Redhead1998,Clifton1998Superentangled} and that measurements on field observables of spacelike separated regions can violate Bell’s inequalities~\cite{Summers1985TheVV,Summers1987}. When performing entanglement harvesting, the probes used to extract the entanglement from the quantum field are commonly modeled as \textit{particle detectors}: internally simple quantum systems that couple locally to the quantum field. The most commonly used example of such a model is the Unruh-DeWitt (UDW) detector~\cite{Unruh1976,DeWitt,Unruh-Wald,Schlicht2004,Jorma2006}, which has also been analyzed as a simplified model of the light-matter interaction (see, e.g., ~\cite{Eduardo2013Wavepacket,Pozas2016Electromagnetic,Richard2021LightMatter}). 

Entanglement harvesting holds promise both as a way to understand the entanglement structure in quantum field theory~\cite{Klco2021Spheres, Klco2022Structures, Klco2023StructuresII, Bruno2023EntanglementStructure, Ivan2023Ubiquitous, Gao2024spacelike, Ivan2024multimode} and as a way to harness non-locality resources in quantum information~\cite{Eduardo2013Farming}.

Despite its significance, entanglement harvesting has not been experimentally realized as of the time of the writing of this paper. Implementations have been proposed in linear ion traps~\cite{Retzker2005}, superconducting circuits~\cite{Sabin2010microwave,Sabin2011Fermi,Sabin2012PastFuture}, in graphene \cite{Ardenghi2018}, in Bose-Einstein condensates~\cite{Cisco2020Interferometric,Cisco2023vacuum} and using the vacuum state of the electromagnetic field inside a non-linear crystal~\cite{Lindel2020Detection,Settembrini2022Correlations,Lindel2023probing,Lindel2023entanglement}, but so far no direct experimental test of entanglement harvesting has been performed.

In this paper we will focus on the superconducting circuit platform. The implementation of entanglement harvesting in superconducting circuits leverages the ability of these circuits to create ``artificial atoms'' that can strongly couple to the electromagnetic field~\cite{FriskKockum2019,Forn-Diaz2019USCreview}. Specifically, the interaction can reach the ultra-strong coupling regime. At such strong couplings, the rotating wave approximation, commonly used to describe light-matter interaction~\cite{JaynesCummings1963}, breaks down~\cite{Peropadre2013PRL}. Notably, this approximation is not present in UDW detector models used in RQI.  A key advantage of using superconducting circuits for RQI experiments is their high tunability, allowing for circuits with switchable coupling that can reach the ultra-strong regime~\cite{Peropadre2010switchable,Romero2012Ultrafast,Tunable_2023}. This enables superconducting circuit platforms to access regimes where the amounts of harvested entanglement become significant. In these setups, the interaction of superconducting qubits with the field inside a microwave waveguide can be switched on and off within fractions of a nanosecond, allowing these detectors to probe fields in (or close to) spacelike separated regions.

The first goal of this study is to strengthen the connection between experimentally implemented particle detectors in superconducting circuits and the idealized UDW detector models commonly employed in RQI. We begin with a complete circuit model of the superconducting implementation. Then, we review the series of approximations upon it that result in a model that resembles a UDW detector coupled to a 1+1D scalar massless field, but that retains crucial implementation-specific features of the superconducting circuit platform. These features, which allow to better model realistic experiments, include: a variable energy gap, coupling to the derivative of the field amplitude, and a soft UV cutoff. Among these features, derivative coupling has been previously explored, since it provides a natural way to remove IR divergences~\cite{Raine1991,Raval1996,Eduardo2014Zero,Juarez-Aubry_2014,Wang2014,Juarez-Aubry2018,Tjoa2020,Juarez-Aubry2022,DBunney_2023}, and is a better model for the light-matter interaction in some regimes \cite{Pozas2016Electromagnetic,Richard2021LightMatter,McKay_2017}. Moreover, derivative coupling and the more commonly explored amplitude coupling are related by the duality discussed in~\cite{Matheus2023Duality}. The impact of the cutoff and its relation to the detector spatial localization has been previously analyzed in~\cite{McKay_2017}. However, the variable energy gap has only been explored in very idealized scenarios outside of superconducting circuits and in timelike connection~\cite{Olson2011FuturePast,Olson2012Extraction}. In this study, the variation of the energy gap is dictated by the experimental constraints in implementing the protocols in superconducting circuits. Using the implementation of~\cite{Tunable_2023}, the variation of the gap is linked to the strength of the coupling, which is time dependent. This is a constraint whose effect has not yet been fully explored in the context of RQI protocols.

Once the model is established, we study how these implementation-specific features affect entanglement harvesting. We examine entanglement harvesting both for spacelike and causally connected detectors. Detectors in causal contact acquire entanglement from two sources: communication through the field or harvesting of pre-existing field entanglement. To distinguish these two contributions we use the methods developed in~\cite{Erickson2021_When}, which allow us to identify situations where entanglement is genuinely harvested even in causal contact~\cite{adam2024derivative}.

This article is organized as follows: Section~\ref{sec:SC_CouplerModel} provides the circuit model for implementing particle detectors as a superconducting device. Section~\ref{sec:circuit_to_detector} describes the simplifications that turn the detector implementation model into a UDW-like detector with implementation-specific features such as a variable gap and derivative coupling. Section~\ref{sec:time_evolution} illustrates the entanglement harvesting protocol for a pair of variable gap, derivatively coupled detectors. Section~\ref{sec:genuine_harvesting} delineates how to separate the communication and genuine harvesting contributions to the entanglement acquired by causally connected detectors. Section~\ref{sec:harvesting_results} shows how the implementation-specific features such as gap variation affect entanglement harvesting.

\section{Review of superconducting tunable couplers}
\label{sec:SC_CouplerModel}

In this section we will analyze the superconducting circuit model behind implementing tunable coupling of a superconducting qubit to a transmission line~\cite{Tunable_2023}. Namely, in this implementation, we can tune the coupling between superconducting qubits and a transmission line from zero coupling to the ultra-strong coupling regime in a matter of fractions of nanosecond. We will review the derivation of the Hamiltonian of this device, which will be useful for Section~\ref{sec:circuit_to_detector} where we relate it to the typical particle detector models employed in the literature of entanglement harvesting. To model the system we will follow a circuit model approach, similarly to the Supplementary Material of \cite{Peropadre2013PRL} and to \cite{ShiJiahao2019,RenShaun2022}.

\subsection{Quantized transmission line}
\label{sec:TL_quantization}
Superconducting transmission lines provide an experimental system where dynamical degrees of freedom of the electromagnetic field can be simplified into a one-dimensional real scalar quantum field theory. Although it is well-known in the literature of superconducting quantum devices, for convenience and notation setting purposes, we include here a brief  derivation of the Hamiltonian of the transmission line from its lumped circuit model depicted in Figure \ref{fig:lumpedTL}.
\ctikzset{bipoles/length=1.1cm}
\begin{figure}[ht]
    \centering
    \begin{center}
    \begin{circuitikz}
    \draw 
    (1.75,0) to (2,0) 
    to[L, l=$\ell_0 \Delta x$,*-*] (4,0)
    to[L, l=$\ell_0 \Delta x$,-*] (6,0)
    to (6.25,0);
    \draw[dashed] (6.25,0) -- (7,0) (1.75,0) -- (1,0);
    \draw (2,0) to (2,-0.3) to [C, l=$c_0 \Delta x$] (2,-1.6) node[ground]{};
    \draw (4,0) to (4,-0.3) to [C, l=$c_0 \Delta x$] (4,-1.6) node[ground]{};
    \draw (6,0) to (6,-0.3) to [C, l=$c_0 \Delta x$] (6,-1.6) node[ground]{};
    \node[anchor=south] at (2,0) {$\Phi_{-1}$}; 
    \node[anchor=south] at (4,0) {$\Phi_{0}$}; 
    \node[anchor=south] at (6,0) {$\Phi_{1}$}; 
    \draw[>=latex, <->] (2,-2.3) -- (4,-2.3);
    \draw[>=latex, <->] (4,-2.3) -- (6,-2.3);
    \draw [anchor=south] (3,-2.3) node {$\Delta x$};
    \draw [anchor=south] (5,-2.3) node {$\Delta x$};
    \end{circuitikz}
    \end{center}
    \caption{Lumped circuit model of a resistance-less transmission line.}
    \label{fig:lumpedTL}
\end{figure}

The Hamiltonian of this lumped circuit model is
\begin{equation}
    H_{\textsc{tl},\Delta x}=\frac{1}{2}\sum_{i} \frac{(\Phi_{i+1}-\Phi_{i})^2}{\ell_0\Delta x}  + \frac{q_i^2}{c_0\Delta x}, 
\end{equation}
where $\ell_0$ and $c_0$ respectively are the inductance and capacitance per unit length. The variables $\Phi_{i}$ and $q_{i}$ respectively are the flux and charge for the node $i$, and are defined as
\begin{equation}
    \Phi_{i}(t) = \int_{-\infty}^t\diff{t'}V_i(t'),\ q_{i}(t) = \int_{-\infty}^t\diff{t'}I_i(t'),
\end{equation}
where $V_i$ and $I_i$ respectively are the voltage and current from the node $i$ to ground.

To get the Hamiltonian of the transmission line, we take the continuum limit, $\Delta x\to 0$, obtaining
\begin{equation}
    H_{\textsc{tl}}=\frac{1}{2}\int \diff x \frac{ \partial_x \Phi(x)^2}{\ell_0} + \frac{q(x)^2}{c_0},\label{eq:H_TL}
\end{equation}
where 
\begin{equation}
    \partial_t \Phi(x) = \frac{\delta H_{\textsc{tl}}}{\delta q(x)} = \frac{q(x)}{c_0}.
\end{equation}
After quantization, $\hat{H}_{\textsc{tl}}$ can be recognized as the Hamiltonian of a 1D real massless scalar quantum field. Its Heisenberg picture amplitude operator $\hat\Phi$ is given by 
\begin{align}
    \hat\Phi(t,x)=\sqrt{\hbar Z_0}\int\frac{\diff k}{\sqrt{4\pi |k|}} (e^{\ii (\omega_k t-kx)} \ad{k}  + \text{H.c.})\,\label{eq:TLfield}.
\end{align}
Here, $Z_0=\sqrt{\frac{\ell_0}{c_0}}$ is the characteristic impedance of the transmission line, $\omega_k = v|k|$ is the frequency of the mode of wavenumber $k$ and $v=1/\sqrt{c_0\ell_0}$ is the speed of light in the transmission line. The $\a{k}$ and $\ad{k}$ are the annihilation and creation operators of the mode $k$, which fulfill $[\a{k},\ad{k'}]=\delta(k-k')$. Using these definitions, one finds that
\begin{equation}
    \hat H_{\textsc{tl}}=\frac{1}{2}\int \diff k \hbar \omega_k \Big(\ad{k}\a{k}+\a{k}\ad{k}\Big).\label{eq:H_boson}
\end{equation}

To give an idea about the physical scales of the system, typical parameters for the transmission line are (see e.g.,~\cite{Goppl2008,Forn-Diaz2017})
\begin{equation}
   v\approx 1.2\times 10^8 \,\unit{\meter\per\second}, \ Z_0\approx 50\,\unit{\ohm}.\label{eq:TLparameters}
\end{equation}
We will use these values throughout this work.

There is a frequency scale beyond which electromagnetic signals in the transmission line get attenuated. This attenuation increases drastically beyond the superconducting gap. For aluminum, commonly used for superconducting transmission lines, the superconducting gap is $75\,\unit{\giga\hertz}$ \cite{Douglass1964}. Therefore, the experimental implementation introduces an effective UV cutoff scale. The effects of the cutoff and finite size of the superconducting qubits were explored in \cite{McKay_2017}, and become more relevant for shorter interactions. 

In this work, we consider an exponential cutoff at a frequency scale of $\Omcut/(2\pi)= 50 \,\unit{\giga\hertz}$. This cutoff value was found in \cite{Forn-Diaz2017} by matching experimental measurements with the renormalized frequency of the coupled qubit (which is cutoff dependent), under the renormalization model of \cite{Camacho2016, Legget1987}. To implement the cutoff, we modify the field operator that interacts with the qubit as follows,
\begin{align}
    \hat\Phi_{\cutoff}(t,x) &= \sqrt{\hbar Z_0}\int\diff k \frac{\cutoff(\omega_k)}{\sqrt{4\pi |k|}}(e^{\ii (\omega_k t-kx) } \ad{k} + \text{H.c.})\,. \label{eq:TLfield_cutoff}
\end{align}
Here, we added the exponentially decreasing weight
\begin{equation}
    \cutoff(\omega) = e^{-\frac{|\omega|}{2\Omcut}}.\label{eq:defCutoff}
\end{equation}

As discussed in~\cite{McKay_2017}, this implementation of the cutoff is equivalent to the qubit interacting with a spatially smeared version of the field. Specifically,
\begin{equation}
    \hat\Phi_{\cutoff}(t,x) = \int\diff{x'} F_\text{eff}(x'-x)\hat\Phi(t,x'),\label{eq:smeared_field}
\end{equation}
where the effective smearing function $ F_\text{eff}(x)$ is
\begin{equation}
    F_\text{eff}(x) = \frac{1}{2\pi}\int \diff k \cutoff(\omega_k)e^{\ii k x} = \frac{1}{2\pi v}\widetilde{\cutoff}\bigg(\frac{x}{v}\bigg),\label{eq:effSmearing}
\end{equation}
with the following Fourier transform convention 
\begin{equation}
    \widetilde{f}(k) = \int \diff x f(x)e^{\ii k x}.\label{eq:fourier_convention}
\end{equation}
The expression for $F_\text{eff}(x)$ follows from comparing Eqs.~\eqref{eq:TLfield} and \eqref{eq:TLfield_cutoff} to see that $\widetilde{F}_\text{eff}(k) = \cutoff(\omega_k)$, and then using the inverse Fourier transform together with \mbox{$\omega_k =v|k|$}. Notice that $F_\text{eff}(x)\in \mathbb R$ and \mbox{$F_\text{eff}(x)=F_\text{eff}(-x)$}, due to $\widetilde{F}_\text{eff}(k)\in \mathbb R$ and $\widetilde{F}_\text{eff}(k)=\widetilde{F}_\text{eff}(-k)$. 

In particular, for the exponential cutoff chosen in Eq.~\eqref{eq:defCutoff},
\begin{equation}
    F_\text{eff}(x) = \frac{2\Omcut}{\pi v}\frac{1}{1+\big(\frac{2\Omcut}{v}x\big)^2}.\label{eq:effSmearing_ExpCutoff}
\end{equation}

\subsection{Tunable coupler and flux qubit}
Here we describe the superconducting circuit which performs the role of the detector in the entanglement harvesting implementation. Following the proposal of \cite{Tunable_2023}, this circuit consists of a flux qubit tunably coupled to the transmission line. This circuit is made out of Josephson junctions, which can be implemented as a small insulating gap between superconducting materials. In circuit diagrams, Josephson junctions are indicated as crosses. In practice, Josephson junctions always have a capacitance in parallel, indicated by drawing the crosses in a box.

Consider the flux qubit depicted in Figure \ref{fig:fluxQubit}. The qubit subspace consists of the two lowest energy levels of the circuit. The flux qubit consists of a superconducting loop with Josephson junctions. The loop is threaded by an external magnetic flux $f_\varepsilon$, which can be tuned arbitrarily and is chosen so that the qubit is in the symmetry point, i.e. in a minimum of the qubit frequency.
\ctikzset{bipoles/length=0.6cm}
\begin{figure}[ht]
    \centering
    \begin{center}
    \begin{circuitikz}
      \draw (0,0) to[josephson,l=$\gamma_1$] (1,0) to[josephson,l=$\gamma_2$] (2,0) to[josephson,l=$\gamma_3$] (3,0) to (3,1.8) to (0,1.8) to (0,0);
      \node[anchor=west] at (1.54,1) {$f_\varepsilon$};
      \node[anchor=center,draw,circle,inner sep=0.04cm] at (1.5,1) {\tiny\textbullet};
      \draw[-latex] (2.25,-0.3) -- (2.75,-0.3);
      \draw[-latex] (1.25,-0.3) -- (1.75,-0.3);
      \draw[-latex] (0.25,-0.3) -- (0.75,-0.3);
    \end{circuitikz}
    \end{center}
    \caption{Circuit model of a flux qubit. The $\gamma_i$ indicate the phase variable of the $i$-th Josephson junction.}
    \label{fig:fluxQubit}
\end{figure}

The tunable coupler is an additional loop with Josephson junctions, threaded by an external flux $f_\beta$, shown in Figure \ref{fig:tunableCoupler}. From now on, we will refer to the two lowest energy levels of the combined circuit presented in Figure \ref{fig:tunableCoupler} as the \textit{qubit subspace}, which differs from the qubit subspace of the flux qubit.

The transmission line is connected at $a$ and $b$, which makes the phase operator $\hat\gamma_5$ couple to the transmission line. Then, the parameter $f_\beta$ allows to tune the coupling strength between the qubit and the transmission line, by changing the size of $\hat\gamma_5$ in the qubit subspace.

\begin{figure}[ht]
    \centering
    \begin{center}
    \begin{circuitikz}
      \draw (0,0) to[josephson,l=$\gamma_1$] (1,0) to[josephson,l=$\gamma_2$] (2,0) to[josephson,l=$\gamma_3$] (3,0) to (3,2) to[josephson,l_=$\gamma_4$] (0,2) to (0,0);
      \node[anchor=west] at (1.54,1) {$f_\varepsilon$};
      \node[anchor=center,draw,circle,inner sep=0.04cm] at (1.5,1) {\tiny\textbullet};

      \draw (3,2) to (3,4) to[josephson,l_=$\gamma_5$,*-*] (0,4) to[josephson,l=$\gamma_6$] (0,2);
      \node[anchor=south] at (0,4) {$a$}; 
      \node[anchor=south] at (3,4) {$b$}; 
      \node[anchor=west] at (1.54,3) {$f_\beta$};
      \node[anchor=center,draw,circle,inner sep=0.04cm] at (1.5,3) {\tiny\textbullet};
      \draw[-latex] (2.25,-0.3) -- (2.75,-0.3);
      \draw[-latex] (1.25,-0.3) -- (1.75,-0.3);
      \draw[-latex] (0.25,-0.3) -- (0.75,-0.3);
      \draw[-latex] (1.75,2-0.3) -- (1.25,2-0.3);
      \draw[-latex] (1.25,4-0.3) -- (1.75,4-0.3);
      \draw[-latex] (-0.3,2.75) -- (-0.3,3.25);
    \end{circuitikz}
    \end{center}
    \caption{Circuit model of the tunable coupler connected to the flux qubit. The tunable coupler connects to the transmission line through the points $a$ and $b$. }
    \label{fig:tunableCoupler}
\end{figure}

The tunable coupler + flux qubit circuit is quantized by using the following conjugate sets of variables 
\begin{equation}
    \bm \gamma = 
    \begin{bmatrix}
    \gamma_1\\
    \gamma_2\\
    \gamma_5\\
    \gamma_6
    \end{bmatrix}, \quad
    \bm  p = \hbar\bm  N =
    \hbar\begin{bmatrix}
    N_1\\
    N_2\\
    N_5\\
    N_6
    \end{bmatrix},
\end{equation}
where the phase degrees of freedom, $\gamma_i$, are $2\pi$-periodic. The resulting Hamiltonian of the tunable coupler and the flux qubit is
\begin{align}
    \hat H_\textsc{tc+fq} &= 2e^2 \hat{\bm { N}}^\top  C^{-1} \hat{\bm {N}} - \sum_{i\in\{1,2,5,6\}} E_{J_i} \cos(\hat\gamma_i) \nonumber\\
    &\quad - E_{J_3} \cos(\hat\gamma_1+\hat\gamma_2-\hat\gamma_5-\hat\gamma_6 +2\pi f_\varepsilon+2\pi f_\beta) \nonumber\\
    &\quad - E_{J_4} \cos(\hat\gamma_5+\hat\gamma_6 - 2\pi f_\beta),\label{eq:H_TC+FQ}
\end{align}
where the $C$ is a matrix of capacitances, the $E_{J_i}$ is the Josephson energy of the $i$-th junction\footnote{Ref.~\cite{Tunable_2023} obtained the matrix of capacitances from an electromagnetic simulation of the superconducting circuit. Then, they found the value of the critical currents $I_{c_i}=E_{J_i}/\varphi_0$ by fitting the numerically computed qubit frequencies to their measurements for multiple values of $f_\beta$ and $f_\varepsilon$. Here, $\varphi_0=\hbar/(2e)$ denotes the reduced magnetic flux quantum.}, and the $f_\varepsilon$ and $f_\beta$ are external magnetic fluxes, which can be controlled during the experiment. For a review of the circuit theory for superconducting qubits, which was used to derive this Hamiltonian, see \cite{Vool2017Introduction}. 

Notice that $\hat H_\textsc{tc+fq}$ plays the role of the free Hamiltonian in the usual detector models, and importantly, it has non-linear cosine terms, which come from the Josephson junctions and do not appear in most common detector models, such as qubits or harmonic oscillators. These terms are key to create superconducting qubits and superconducting devices in general. Moreover, the presence of the external magnetic fluxes $f_\varepsilon$ and $f_\beta$ in the non-linear terms is what allows to tune the coupling to the transmission line. However, relating the Hamiltonian of this circuit to the common detector models requires some work that will be done in section \ref{sec:circuit_to_detector}. It is there that we will show how this circuit can be approximated under certain regimes as a variable gap qubit detector.

\subsection{Coupling the superconducting circuit to the transmission line}
Here, we derive the interaction between a superconducting circuit, such as the tunable coupler + flux qubit (TC+FQ) circuit, and the transmission line. To do so, we start from the lumped circuit model depicted in Figure \ref{fig:QubitInTL}, which includes both the TC+FQ circuit and the transmission line, and take the limit $\Delta x\to 0$.

\ctikzset{bipoles/length=1.1cm}
\begin{figure}[ht]
    \centering
    \begin{center}
    \begin{circuitikz}
    \draw 
    (-0.25,0) to
    (0,0) to[L, l=$\ell_0 \Delta x$,*-*] (2,0)
    to[L, l=$\ell_0 \frac{\Delta x}{2}$,-*] (3.5,0);
    \draw (5,0) to[L, l=$\ell_0 \frac{\Delta x}{2}$,*-*] (6.5,0) -- (6.75,0);
    \draw[dashed] (6.75,0) -- (7.25,0) (-0.25,0) -- (-0.75,0);
    \draw (0,0) to (0,-0.3) to [C, l=$c_0 \Delta x$] (0,-1.6) node[ground]{};
    \draw (2,0) to (2,-0.3) to [C, l=$c_0 \Delta x$] (2,-1.6) node[ground]{};
    
    \draw (6.5,0) to (6.5,-0.3) to [C, l=$c_0 \Delta x$] (6.5,-1.6) node[ground]{};
    \node[anchor=south] at (0,0) {$\Phi_{-1}$}; 
    \node[anchor=south] at (2,0) {$\Phi_{0}$}; 
    \node[anchor=south] at (6.5,0) {$\Phi_{1}$}; 
    \draw[>=latex, <->] (0,-2.3) -- (2,-2.3);
    \draw [anchor=south] (1,-2.3) node {$\Delta x$};

    \draw[dashed] (3.5,-0.7) rectangle (5,0.8);
    
    \node at (4.25,0.45) {TC+FQ};

    \draw [anchor=west] (3.5,0) node {$\Phi_a$};
    \draw [anchor=east] (5,0) node {$\Phi_b$};
    \end{circuitikz}
    \end{center}
    \caption{Lumped circuit model of a transmission line coupled to the TC+FQ circuit of Figure~\ref{fig:tunableCoupler}, depicted as a dashed box. }
    \label{fig:QubitInTL}
\end{figure}

The circuit in Figure~\ref{fig:QubitInTL} has the following Hamiltonian:
\begin{align}
    \hat H_{\Delta x} &= 
    \hat H_\textsc{tc+fq} +  \sum_{i}\frac{\hat{q}_i^2}{2c_0\Delta x}+\sum_{i\neq0} \frac{1}{2\ell_0\Delta x} (\hat{\Phi}_{i+1}-\hat{\Phi}_{i})^2\nonumber\\
    & + \frac{1}{\ell_0\Delta x} (\hat{\Phi}_a-\hat{\Phi}_0)^2 + \frac{1}{\ell_0\Delta x} (\hat{\Phi}_1-\hat{\Phi}_b)^2.
\end{align}
Here, $\hat\Phi_a$ and $\hat\Phi_b$ respectively denote the fluxes at the nodes $a$ and $b$ of the TC+FQ circuit depicted in Figure~\ref{fig:tunableCoupler}. These fluxes relate to the branch variable $\hat\gamma_5$ as follows, $\hat{\Phi}_b-\hat{\Phi}_a=\varphi_0\hat\gamma_5$, with $\varphi_0=\frac{\hbar}{2e}$ the reduced magnetic flux quantum.
Performing the change of variables \mbox{$\hat{\Phi}_{\pm} = \hat{\Phi}_b\pm\hat{\Phi}_a$} and a few algebraic manipulations, 
\begin{align}
     &(\hat{\Phi}_0-\hat{\Phi}_a)^2 +  (\hat{\Phi}_1-\hat{\Phi}_b)^2\nonumber\\
     &= \frac{1}{2}(\hat{\Phi}_1 - \hat{\Phi}_0)^2  - (\hat{\Phi}_1 - \hat{\Phi}_0)\hat{\Phi}_- + \frac{\hat{\Phi}_-^2}{2} \nonumber\\ 
     &\quad +\frac{1}{2}(\hat{\Phi}_1+\hat{\Phi}_0 - \hat{\Phi}_+)^2 .
\end{align}
Substituting back into $\hat H_{\Delta x}$,
\begin{align}
    \hat H_{\Delta x} &= \hat H_\textsc{tc+fq} + \hat H_{\textsc{tl},\Delta x} - \frac{\hat{\Phi}_1 -\hat{\Phi}_0}{\ell_0\Delta x}\hat{\Phi}_- \nonumber\\
    & \textcolor{white}{=}  + \frac{\hat{\Phi}_-^2}{2\ell_0\Delta x} + \frac{(\hat{\Phi}_1+\hat{\Phi}_0 - \hat{\Phi}_+)^2}{2\ell_0\Delta x} .
\end{align}
Now, taking $\Delta x \to 0$, the Hamiltonian becomes
\begin{align}
    &\hat H = \hat H_\textsc{tc+fq} + \hat H_\textsc{tl} + \hat H_\textsc{int},\ \nonumber\\
    &\hat H_\textsc{int} = - \frac{1}{\ell_0} \hat{\Phi}_- \partial_x\hat{\Phi}(x_\textsc{tc+fq}).
\end{align}
Here, $x_\textsc{tc+fq}$ is the position of the TC+FQ circuit, which we effectively treat as point-like. This treatment is justified because the field cutoff given in Eq.~\eqref{eq:defCutoff} (with scale \mbox{$\Omcut/(2\pi) = 50\,\unit{\giga\hertz}$}) dominates the TC+FQ circuit physical size\footnote{The cutoff makes it so that distances smaller than \mbox{$\frac{v}{\Omcut}=0.38 \,\unit{\milli m}$} are not resolved by the field.}, which is approximately $100\, \unit{\micro m}$ in the proposal of~\cite{Tunable_2023}. For a more detailed study of how the effects of the cutoff and the superconducting circuit size combine, see \cite{McKay_2017}. 

Moreover, we used that $(\hat{\Phi}_1+\hat{\Phi}_0 - \hat{\Phi}_+)^2/\Delta x \to 0$ when taking $\Delta x \to 0$. This is a consequence of the following inductor relationships,
\begin{align}
    &\hat{\Phi}_0-\hat{\Phi}_a=\frac{\ell_0\Delta x}{2}\hat{I}_{0a},\quad 
    \hat{\Phi}_b-\hat{\Phi}_1=\frac{\ell_0\Delta x}{2}\hat{I}_{b1}.\label{eq:inductors}
\end{align}
Here, $\hat{I}_{0a}$ and $\hat{I}_{b1}$ respectively are the currents from nodes $0$ to $a$ and from nodes $b$ to $1$. Then, from Eq.~\eqref{eq:inductors} and
\mbox{$\hat\Phi_+=\hat\Phi_a+\hat\Phi_b$},
\begin{align}
    \frac{(\hat{\Phi}_1+\hat{\Phi}_0 - \hat{\Phi}_+)^2}{\Delta x}  &= 
    \frac{\ell_0^2}{4}(\hat{I}_{0a}-\hat{I}_{b1})^2  \Delta x = \order{\Delta x},
\end{align}
which implies that $(\hat{\Phi}_1+\hat{\Phi}_0 - \hat{\Phi}_+)^2/\Delta x\to 0$ as $\Delta x\to 0$. 

Additionally, to obtain $\hat H$, we neglected the renormalization term $\hat{\Phi}_-^2/(2\ell_0\Delta x)$. This approximation is common for superconducting circuits coupled to transmission lines. The renormalization term diverges in the limit $\Delta x \to 0$. This divergence could be avoided by not fully taking the continuum limit, and instead picking a finite value of $\Delta x$. Such a discretized transmission line has a dispersion relation $\omega(k)$ that is linear at small $k$, but nonlinearities kick in as $k$ approaches $1/\Delta x$, see e.g.~\cite{Camacho2016}. A reasonable choice of $\Delta x$ would be the one associated to the smallest resolvable length given the measured cutoff, $v/\Omcut=0.38\,\unit{\milli m}$. For this value of $\Delta x$, the renormalization term would significantly modify the energy levels of the TC+FQ circuit \footnote{If we choose $\Delta x=0.38\,\unit{\milli m}$, then $\ell_0\Delta x = 0.16\,\unit{\nano\henry}$, which is comparable to the Josephson inductance of the junction that lies in the transmission line path. This means that the renormalization term would contribute significantly to $\hat H_\textsc{tc+fq}$.}.
Nonetheless, in this work we \emph{do} take the continuum limit, neglecting the renormalization term, for both ease of computation and ease of comparison with common UDW detector models. We do not expect that adding back the renormalization term would change the main conclusions of our work. Notably, $\hat H_\textsc{int}$ remains the same regardless of whether the renormalization term is included. The effect of the renormalization term will be analyzed in future work.

Finally, notice that the interaction term $\hat H_\textsc{int}$ is proportional to the flux jump $\hat{\Phi}_-$ across the TC+FQ circuit, which fulfills $\hat{\Phi}_-=\varphi_0\hat\gamma_5$. Moreover, after replacing $\hat{\Phi}$ by the $\hat{\Phi}_{\cutoff}$ of Eq.~\eqref{eq:TLfield_cutoff}, which includes the cutoff, the final expression for $\hat H_\textsc{int}$ becomes
\begin{equation}
    \hat H_\textsc{int} = - \frac{\varphi_0}{\ell_0} \hat{\gamma}_5 \partial_x\hat{\Phi}_{\cutoff}(x_\textsc{tc+fq}).\label{eq:Hint_gamma5}
\end{equation}
While this interaction Hamiltonian is constant, tunability is achieved by controlling the parameters $f_\beta$ and $f_\varepsilon$, which change the projection of $\hat\gamma_5$ on the qubit subspace (i.e., the space spanned by the two lowest energy eigenvectors of~\eqref{eq:H_TC+FQ}), effectively reducing or increasing the intensity of the interaction.

\section{A variable gap particle detector as a model of flux qubit and tunable coupler}
\label{sec:circuit_to_detector}
By means of an adiabatic approximation, we will relate the TC+FQ implementation given in Section \ref{sec:SC_CouplerModel} to a variable gap detector model, which resembles the most common particle detector used in RQI protocols, the UDW detector.

\subsection{Simplifying the TC+FQ circuit model}
\label{ssec:approximations}
\subsubsection{Two-level approximation}
To reduce the tunable coupler + flux qubit circuit into a qubit, we need to truncate the energy levels, keeping only the two lowest-energy eigenvectors of $\hat H_\textsc{tc+fq}$. Neglecting higher energy levels can be justified when their transition frequencies are large compared to the qubit transition frequency. Then, if the coupling is switched on and off slowly enough, the adiabatic theorem forbids transitions that leave the qubit subspace. The exact conditions under which this adiabatic approximation in the ultra-strong coupling regime is justified are subtle and will be explored elsewhere.  Here we will operate under the assumption that they hold, as experimentally validated in cases where the coupling is not too strong~\cite{Forn-Diaz2017,Yoshihara2017,Astafiev2010}. All the quantities of interest for the qubit, such as the energy gap and coupling strength, can be computed from the numerical diagonalization of the Hamiltonian $\hat H_\textsc{tc+fq}$ in a suitably truncated charge basis. 
In the TC+FQ circuit, for each $f_\beta$, the value of $f_\varepsilon$ is picked to minimize the qubit transition frequency~\cite{Tunable_2023}. This choice corresponds to a coupling between the qubit and the transmission line that best approximates the transverse coupling model in UDW detectors. Then, the only remaining controllable parameter in the qubit Hamiltonian is $f_\beta$,  
\begin{equation}
    \hat H_\textsc{qb}(f_\beta)= \hbar\Omega(f_\beta)\ket{1_{f_\beta}}\!\bra{1_{f_\beta}}.\label{eq:Hqubit}
\end{equation}
Here, we denoted as $\ket{0_{f_\beta}}$ and $\ket{1_{f_\beta}}$ the ground and first excited states, and as $\hbar \Omega(f_\beta)$ the qubit energy gap. For the qubit approximation it is important to mention that the energy levels do not cross between the different eigenvectors of the TC+FQ system. 

Notice that the reduction to two levels is not necessary to connect the UDW model to the TC+FQ system. Indeed UDW detectors with multiple levels are commonplace in the literature, but for convenience and simplicity of this analysis we will keep it as two levels.

\subsubsection{Taking the adiabatic approximation on the free qubit evolution}
\label{sssec:adiabatic}
Changing the parameter $f_\beta$ over time can induce transitions in the qubit and even cause the TC+FQ circuit of Figure~\ref{fig:tunableCoupler} to leave the qubit subspace. These transitions occur even if the circuit is not connected to the transmission line, and are a consequence of the fact that the energy eigenstates of the TC+FQ circuit change with $f_\beta$. To avoid this phenomenon, we assume that $f_\beta$ is tuned slowly enough to apply the adiabatic theorem to the free qubit evolution.

The adiabatic approximation implies that for a qubit state of the form
\begin{equation}
    \ket{\psi_\textsc{qb}(t)}=a_0(t) \ket{0_{t}} +a_1(t) \ket{1_{t}},
\end{equation}
where $\ket{0_t}$ and $\ket{1_t}$ are the ground and first excited state at time $t$ (which are determined by the choice of $f_\beta(t)$), coefficients remain the same under time evolution modulo a relative phase:
\begin{equation}
    a_0(t) =e^{\ii \theta_{0}(t)}a_0(0),\ a_1(t) =e^{-\ii\varphi(t)+\ii \theta_{1}(t)}a_1(0).\label{eq:adiabatic}
\end{equation} 
This is a good approximation for the free evolution of the qubit when the Hamiltonian $\hat H_\textsc{qb}(t)$ changes slowly enough and the energy gap is non-zero, i.e. $\hbar\Omega(t)\neq0$. The phases $\varphi(t)$ and $\theta_i(t)$ in Eq.~\eqref{eq:adiabatic} are given by
\begin{equation}
    \varphi(t)=\int_{0}^t \diff t' \Omega(t'),\ \theta_i(t)=\ii\int_{0}^{t}\diff {t'}  \braket{i_{t'}|\partial_{t'}|i_{t'}}\!,\ i \in \{0,1\}
\end{equation}
with $\theta_i(t)$ called the geometric phase. We choose to set \mbox{$\theta_0(t)=\theta_1(t)=0$}, which can always be done when the Hamiltonian depends on a single parameter, by appropriately changing the basis.\footnote{To impose \mbox{$\theta_0(t)=\theta_1(t)=0$}, we use that the geometric phase can be rewritten as $\theta_i(f_\beta)$ and absorb it into the definition of the $\ket{i_{f_\beta}}$.}

In summary, the adiabatic limit ensures that there are no transitions between energy levels due to the free dynamics, since their probabilities $p_i=|a_i|^2$ remain constant. Moreover, we can guarantee that the circuit stays in the qubit subspace during its free evolution by similarly applying the adiabatic theorem to the complete multilevel Hamiltonian given in Eq.~\eqref{eq:H_TC+FQ}. Therefore, after the adiabatic approximation, the circuit behaves exactly as a qubit with constant energy eigenbasis $\{\ket{0}$, $\ket{1}\}$ and time-dependent energy gap $\hbar\Omega(t)$. 

\subsubsection{Transversal coupling approximation}
\label{sssec:transversal_approximation}
The interaction Hamiltonian obtained in Eq.~\eqref{eq:Hint_gamma5} shows that the coupling occurs through $\hat{\gamma}_5$. Consider $\hat{\gamma}_5^{\textsc{qb}}$ to be the restriction of $\hat{\gamma}_5$ onto the qubit subspace. Then, $\hat{\gamma}_5^{\textsc{qb}}$ can be expressed as a linear combination of identity and Pauli operators,
\begin{equation}
    \hat{\gamma}_5^{\textsc{qb}} = \gamma_x \hat\sigma_x + \gamma_y \hat\sigma_y + \gamma_z \hat\sigma_z + \gamma_{id}\hat \openone,\label{eq:gamma5QB}
\end{equation}
where $\hat \sigma_z=\ket{0_{f_\beta}}\!\bra{0_{f_\beta}} - \ket{1_{f_\beta}}\!\bra{1_{f_\beta}}$.
Notice that, for simplicity of notation we have omitted writing the dependency of all terms on $f_\beta$.

In this paper we are going to assume that the longitudinal coupling $\gamma_z$ and the term $\gamma_{id}$ are zero. For the experimental implementation presented in section \ref{sec:SC_CouplerModel}, the longitudinal coupling $\gamma_z$ was shown to be negligible in \cite{Tunable_2023}.  Moreover, neglecting $\gamma_z$ and $\gamma_{id}$ does not affect entanglement harvesting at leading order in the coupling strength under the following sufficient conditions: 1) the qubits are prepared in an eigenstate of their free Hamiltonians, 2) we are under the adiabatic approximation for the free evolution of the qubits. A proof for this has recently been finished and will appear elsewhere~\cite{adamLongitudinalcoupling}.\footnote{This work in preparation~\cite{adamLongitudinalcoupling} also contains corrections to entanglement harvesting beyond leading order due to the longitudinal couplings, $\gamma_z$ and $\gamma_{id}$.} For the present work, condition 1) applies because we consider detectors to be initialized in their ground states. Additionally, condition 2) is an assumption that we already took in subsection \ref{sssec:adiabatic} in order to derive a variable gap detector model. Therefore, $\gamma_z$ and $\gamma_{id}$ do not affect the leading order contribution to entanglement computed in the next sections.

The transversal coupling coefficients $\gamma_x$ and $\gamma_y$ as a function of $f_\beta$ are
\begin{equation}
    \gamma_x = \Re \braket{1_{f_\beta}|\hat{\gamma}_5|0_{f_\beta}},\ \gamma_y = \Im \braket{1_{f_\beta}|\hat{\gamma}_5|0_{f_\beta}}.
\end{equation}
We obtain the states $\ket{0_{f_\beta}}$ and $\ket{1_{f_\beta}}$, up to an overall phase, by diagonalizing the Hamiltonian in Eq.~\eqref{eq:H_TC+FQ}. We fix this phase freedom by choosing the energy eigenfunctions to be real in the phase representation\footnote{In the phase representation, the states are represented by wavefunctions such as $\psi(\gamma_1,\gamma_2,\gamma_5,\gamma_6)$. For $i\in\{1,2,5,6\}$, the phase operators $\hat{\gamma}_i$ act as a multiplication by $\gamma_i$, and the number operators $\hat{N}_i$ act as the partial derivative $-\ii\partial_{\gamma_i}$.}. This can be done because \mbox{$\hat H_\textsc{tc+fq}\ket{i_{f_\beta}}=E_i(f_\beta) \ket{i_{f_\beta}}$} is a real differential equation in this representation. With this phase choice, the geometric phases automatically vanish, i.e. \mbox{$\braket{i_{f_\beta}|\partial_{f_\beta}|i_{f_\beta}}=0$}. To see why, notice that since the wavefunction of $\ket{i_{f_\beta}}$ is real, then $\braket{i_{f_\beta}|\partial_{f_\beta}|i_{f_\beta}}$ must be real. However, the normalization condition $\braket{i_{f_\beta}|i_{f_\beta}}=1$ implies that $\braket{i_{f_\beta}|\partial_{f_\beta}|i_{f_\beta}}$ must be imaginary. This only leaves the possibility that $\braket{i_{f_\beta}|\partial_{f_\beta}|i_{f_\beta}}=0$, which means that there is no geometric phase for our choice. 

Fixing the phase freedom as described above and using the parameters that match the design proposal of Ref.~\cite{Tunable_2023}\footnote{ Specifically, we use for the critical currents: $I_{c_1}=0.236\,\unit{\micro\ampere}$, $I_{c_2}=0.131\,\unit{\micro\ampere}$, \mbox{$I_{c_3}=0.236\,\unit{\micro\ampere}$}, $I_{c_4}=0.411\,\unit{\micro\ampere}$, $I_{c_5}=0.584\,\unit{\micro\ampere}$, $I_{c_6}=0.185\,\unit{\micro\ampere}$, and for the inverse capacitance matrix:
    $$
    C^{-1}=\left(
        \begin{array}{cccc}
        144. & -81.5 & 2.94 & 18.6 \\
        -81.5 & 196. & 6.18 & 24.8 \\
        2.94 & 6.18 & 46.1 & -32.4 \\
        18.6 & 24.8 & -32.4 & 87.7 \\
        \end{array}
    \right) (\unit{\pico\farad})^{-1}.$$
}, we obtain the following: for $f_\beta\in[0.3,0.5]$, $\gamma_x(f_\beta)\in[-0.02,0.22]$ and \mbox{$\gamma_y(f_\beta)=0$}, with $\gamma_x(f_\beta)$ an increasing function. Therefore, controlling $f_\beta$ tunes the transversal coupling strength $\gamma_x$. The reason for $\gamma_y(f_\beta)=0$ is that $\braket{1_{f_\beta}|\hat{\gamma}_5|0_{f_\beta}}$ is real. To see this, consider that in the phase representation, $\braket{1_{f_\beta}|\hat{\gamma}_5|0_{f_\beta}}$ is the integral of the product of three real functions. Specifically, the real wavefunctions of $\ket{0_{f_\beta}}$ and $\ket{1_{f_\beta}}$, and the real quantity $\gamma_5$.


\subsubsection{Approximated linear dependence of the gap on the instantaneous transversal coupling strength}
\begin{figure}[ht]
   \centering
   \includegraphics[width=0.9\columnwidth]{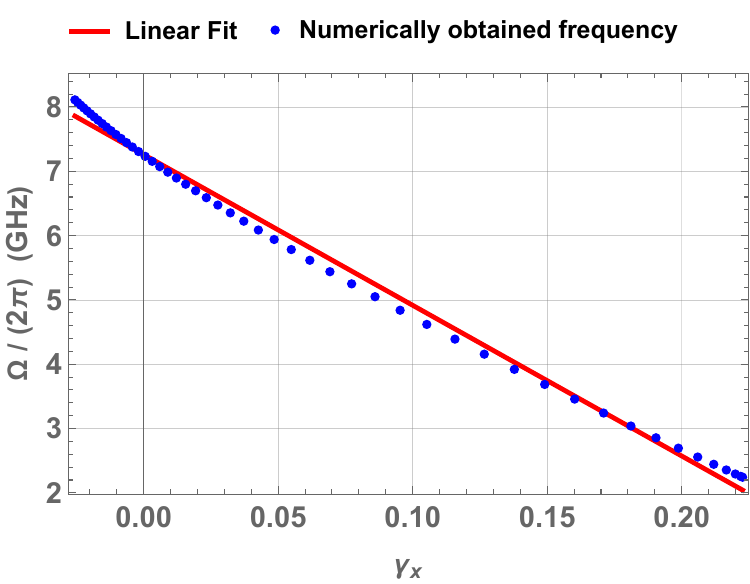}
    \caption{The blue dots show the superconducting qubit transition frequency $\Omega/2\pi$ for each transversal coupling strength $\gamma_x$. This frequency is computed from the numerical diagonalization of the Hamiltonian~\eqref{eq:H_TC+FQ}, with parameters chosen to match \cite{Tunable_2023}. The red line is the linear best fit. }
    \label{fig:gap_line_new}
\end{figure}

The energy gap of the qubit $\hbar\Omega$ varies with $f_\beta$, which can be rewritten as a dependence of $\Omega$ on the instantaneous transversal coupling strength $\gamma_x$. The exact form of this dependence is also obtained using the numerical diagonalization of the Hamiltonian of the tunable coupler + flux qubit. Using the best fit parameters found in \cite{Tunable_2023} for the Hamiltonian~\eqref{eq:H_TC+FQ} results in the data in Figure~\ref{fig:gap_line_new}.

The figure shows that the dependency $\Omega(\gamma_x)$ is approximately linear, which implies that
\begin{equation}
    \Omega(t) \approx \Omu{} + \Omv{} \chi(t).\label{eq:Omega_linear}
\end{equation}

Here, we chose $\gamma_x(t) =\gamma \chi(t)$, with $\chi(t)$ a switching function which takes values between zero and one, and $\gamma$ the maximum value of $\gamma_x(t)$ reached during the interaction. Moreover, $\hbar\Omu{}$ is the energy gap of the free qubit, or in other words, the energy gap when the interaction is turned off (when $\chi(t) = 0$). The product $\hbar\Omv{}$ determines the strength of the gap variation due to the coupling. Specifically, $\hbar\Omv{}$ is the increase in energy gap due to turning fully on the coupling. This is so because the gap changes to $\hbar(\Omu{} + \Omv{})$ when the coupling is fully on (when $\chi(t) = 1$). From Fig.~\ref{fig:gap_line_new}, we estimate
\begin{equation}
    \frac{\Omu{}}{2\pi}\approx 7.3 \,\unit{\giga\hertz},\ \frac{\Omv{}}{2\pi}\approx - 23\gamma \,\unit{\giga\hertz}. \label{eq:OmegaExp}
\end{equation}

Here, we focus on the linear dependence of the gap variation because it adequately models the presented superconducting implementation and is the simplest dependency model. While there is a wide range of other gap dependency models to be explored, doing so would go beyond the scope of this paper and is therefore left for future work.

\subsection{Result of the simplifications of the TC+FQ circuit model}
\label{ssec:variablegap_detector}
In summary, section \ref{ssec:approximations} provided the following approximations to simplify the TC+FQ circuit:
\begin{enumerate}
    \item The two-level approximation.
    \item The adiabatic approximation of the free qubit evolution.
    \item The transversal coupling assumption, \mbox{$\gamma_{id}=\gamma_z=0$}.
    \item The linear approximation of the dependence of $\Omega$ on $\gamma_x$.
\end{enumerate}
The resulting simplified model resembles a qubit particle detector with variable gap and spatial derivative coupling. This detector model has the following interaction picture interaction Hamiltonian,
\begin{align}
    \hat H_\textsc{int}(t) = - \frac{\varphi_0}{\ell_0}\gamma\chi(t)\hat{\mu}(t) \partial_x\hat{\Phi}_{\cutoff}(t,x_\textsc{d}).
     \label{eq:HintApprox}
\end{align}
Here, remember that $\varphi_0=\frac{\hbar}{2e}$ is the reduced magnetic flux quantum, $\ell_0$ the inductance per unit length of the transmission line, $\gamma$ the adimensional coupling strength, 
\mbox{$\chi(t)\in[0,1]$} the switching function, and $x_\textsc{d}$ the detector position. Moreover, $\hat{\Phi}_{\cutoff}$ is the field with cutoff of Eq.~\eqref{eq:TLfield_cutoff}, and $\hat\mu(t)$ is the monopole moment in the interaction picture, given by
\begin{equation}
    \hat\mu(t)= e^{\ii \varphi(t)}\hat\sigma^+ + \text{H.c.},\label{eq:monopoleNew}   
    \  \varphi(t)=\Omu{}t + \Omv{}\int_{0}^t \diff t' \chi(t'),
\end{equation}
with $\hbar\Omu{}$ the energy gap of the free qubit, $\hbar\Omv{}$ the strength of the energy gap variation due to the coupling, and \mbox{$\hat\sigma^+=\ketbra{1}{0}$}, $\hat\sigma^-=\ketbra{0}{1}$ the qubit ladder operators.

It is convenient to relate this detector model to the spin-boson model, commonly used to model superconducting qubits coupled to transmission lines, for comparison with other works. Appendix~\ref{apx:spin-boson} is devoted to showing this relation. When the interaction is fully switched on ($\chi(t)=1$), $\gamma$ relates to the dimensionless spin-boson coupling constant $\alpha$ as follows,
\begin{equation}
     \alpha=\frac{R_\textsc{k}}{8\pi^2 Z_0}\gamma^2\approx 6.54 \gamma^2,\label{eq:alpha_from_gamma}
\end{equation}
where $R_\textsc{k}=\frac{h}{e^2}$ is the von Klitzing constant, and we used $Z_0\approx50\,\unit{\ohm}$ (see Eq.~\eqref{eq:TLparameters}).

\subsection{Comparison of the variable gap detector to UDW detectors}
Even with the simplifications of the TC+FQ circuit in Subsection~\ref{ssec:variablegap_detector}, the resulting detector model (see Eq.~\eqref{eq:HintApprox}) has features that have been largely unexplored in the context of RQI. A very usual choice of interaction Hamiltonian for particle detectors in the RQI literature is the simple Unruh-DeWitt model, whose interaction with the field (generating time evolution with respect to the proper time of the detector centre-of-mass) in a 1+1 dimensional Minkowski spacetime is given by
\begin{equation}
    \hat H_\textsc{udw}(t) = \lambda \chi(t)\hat\mu_\textsc{udw}(t) \int \diff{x} F(x) \hat\phi(t,x)\,.
\end{equation}
Here, $\lambda$ is the coupling strength, $\chi(t)$ is the switching function, $F(x)$ is the smearing function, $\hat\phi(t,x)$ is the field amplitude operator, which admits the plane-wave expansion
\begin{align}
    \hat\phi(t,x) &= \int\frac{\diff k }{\sqrt{4\pi |k|}}(e^{\ii (\omega_k t-kx) } \ad{k} + \text{H.c.})\,.
\end{align}
The monopole operator is
\begin{equation}
    \hat\mu_\textsc{udw}(t)= e^{\ii \Omega t}\hat\sigma^+ + \text{H.c.},
\end{equation}
where the constant qubit energy gap is $\Omega$ and typically natural units $\hbar=c=1$ are chosen.

Notice the similarities with the detector that we provided in Eq.~\eqref{eq:HintApprox}, which therefore sits in between the full TC+FQ model of the experiment and the simpler UDW model used in RQI theory. The additional features of our proposed intermediate model are:
\begin{enumerate}[(i)]
    \item The gap $\Omega$ varies over time, with the variation being proportional to $\chi(t)$ (see Eq.~\eqref{eq:Omega_linear}).
    \item The detector couples to the spatial derivative of a 1D real massless scalar quantum field (see Eq.~\eqref{eq:HintApprox}).
    \item The detector is point-like, but the field has a cutoff (see Eq.~\eqref{eq:TLfield_cutoff}). This corresponds to effectively having a smearing that fulfills $\widetilde{F}(k)=\cutoff(\omega_k)$, as shown in Eq.~\eqref{eq:effSmearing}.
\end{enumerate}
These features will impact entanglement harvesting in future experiments that use the TC+FQ circuit design from \cite{Tunable_2023}, but they have not yet been studied in detail in the literature on entanglement harvesting. Doing precisely this is the object of the following sections.

\section{Entanglement Harvesting with detectors of variable gap}
\label{sec:time_evolution}

This section presents a common protocol to perform entanglement harvesting in RQI, with the modification that we use variable gap detectors with the features highlighted in Section~\ref{ssec:variablegap_detector} instead of the usual UDW detectors. We will refer to this model as VGSD detector (variable gap, spatial-derivative coupling). It is useful to write this section with the most generality possible in order to be able to compare with previously existing RQI literature. Therefore, we will work with an arbitrary coupling strength $\lambda$ and only make the substitutions related to the implementation in superconducting circuits in section \ref{sec:harvesting_results}. However, we will still keep full dimensional $\hbar$ and $c$ in the protocol presented below, to make the translation easier. 

Consider a massless scalar 1+1D field. The field amplitude operator, expanded in terms of plane waves and including a cutoff function, is\footnote{For convenience, we choose the convention of taking the field amplitude for 1+1D fields to be dimensionless, even after including $\hbar$ and $c$.}
\begin{align}
    \hat\phi_{\cutoff}(t,x) &= \int\diff k \frac{\cutoff(\omega_k)}{\sqrt{4\pi |k|}}(e^{\ii (\omega_k t-kx) } \ad{k} + \text{H.c.})\,. \label{eq:field_cutoff}
\end{align}
Here, $\omega_k=c|k|$, and $\cutoff(\omega_k)$ is the weight function that implements the cutoff.

In order to harvest entanglement, we couple two VGSD detectors to the field, labeled by $\nu\in \{A,B\}$ according to the following interaction Hamiltonian in the interaction picture,
\begin{align}
    \hat H_\text{I}(t)&=\hbar c \sum_{\nu}\lambda_{\nu}\chi_{\nu}(t)\hat\mu_{\nu}(t)\partial_x\hat\phi_{\cutoff}(t,x_{\nu}).\label{eq:Hint}
\end{align}
For the detector $\nu$, $\lambda_{\nu}$ is the coupling strength, $\chi_\nu(t)$ is the  switching function, $x_{\nu}$ is the detector position\footnote{Recall that even though we consider the detector to be pointlike, the spatial smearing of the detector can just be reabsorbed in the cutoff function. See~\cite{McKay_2017}.}, and $\hat\mu_{\nu}(t)$ is the monopole operator in the interaction picture,
\begin{align}
    &\hat\mu_\nu(t)= e^{\ii \varphi_\nu(t)}\hat\sigma_\nu^+ + \text{H.c.},\nonumber\\ 
    &\varphi_\nu(t)= \int_{0}^t \diff t' \Omega_\nu(t'),\label{eq:monopole_vargap_nu}  
\end{align}
with \mbox{$\hat\sigma^+_\nu=\ketbra{1_\nu}{0_\nu}$}, $\hat\sigma^-_\nu=\ketbra{0_\nu}{1_\nu}$ the qubit ladder operators and $\hbar\Omega_\nu(t)$ the variable energy gap.

Using the model above, we compute the final state $\hat\rho$. Given an initial state $\hat\rho_0$, 
\begin{equation}
    \hat\rho = \hat U \hat\rho_0 \hat U^\dagger,\quad \hat U = \mathcal T e^{-\ii\int\diff{t}\hat H_I(t)}, \label{eq:time_ordered_evolution}
\end{equation}
with the $\mathcal T$ in the second term is there to denote the time-ordered exponential. We will assume that $\lambda=\lambda_\textsc{a}=\lambda_\textsc{b}$ for simplicity and perform a Dyson expansion of $\hat U$ in $\lambda$,
\begin{equation}
    \hat U = \hat U^{(0)}+\hat U^{(1)}+\hat U^{(2)}+\mathcal{O}(\lambda^3), 
\end{equation}
\begin{align}
    &\hat U^{(0)} = \hat\openone, \nonumber\\
    &\hat U^{(1)} =-\frac{\ii}{\hbar}\int\diff {t} \hat H_I(t), \nonumber\\
    &\hat U^{(2)} =-\frac{1}{\hbar^2} \int_{t>t'}\diff {t} \diff {t'} \hat H_I(t)\hat H_I(t').
    \label{eq:DysonSeries}
\end{align}
Then, the time-evolved density operator for the state of the detectors and the field is
\begin{equation}
    \hat \rho = \hat\rho_0+\hat \rho^{(1,0)}+\hat \rho^{(0,1)}+\hat \rho^{(2,0)}+\hat \rho^{(1,1)}+\hat \rho^{(0,2)}+\mathcal{O}(\lambda^3)\,,
\end{equation}
where the different corrections are
\begin{equation}
    \hat \rho^{(i,j)}=\hat U^{(i)}\hat\rho_0\hat U^{(j)\dagger}\,.
\end{equation}

For our study, let us assume that the field and detectors start uncorrelated, 
\begin{equation}
    \hat\rho_0 = \hat\rho_{\textsc{ab},0}\otimes\hat\rho_{\phi,0}\,.
\end{equation}
We also assume that the initial field states $\hat\rho_{\phi,0}$ have vanishing odd-point functions, that is, \mbox{$\forall n\in \{0,1,2,\ldots\}$},
\begin{align}
&\Tr(\hat\rho_{\phi,0}
\hat \phi(t_0,x_0)\ldots \hat \phi(t_{2n+1},x_{2n+1}))=0.
\end{align}
For example, that is the case for the field vacuum, and any Fock states as well as any zero mean Gaussian state such as thermal states, squeezed vacuum, etc. With this assumption, the odd order corrections cancel and the final state for the two detectors (after tracing over the field) becomes
\begin{equation}
    \hat \rho_{\textsc{ab}} = \Tr_\phi  \hat \rho=\hat\rho_{\textsc{ab},0}+\hat \rho^{(2,0)}_{\textsc{ab}}+\hat \rho^{(1,1)}_{\textsc{ab}}+\hat \rho^{(0,2)}_{\textsc{ab}}+\mathcal{O}(\lambda^4)\,.\label{eq:GenericFinalStateDetector}
\end{equation}
Finally, we assume that the \textit{detectors start from the ground state} 
\begin{equation}
    \hat\rho_\textsc{ab,0}=\ketbra{0_\textsc{a}0_\textsc{b}}{0_\textsc{a}0_\textsc{b}}.
\end{equation}
The final state has the usual form of the final state in a vacuum entanglement harvesting setting (see, e.g., \cite{Pozas2015}). The details of the calculation are also included in Appendix \ref{apx:final_state}. The state of the detectors after the interaction, represented in the basis $\ket{0_\textsc{a}0_\textsc{b}}$, $\ket{1_\textsc{a}0_\textsc{b}}$, $\ket{0_\textsc{a}1_\textsc{b}}$, $\ket{1_\textsc{a}1_\textsc{b}}$, is
\begin{equation}
    \hat\rho_\textsc{ab} = \begin{pmatrix}
    1-\mathcal L_{\textsc{a}\textsc{a}}-\mathcal L_{\textsc{b}\textsc{b}} & 0 & 0 & \mathcal{M}^*\\
    0 & \mathcal L_{\textsc{a}\textsc{a}} & \mathcal L_{\textsc{a}\textsc{b}} & 0 \\
    0 & \mathcal L_{\textsc{b}\textsc{a}} & \mathcal L_{\textsc{b}\textsc{b}} & 0 \\
    \mathcal{M} & 0 & 0 & 0 \\
    \end{pmatrix} + \mathcal{O}(\lambda^4),\label{eq:rho_qubit}
\end{equation}
with 
\begin{align}
    &\mathcal L_{\mu\nu}= c^2\lambda^2\int\diff{t} \diff{t'} \mathcal W_{xx'}(t',x_{\nu},t,x_{\mu})\chi_{\cpx_\mu}(t)\chi^*_{\cpx_\nu}(t'),
    \nonumber \\
    &\mathcal M=-c^2\lambda^2\int\diff{t} \diff{t'} G_{xx'}(t,x_{\textsc{a}},t',x_{\textsc{b}})\chi_{\cpx_\textsc{a}}(t) \chi_{\cpx_\textsc{b}}(t'),\nonumber\\
    &\chi_{\cpx_\nu}(t)=
    e^{\ii\varphi_\nu(t)}\chi_{\nu}(t).\label{eq:LMGeneral_complexchi}
\end{align}
The phase $\varphi_\nu(t)$ is given in Eq.~\eqref{eq:monopole_vargap_nu} and
\begin{align}
    &\mathcal W_{xx'}(t,x,t',x') =  \langle\partial_x\hat\phi_{\cutoff}(t,x)\partial_{x'}\hat\phi_{\cutoff}(t',x')\rangle_{\hat\rho_{\phi,0}},\nonumber\\
    &G_{xx'}(t,x,t',x') = \Heaviside(t-t')\mathcal W_{xx'}(t,x,t',x')\nonumber\\&\qquad\qquad\qquad\quad\ +\Heaviside(t'-t)\mathcal  W_{xx'}(t',x',t,x)\,, \label{eq:WGdef}
\end{align}
where $\Heaviside$ is the Heaviside step function.

After the interaction the detectors will be generically entangled. As a measure of the entanglement acquired by the detectors we will use the negativity. This quantity is a faithful entanglement monotone for systems of two qubits~\cite{VidalNegativity,Plenio2005LogNeg}.
The negativity of $\hat \rho_\textsc{ab}$ amounts to
\begin{align}
    &\mathcal{N} = \max(\eta,0)+ \mathcal{O}(\lambda^4),\nonumber\\
    &\eta = \sqrt{|\mathcal{M}|^2 - \frac{(\mathcal L_{\textsc{a}\textsc{a}}-\mathcal L_{\textsc{b}\textsc{b}})^2}{2}} - \frac{\mathcal L_{\textsc{a}\textsc{a}}+\mathcal L_{\textsc{b}\textsc{b}}}{2}.\label{eq:negativity}
\end{align}
This expression simplifies, for $\mathcal L=\mathcal L_\textsc{aa}=\mathcal L_\textsc{bb}$, to
\begin{equation}
    \mathcal{N} = \max(|\mathcal{M}|-\mathcal{L},0)+ \mathcal{O}(\lambda^4).
\end{equation} 

\subsection{Harvesting from the vacuum}
Let us consider that the quantum field is prepared in the vacuum state, given by
\begin{equation}
    \hat\rho_{\phi,0}=\ketbra{0_\phi}{0_\phi}.
\end{equation}
For the vacuum, the correlators defined in Eq.~\eqref{eq:WGdef} only depend on $t_-= t-t'$ and $x_-= x-x'$:
\begin{align}
    &\mathcal W^{\text{vac}}_{xx'} (t_-,x_-) = \frac{1}{2\pi c^2}\int_{0}^\infty\!\diff \omega \omega \cutoff(\omega)^2 \cos\left(\omega \frac{x_-}{c}\right)e^{-\ii\omega t_-}
    \,,\nonumber\\
    &G^{\text{vac}}_{xx'} (t_-,x_-) = \mathcal W^{\text{vac}}_{xx'} (|t_-|,x_-). \label{eq:WG_vac}
\end{align}
Next, we substitute $\mathcal W^{\text{vac}}_{xx'}$ and $G^{\text{vac}}_{xx'}$ back into $\mathcal L_{\mu\nu}$ and $\mathcal M$ and provide two ways to simplify the resulting integrals, each of them helpful in different scenarios.

\subsubsection{Integrating over the field modes last}
Substituting Eq.~\eqref{eq:WG_vac} into Eq.~\eqref{eq:LMGeneral_complexchi} leads to
\begin{align}
    &\mathcal L_{\nu\nu}= \frac{\lambda^2}{2\pi} \int_{0}^\infty\!\diff \omega \omega \cutoff(\omega)^2 |\widetilde{\chi_{\cpx_\nu}}(\omega)|^2,\nonumber\\
    &\mathcal L_\textsc{ab}= \frac{\lambda^2}{2\pi}\int_{0}^\infty\!\diff \omega \omega \cutoff(\omega)^2 \cos(\omega t_d)\widetilde{\chi_{\cpx_\textsc{a}}}(\omega)\widetilde{\chi_{\cpx_\textsc{b}}}(\omega)^*,\nonumber\\
    &\mathcal M=-\frac{\lambda^2}{2\pi}\int_{0}^\infty\!\diff \omega\omega \cutoff(\omega)^2\cos(\omega t_d)
    \nonumber\\&\qquad\quad\cross\int\diff{t} \diff{t'}e^{-\ii\omega|t-t'|}\chi_{\cpx_\textsc{a}}(t) \chi_{\cpx_\textsc{b}}(t')\label{eq:LMfinal}.
\end{align}
Here, we used the Fourier transform convention stated in Eq.~\eqref{eq:fourier_convention}. Notice that the expression for $\mathcal L_\textsc{ba}$ follows from $\mathcal L_\textsc{ba}=\mathcal L_\textsc{ab}^*$. Moreover, $t_d$ stands for the time it takes for the detectors to send a signal to each other through the field,
\begin{equation}
    t_d=\frac{|x_\textsc{a}-x_\textsc{b}|}{c}.
\end{equation}

The expressions for $\mathcal L_{\mu\nu}$ and $\mathcal M$ are similar to the ones obtained for UDW detectors in (3+1)D that couple to the field amplitude (see, e.g.~\cite{Pozas2015}). Specifically, one can modify the $\mathcal L_{\mu\nu}$ and $\mathcal M$ in Eq.~\eqref{eq:LMfinal} to get the version for UDW detectors in (3+1)D with amplitude coupling and radially symmetric smearing functions\footnote{To obtain $\mathcal L_{\mu\nu}$ and $\mathcal M$ for the 3+1D amplitude coupling UDW model, the modifications of Eqs.~\eqref{eq:LMfinal} are as follows: 1) replace all $\cos(\omega t_d)$ by $\sin(\omega t_d)/(\omega t_d)$, 2) fix $\hbar\Omega_\nu(t)=\hbar\Omega_\nu$, which implies \mbox{$\chi_{\cpx_\nu}(t)=e^{\ii\Omega_\nu t}\chi_{\nu}(t)$}, 3) replace $\cutoff(\omega)$ by $|\widetilde{F}(\bm{k})|$ (with the Fourier transform convention of Eq.~\eqref{eq:fourier_convention}) evaluated at a radius $|\bm{k}|=\omega/c$, 4) change the overall constants to account for the different model and geometry. For example, for the setup of \cite{Pozas2015}, the constants in front of the equations in~\eqref{eq:LMfinal} would be $\frac{\lambda^2}{4\pi^2}$ instead of $\frac{\lambda^2}{2\pi}$.} $F(\bm{x})$. 

\subsubsection{Integrating over the field modes first}\label{sec:ModesFirst}
An alternative to Eq.~\eqref{eq:LMfinal} and a quite helpful way of evaluating these integrals is to integrate over $\omega$ before performing the time integrals. Doing this, $\mathcal W^{\text{vac}}_{xx'}$ becomes
\begin{align}
    &\mathcal W^{\text{vac}}_{xx'} (t_-,x_-) = \frac{\Jmodes \Big(t_- +\frac{|x_-|}{c}\Big)\!+\Jmodes\Big(t_- -\frac{|x_-|}{c}\Big)}{4\pi c^2},\nonumber\\
    &\Jmodes(t)=\int_{0}^\infty\!\diff \omega\omega \cutoff(\omega)^2 e^{-\ii\omega  t} \label{eq:defJ}. 
\end{align}
For the exponential cutoff of Eq.~\eqref{eq:defCutoff}, $\Jmodes(t)$ has the following analytical expression,
\begin{equation}
    \Jmodes(t)=\frac{\Omcut^2}{(1+\ii \Omcut t)^{2}}.\label{eq:J_expcutoff}
\end{equation}
Substituting Eq.~\eqref{eq:defJ} back into Eq.~\eqref{eq:LMGeneral_complexchi},
\begin{align}
    &\mathcal L_{\nu\nu}=\frac{\lambda^2}{2\pi}\int\diff{t} \diff{t'} \Re\bigl(\Jmodes(t'-t)\chi_{\cpx_\nu}(t)\chi^*_{\cpx_\nu}(t')\bigr),\nonumber\\
    &\mathcal L_{\textsc{ab}}=\frac{\lambda^2}{2\pi}\int\diff{t} \diff{t'}\Imodes(t'-t) \chi_{\cpx_\textsc{a}}(t)\chi_{\cpx_\textsc{b}}^*(t'),\nonumber\\
    &\mathcal M=-\frac{\lambda^2}{2\pi}\int\diff{t} \diff{t'}\Imodes(|t-t'|)\chi_{\cpx_\textsc{a}}(t) \chi_{\cpx_\textsc{b}}(t'),\nonumber\\
    &\Imodes(t)=\frac{1}{2}\big(\Jmodes(t+t_d)+\Jmodes(t-t_d)\big),\label{eq:ModesFirstLM}
\end{align}
where we used $\Jmodes(-t)=\Jmodes^*(t)$ and $t_d=|x_\textsc{a}-x_\textsc{b}|/c$. 

Further simplifications of the $\mathcal L_{\mu\nu}$ and $\mathcal M$ integrals, used to ease the numerical calculations in the next sections, are explored in Appendix~\ref{apx:simplifyLM}.

\section{Genuine entanglement harvesting}
\label{sec:genuine_harvesting}

When two detectors are in causal contact they can acquire entanglement in two different ways: they can communicate by exchanging information via the field, but they can also harvest entanglement at the same time. This section outlines how to decompose the entanglement acquired by causally connected detectors into two components: (a) entanglement extracted from pre-existing field correlations and (b) entanglement mediated by communication through the field, following \cite{Erickson2021_When,adam2024derivative}. For simplicity in the equations, we denote spacetime points by $\mathsf{x}$ in this section.

The decomposition of the acquired entanglement is based on the two-point correlator of the field amplitude $\hat\phi$, which for a field state $\hat{\rho}_{\phi}$ is
\begin{equation}
    W(\mathsf{x},\mathsf{x}') = \langle \hat\phi(\mathsf{x}) \hat\phi(\mathsf{x}') \rangle_{\hat{\rho}_{\phi}}.
\end{equation}
This correlator can be split into symmetric and antisymmetric parts
\begin{equation}
W^{\pm}(\mathsf{x},\mathsf{x}') = \frac{W(\mathsf{x},\mathsf{x}') \pm W(\mathsf{x}', \mathsf{x})}{2}.
\end{equation}
$ W^+ $ and $W^-$ respectively are the real and imaginary parts of $W$, due to $ W(\mathsf{x}',\mathsf{x}) = W^*(\mathsf{x}, \mathsf{x}') $. Moreover,
\begin{align}
&W^+(\mathsf{x}, \mathsf{x}') = \frac{1}{2} \langle \{ \hat\phi(\mathsf{x}), \hat\phi(\mathsf{x}') \} \rangle_{\hat{\rho}_{\phi}}, \nonumber\\
&W^-(\mathsf{x}, \mathsf{x}') = \frac{1}{2} \langle [\hat\phi(\mathsf{x}), \hat\phi(\mathsf{x}')] \rangle_{\hat{\rho}_{\phi}}.
\end{align}

The contribution of $ W^+ $ to the acquired entanglement can be associated with genuine harvesting, because of the following reasons, given in \cite{Erickson2021_When}:
\begin{enumerate}
    \item The expectation value of $ [\hat\phi(\mathsf{x}), \hat\phi(\mathsf{x}')] $ does not depend on the field state, while the expectation value of $ \{ \hat\phi(\mathsf{x}), \hat\phi(\mathsf{x}') \} $ does. Therefore, $W^-$ is not affected by the amount of pre-existing entanglement in the field, while $W^+$ is.
    \item $W^+$ does not participate in communication at leading order, which is instead mediated by $W^-$~\cite{cavitySignaling2014,infoWOenergy2015,detectorsSignaling2015,Pipo2023Signaling}. Furthermore, even non-perturbatively, detectors cannot communicate by coupling to commuting field observables in the interaction picture (see, e.g. Appendix of \cite{adam2024derivative}). Since communication cannot occur without $W^-$, the entanglement mediated solely by $W^+$ cannot be associated to communication and hence we conclude it can quantify genuine harvested entanglement from the field.
    \item The commutator $[\hat\phi(\mathsf{x}), \hat\phi(\mathsf{x}')]$ is proportional to the difference between the retarded and advanced Green’s functions (the classical causal propagator~\cite{Tales2023quantumFreedom}). The causal propagator, and thus $ W^- $, vanishes outside the light cone. On the other hand, the field anticommutator has support even for spacelike separated events, meaning that only $ W^+ $ contributes to spacelike entanglement harvesting.
\end{enumerate}
Following~\cite{Erickson2021_When} and based on the arguments above, we decompose the term $ \mathcal{M} $ in Eq.~\eqref{eq:negativity} into two parts associated to pre-existing field correlations ($ \mathcal{M}^+ $) and communication ($ \mathcal{M}^- $),
\begin{equation}
\mathcal{M} = \mathcal{M}^+ + \mathcal{M}^-,
\end{equation}
where $ \mathcal{M}^{\pm} $ contains only the contribution of $ W^{\pm} $. As an estimator of the amount of correlations that are genuinely harvested, we use the following ratio,
\begin{equation}
    \frac{|\mathcal{M}^+|}{|\mathcal{M}^+| + |\mathcal{M}^-|}, \label{eq:harvesting_estimator}
\end{equation}  
which ranges from 0 to 1.

In this article, we explore entanglement harvesting with the variable gap detector models of Eq.~\eqref{eq:Hint}. These detectors do not couple to $\hat\phi$, but rather to $\partial_x\hat\phi_{\cutoff}$, which incorporates the cutoff, see Eq.~\eqref{eq:field_cutoff}. Remember that the cutoff is equivalent to having a smeared field, as shown in Eq.~\eqref{eq:smeared_field}. Nonetheless, the analysis above to separate the communication and genuine harvesting contributions still carries over, see~\cite{adam2024derivative}.

\section{Exploring the effects of the implementation features on entanglement harvesting}
\label{sec:harvesting_results}

In this section, we study entanglement harvesting using the detectors of Eq.~\eqref{eq:HintApprox}, motivated by the experimental implementation. As shown in Section \ref{sec:circuit_to_detector}, these particular VGSD detectors are designed to emulate the TC+FQ superconducting circuit implementation demonstrated in~\cite{Tunable_2023}. Our goal is to evaluate the effect of the combined variable gap and derivative coupling features on the ability of the detectors to harvest entanglement. To do so, we will numerically compute the amount of entanglement acquired by VGSD detectors, using the results of Section~\ref{sec:time_evolution}. We will do so in a variety of experimentally accessible scenarios, and estimate how much of the entanglement is actually harvested from the field (see  Section~\ref{sec:genuine_harvesting}). 

First of all, we apply the calculations of Section~\ref{sec:time_evolution} to the VGSD detector of Eq.~\eqref{eq:HintApprox}. This amounts to substituting the speed of light $c$, the coupling strength $\lambda_\nu$, and the time dependency of the energy gap $\hbar\Omega_\nu(t)$ by expressions that match the superconducting implementation.

Firstly, the speed of light in the transmission line is not $c$, but rather a function of its capacitance $c_0$ and inductance $\ell_0$ per unit length: \mbox{$v=1/\sqrt{c_0\ell_0}$}. We recall that common values are of the order of $v\approx 0.3\,c$. This modification affects the definition of $t_d$, which now becomes the time required for signals to travel between the detectors through the transmission line,
\begin{equation}
    t_d=\frac{|x_\textsc{a}-x_\textsc{b}|}{v}.
\end{equation}

Secondly, to determine $\lambda_\nu$, we need to relate it with $\gamma_\nu$. To do so, we compare the particular interaction Hamiltonian of Eq.~\eqref{eq:HintApprox} with the general VGSD interaction Hamiltonian of Eq.~\eqref{eq:Hint}. The models match when
\begin{equation}
    \lambda_\nu = -\frac{\varphi_0}{\hbar v\ell_0}\gamma_\nu\sqrt{\hbar Z_0} = -
    \sqrt{\frac{R_\textsc{k}}{8\pi Z_0}} \gamma_\nu \approx -4.53  \gamma_\nu,\label{eq:lambda_from_gamma}
\end{equation}
where we used $\varphi_0=\hbar/(2e)$, $Z_0=v \ell_0$, and $R_\textsc{k}=h/e^2$. The numerical value comes from choosing $Z_0\approx50\,\unit{\ohm}$, as in Eq.~\eqref{eq:TLparameters}.
Notably, $\lambda_\nu^2 = \pi\alpha_\nu$, where $\alpha_\nu$ is the dimensionless spin-boson coupling constant of Eq.~\eqref{eq:alpha_from_gamma}, for the  detector $\nu$ (where we recall $\nu\in\{\text{A},\text{B}\}$).

Finally, we link the dependency $\Omega_\nu(t)$ to the switching on and off of the interactions. Specifically, we use the following simple linear dependence that matches well with the superconducting implementation (see Eq.~\eqref{eq:Omega_linear}), 
\begin{equation}
    \Omega_\nu(t) =  \Omu{\nu} + \Omv{\nu}\chi_\nu(t).
\end{equation}
Therefore,
\begin{equation}
    \varphi_\nu(t)= \Omu{\nu}t + \Omv{\nu} \int_{0}^t \diff t' \chi_\nu(t').
\end{equation}

Throughout the present section, the negativity $\mathcal N$ and the harvesting estimator, $|\mathcal M^+|/\big(|\mathcal M^+|+|\mathcal M^-|\big)$ are computed by numerically integrating the simplified expressions given in Appendix \ref{apx:decoupled_timeIntegrals}. For the numerical integration we use Mathematica's \texttt{NIntegrate} function default strategy, which is globally adaptive subdivision combined with local quadrature rules (Gauss–Kronrod rules). The precision of the numerical result is controlled using \texttt{NIntegrate} parameters \texttt{PrecisionGoal} and \texttt{WorkingPrecision}. These parameters are adjusted differently for each type of integral to ensure numerical convergence of the result. The precise values for these \texttt{NIntegrate} parameters are given in Appendix~\ref{apx:decoupled_timeIntegrals}.

\subsection{Fixed vs tunable parameters}
\label{ssec:parameter_choices}
Here, we split the parameters of the particular VGSD model of Eq.~\eqref{eq:HintApprox} into two categories: \emph{fixed parameters} that cannot be changed once the superconducting device is fabricated and \emph{tunable parameters} that can be freely changed without requiring any redesign or fabrication of a new device. For simplicity, let us consider that the two detectors are equal: \mbox{$\lambda=\lambda_\textsc{a}=\lambda_\textsc{b}$}, $\Omu{}=\Omu{\textsc{a}}=\Omu{\textsc{b}}$, $\Omv{}=\Omv{\textsc{a}}=\Omv{\textsc{b}}$.

We give the following \emph{fixed parameters} experimentally realistic values, which best mimic the implementation demonstrated in~\cite{Tunable_2023},
\begin{itemize}
    \item $\Omu{}/(2\pi)\approx 7.3 \,\unit{\giga\hertz}$ for the transition frequency of the free qubit (see Eq.~\eqref{eq:OmegaExp}). 
    \item \mbox{$\Omv{}/(2\pi)\approx 5.2 \lambda \,\unit{\giga\hertz}$} for the difference between the transition frequency of the qubit when the coupling is fully switched on and $\Omu{}/(2\pi)$ (see Eqs.~\eqref{eq:OmegaExp} and~\eqref{eq:lambda_from_gamma}). 
    \item $\Omcut{}/(2\pi)\approx 50 \,\unit{\giga\hertz}$ for the transmission line cutoff frequency, matching \cite{Forn-Diaz2017}.
    \item $Z_0\approx 50\,\unit{\ohm}$ for the transmission line impedance, matching \cite{Forn-Diaz2017}.
    \item $v\approx 1.2\times 10^8 \,\unit{\meter\per\second}$, as measured in \cite{Goppl2008} for a typical superconducting coplanar waveguide.
    \item \mbox{$t_d=\frac{d}{v}=1\,\unit{\nano\second}$} for the time the detectors take to signal to each other. Equivalently, the detectors are separated by \mbox{$d=12\,\unit{\centi\meter}$}, which is feasible with the current fabrication methods~\cite{Noah2022}.
\end{itemize}
These fixed parameters will be kept the same for all the exploration, except for $t_d$. Testing different $t_d$ can advise the choice of distance between detectors in future entanglement harvesting experiments.

The following \emph{tunable parameters} will be varied during the following numerical exploration, to understand their effects on entanglement harvesting,
\begin{itemize}
    \item The coupling strength $\lambda$. We test the $\lambda$ shown in Table~\ref{tb:coupling_strength_scenarios}.  These values are in the achievable range $\lambda\in[-1,0.1]$. This range is obtained from combining Eq.~\eqref{eq:lambda_from_gamma} with the range $\gamma\in[-0.02,0.22]$ obtained from the numerical simulations outlined in subsection~\ref{sssec:transversal_approximation}. 
    \item The switching function $\chi_\nu(t)$. Our choices of switching functions are detailed next, in subsection \ref{ssec:switching_shapes}. 
\end{itemize}

\begin{table}[ht]
\centering
\begin{tabular}{SSSSS} \toprule
    {Scenario} & {$\ \frac{\Omv{}}{2\pi} \,(\unit{\giga\hertz})$} & {$\lambda$} & {$\gamma$} & {$\alpha$}   \\ \midrule
    1  & {$\to 0$} & {$\to 0$} & {$\to 0$} & {$\to 0$} \\
    2  & -0.5  & -0.1  & 0.02  & 0.003  \\
    3  & -1.6  & -0.3  & 0.07  & 0.03 \\
    4  & -3.4  & -0.65 & 0.14  & 0.1   \\
    5  & -5.2  & -1    & 0.22  & 0.3    \\ 
    6  & 0.5   & 0.1   & -0.02 & 0.003  \\
    \bottomrule
\end{tabular}
\caption{This table depicts the scenarios of qubit transition frequency variation $\Omv{}$ that will be explored in the present section. The $\lambda$, $\gamma$ and $\alpha$ quantify the coupling strength that produces $\Omv{}$ according to the model of Eq.~\eqref{eq:OmegaExp}, which best fits the superconducting implementation. See also Eqs.~\eqref{eq:alpha_from_gamma}, \eqref{eq:lambda_from_gamma}, which tie together $\lambda$, $\gamma$ and $\alpha$. The $\to 0$ corresponds to the weak coupling limit, where the frequency variation vanishes. Despite the large value of some of the coupling strengths, perturbation theory is still applicable in some regime of parameters. The validity of the perturbative approximation for these scenarios is discussed in Appendix \ref{apx:fourth_order}. For all plotted perturbative quantities below, we remove the dependency upon $\lambda$ by adequately dividing by a power of $\lambda$. This guarantees that the presented results for each scenario only depend on $\Omv{}$ and not on $\lambda$.}
\label{tb:coupling_strength_scenarios}
\end{table}

\subsection{Explored switching functions}
\label{ssec:switching_shapes}

\begin{figure*}[ht]
    \centering
    \includegraphics[width=0.95\textwidth]{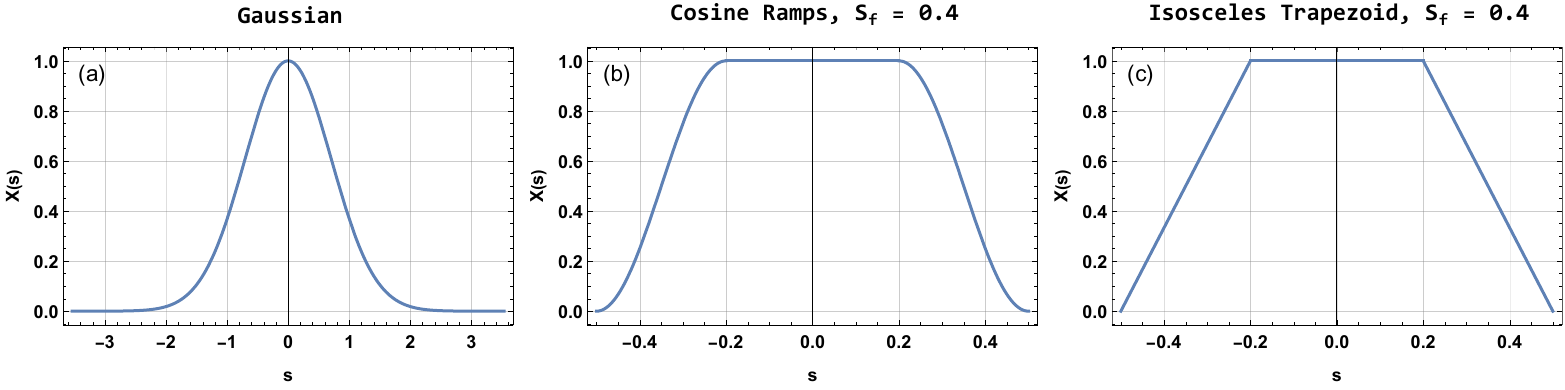}
    \caption{(a)~Gaussian switching shape, with 5 standard deviations shown. (b)~Cosine ramps switching shape with $S_f=0.4$. (c)~Isosceles trapezoid switching shape with $S_f=0.4$. }
    \label{fig:switchings}
\end{figure*}

To simplify our exploration, we will choose switching functions $\chi_\nu(t)$ such that
\begin{equation}
    \chi_\nu(t)=\aswitch\biggl(\frac{t-t_\nu}{\Ti}\biggr),
\end{equation}
where:
\begin{itemize}
    \item $\aswitch(s)$ is a single shape for the switching function of both detectors, which fulfills $\aswitch(s)=\aswitch(-s)$.
    \item $\Ti$ is the duration or time-scale of the interaction.
    \item $t_\nu$ controls the time at which the detector $\nu$ is switched on and off. In our setup, only the delay $t_\Delta = t_\textsc{b}-t_\textsc{a}$ affects the entanglement acquired by the detectors.
\end{itemize}

The explored switching function shapes $\aswitch(s)$ are provided next.

\subsubsection{Gaussian Switching Shape}
The Gaussian switching is common in the literature of entanglement harvesting and we explore it to ease comparisons with established results. Its shape is 
\begin{align}
    \aswitch(\atc)= e^{-s^2},
\end{align}
plotted in Figure~\ref{fig:switchings}(a).

\subsubsection{Cosine Ramps Switching Shape}
The cosine ramps switching function lasts for a finite time and has a continuous derivative. Its shape is
\begin{align}
    &\aswitch(\atc)=\begin{cases}
    1 & |s|\leq \frac{\aTf}{2}, \\
    \left(\cos \left(\frac{\pi}{2}\frac{2 |s|-\aTf}{1-\aTf}\right)\right)^2 &
   \frac{\aTf}{2}<|s|<\frac{1}{2}, \\
     0 & \frac{1}{2}\leq |s|,
\end{cases}\label{eq:cosine_switching}
\end{align}
with an example plotted in Figure~\ref{fig:switchings}(b). $S_f$ is the portion of time that the interaction remains at its maximum. 

\subsubsection{Isosceles Trapezoid Switching Shape}
The isosceles trapezoid switching function lasts a finite time and is continuous. Its shape is
\begin{align}
    &\aswitch(\atc)=\begin{cases}
     1 & |s|\leq \frac{\aTf}{2},\\
     \frac{1-2 |s|}{1-\aTf} & \frac{\aTf}{2}<|s|<\frac{1}{2},\\
     0 & \frac{1}{2}\leq |s|,
    \end{cases}\label{eq:trapezoid_switching}
\end{align}
with an example plotted in Figure~\ref{fig:switchings}(c). $S_f$ is the portion of time that the interaction remains at its maximum.

\subsection{Effect of duration and delay}
\label{ssec:exploration_duration_delay}
\begin{figure*}[p]
    \centering
    \includegraphics[width=.96\textwidth]{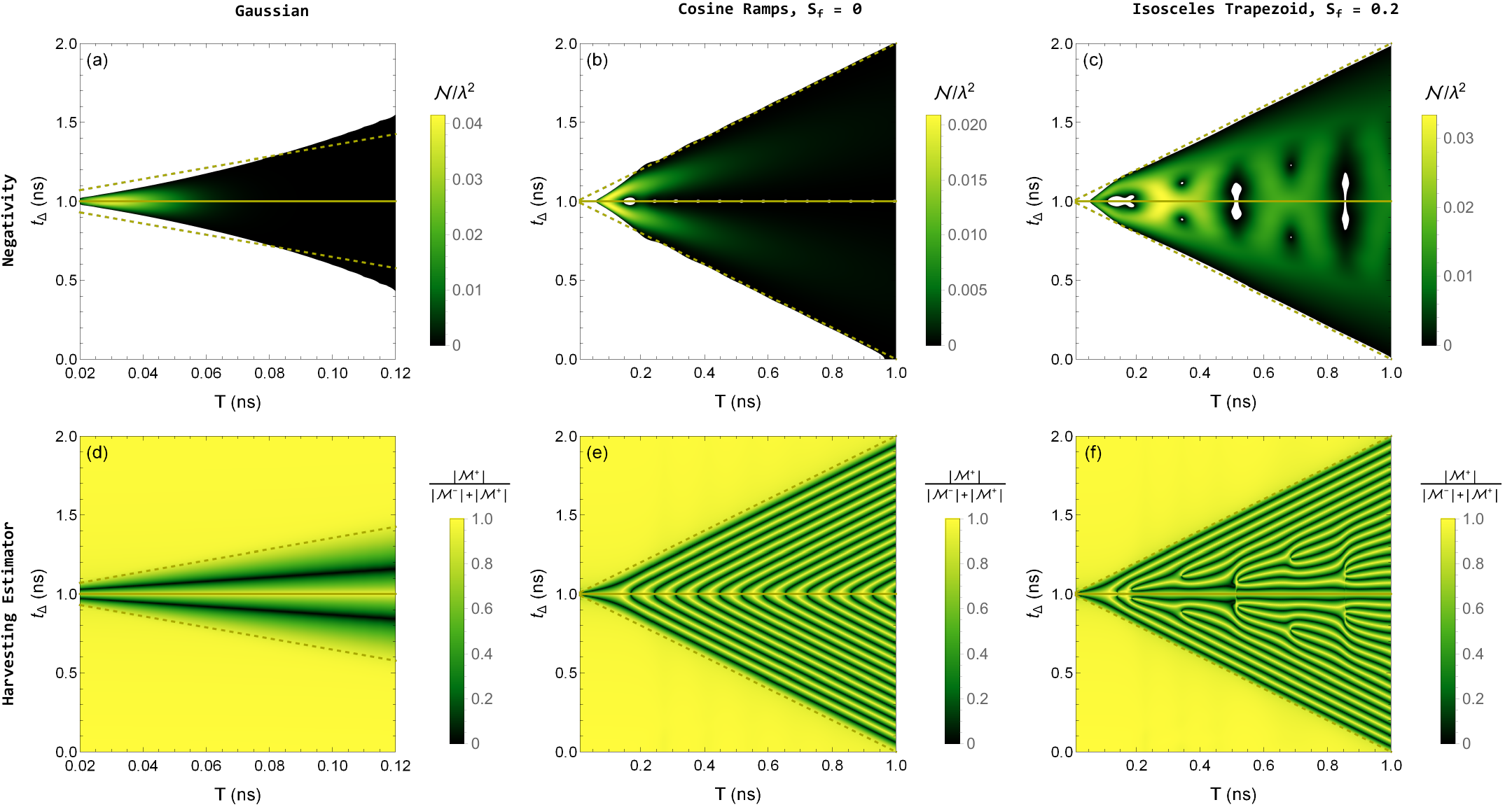}
    \caption{Plots for the scenario 1 of Table~\ref{tb:coupling_strength_scenarios}: $\Omv{}\to0$. \textbf{The following applies to  Figures \ref{fig:NoGV_TtD}, \ref{fig:002_TtD}, \ref{fig:007_TtD}, \ref{fig:014_TtD}, \ref{fig:022_TtD} and \ref{fig:-02_TtD}:}  The first row shows $\mathcal N/\lambda^2$, with white indicating $\mathcal N =0$, while darker colors indicate progressively smaller but non-zero $\mathcal N$. The second row shows the harvesting estimator $|\mathcal M^+|/(|\mathcal M^-|+|\mathcal M^+|)$, which goes from 0 (all entanglement acquired by communication) to 1 (all entanglement from genuine harvesting).
    The horizontal axes indicate the switching timescale or duration $\Ti$ and the vertical axes indicate the delay $t_\Delta$ between switchings. Each column explores a different switching function, which are, from left to right: Gaussian, cosine ramps with $S_f=0$, isosceles trapezoid with $S_f=0.2$. The horizontal solid line indicates full lightlike contact. For the compact switchings, the detectors' interactions are spacelike below the lower dashed lines and timelike above the upper dashed lines. Outside the region enclosed by the dashed yellow lines, only the $5\sigma$ tails of the Gaussian switchings overlap.
    }
    \label{fig:NoGV_TtD}
\end{figure*}

\begin{figure*}[p]
    \centering
    \includegraphics[width=.96\textwidth]{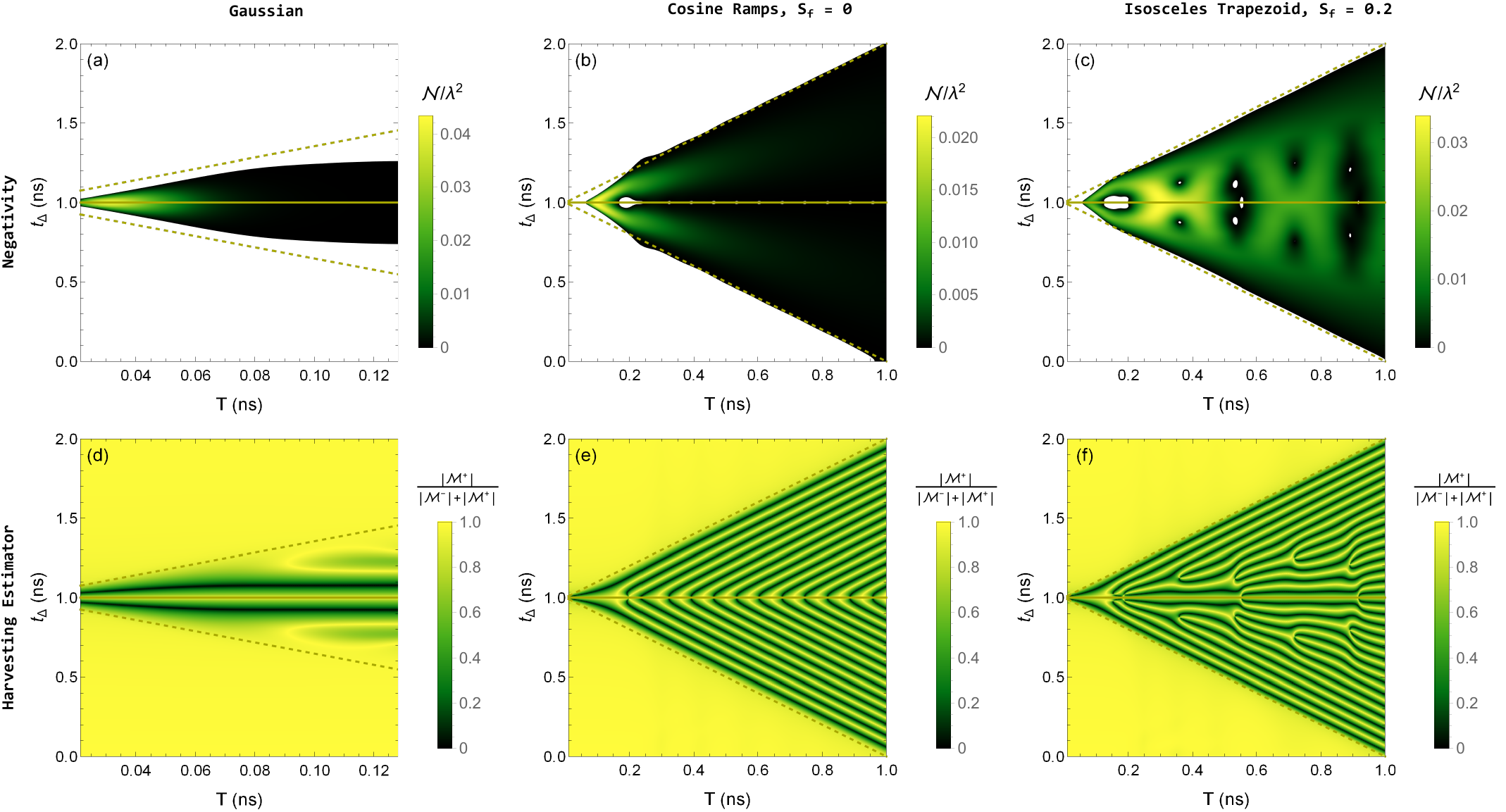}
    \caption{Analogous to Figure \ref{fig:NoGV_TtD}, but for scenario 2 of Table~\ref{tb:coupling_strength_scenarios}, gap variation $\Omv{}/(2\pi)=-0.5$. }
    \label{fig:002_TtD}
\end{figure*}

\begin{figure*}[p]
    \centering
    \includegraphics[width=\textwidth]{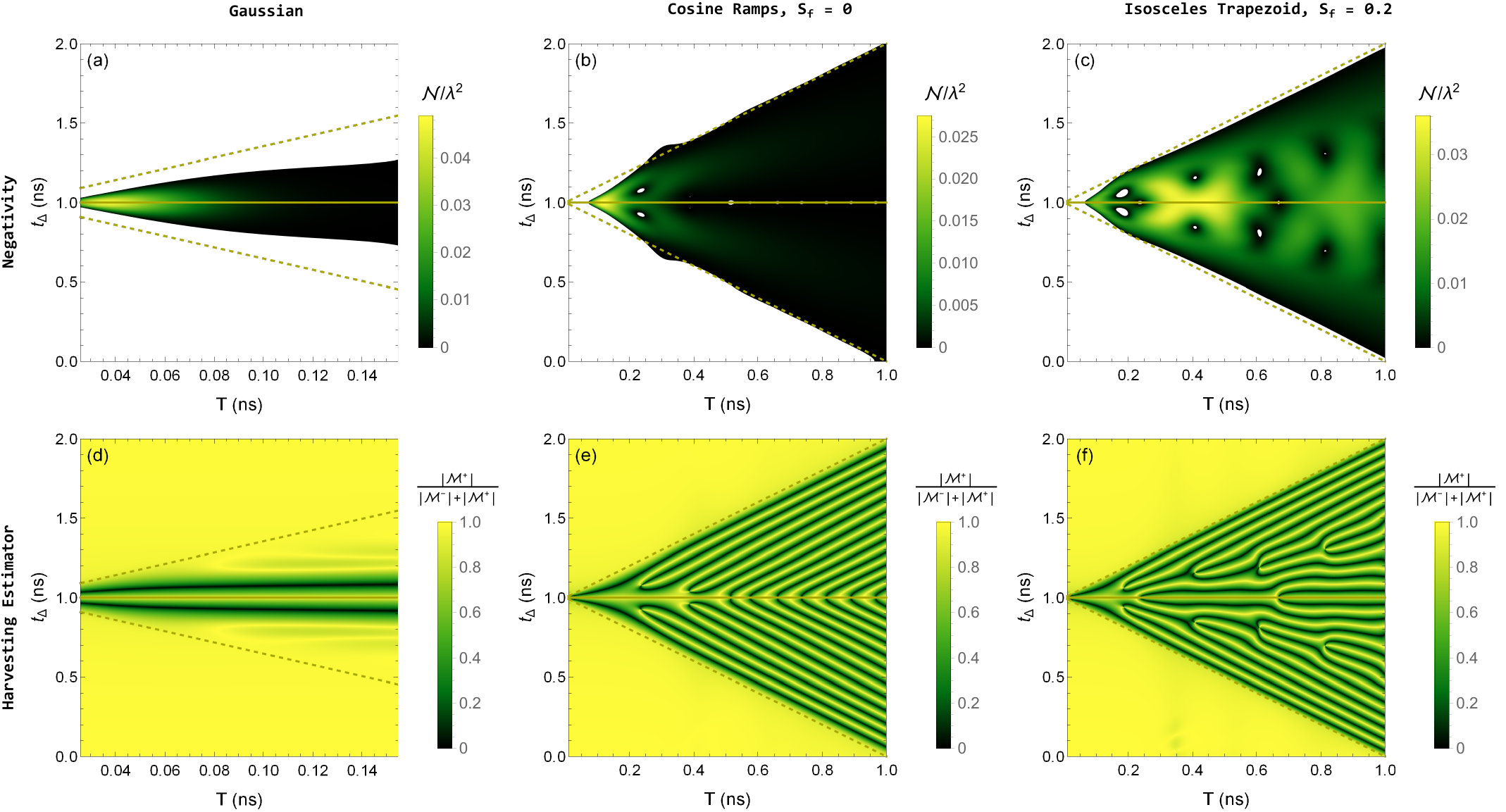}
    \caption{Analogous to Figure \ref{fig:NoGV_TtD}, but for scenario 3 of Table~\ref{tb:coupling_strength_scenarios}, gap variation \mbox{$\Omv{}/(2\pi)=-1.6\text{\,GHz}$}.}
    \label{fig:007_TtD}
\end{figure*}

\begin{figure*}[p]
    \centering
    \includegraphics[width=\textwidth]{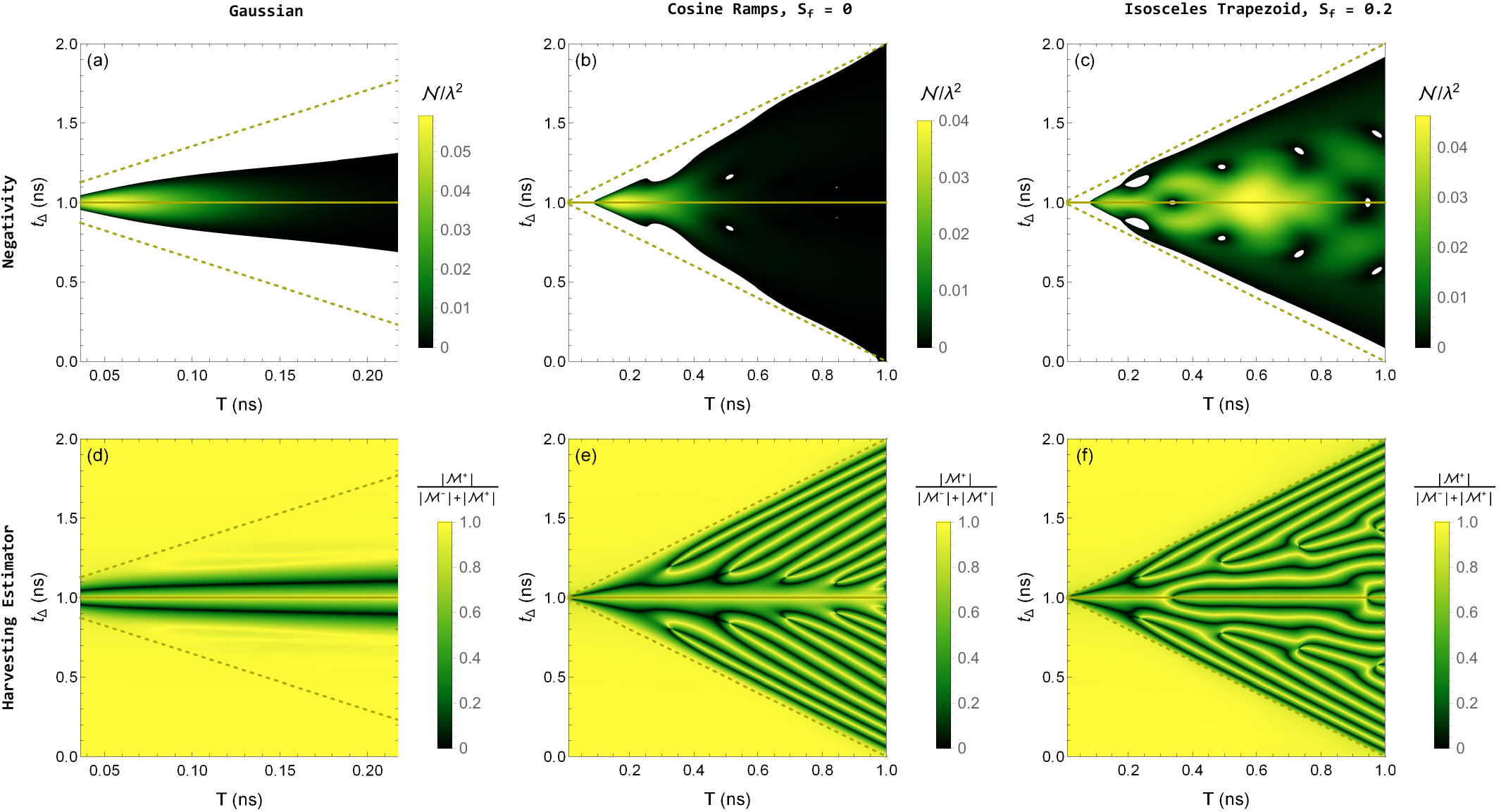}
    \caption{
    Analogous to Figure \ref{fig:NoGV_TtD}, but for scenario 4 of Table~\ref{tb:coupling_strength_scenarios}, gap variation \mbox{$\Omv{}/(2\pi)=-3.4\text{\,GHz}$}.}
    \label{fig:014_TtD}
\end{figure*}

\begin{figure*}[p]
    \centering
    \includegraphics[width=\textwidth]{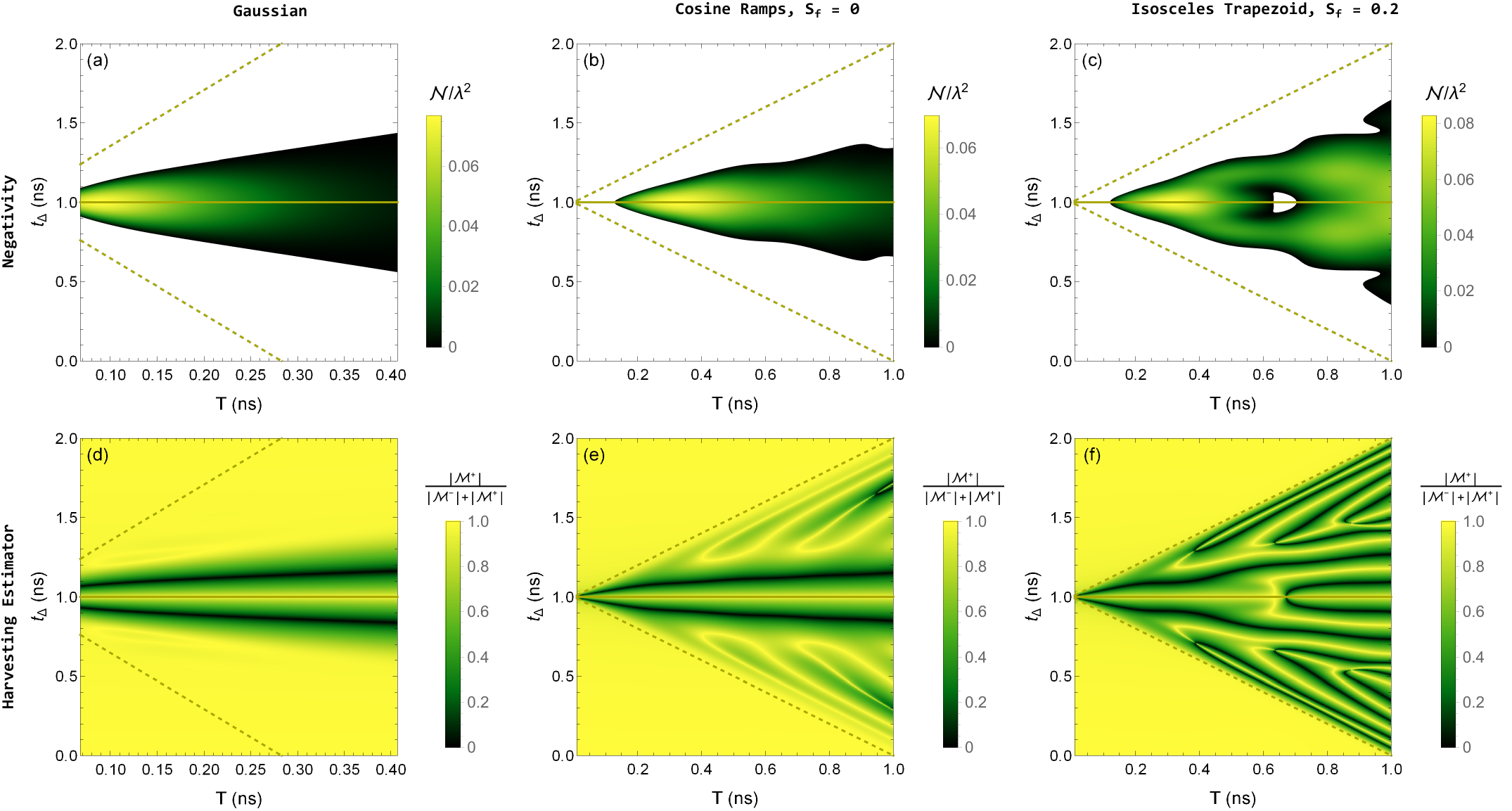}
    \caption{Analogous to Figure \ref{fig:NoGV_TtD}, but for scenario 5 of Table~\ref{tb:coupling_strength_scenarios}, gap variation \mbox{$\Omv{}/(2\pi)=-5.2\text{\,GHz}$}.}
    \label{fig:022_TtD}
\end{figure*}

\begin{figure*}[p]
    \centering
    \includegraphics[width=\textwidth]{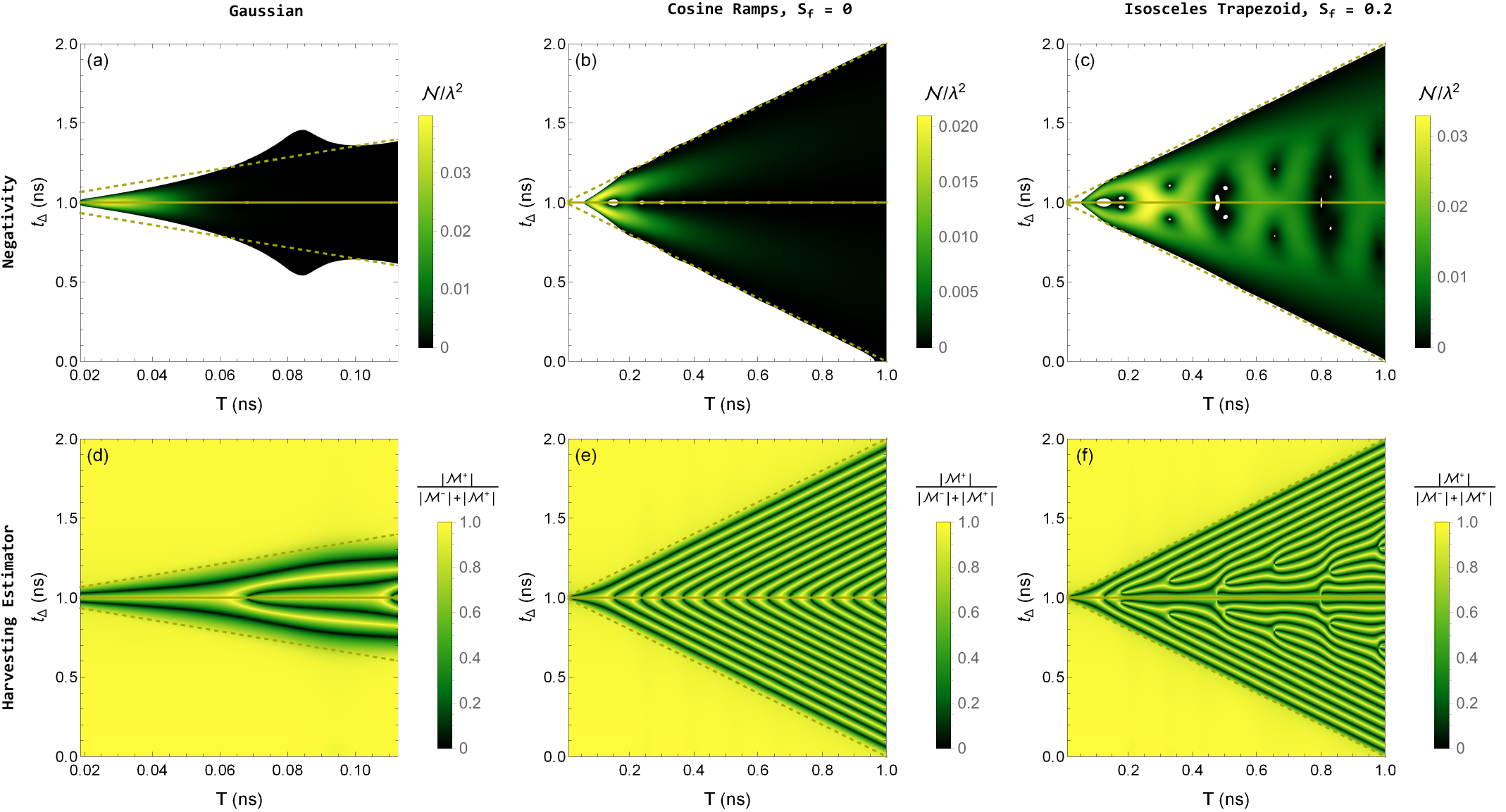}
    \caption{Analogous to Figure \ref{fig:NoGV_TtD}, but for scenario 6 of Table~\ref{tb:coupling_strength_scenarios}, gap variation $\Omv{}/(2\pi)=0.5$. }
    \label{fig:-02_TtD}
\end{figure*}

First, we study the effect of $\Ti$ and $t_\Delta$, which respectively are the interaction duration and the delay between the detector switchings. The fixed parameters are kept as indicated in subsection~\ref{ssec:parameter_choices}, and the results for the scenarios of Table~\ref{tb:coupling_strength_scenarios} are respectively shown in Figures \ref{fig:NoGV_TtD}, \ref{fig:002_TtD}, \ref{fig:007_TtD}, \ref{fig:014_TtD}, \ref{fig:022_TtD}, \ref{fig:-02_TtD}. The scenarios 1 to 5 and the corresponding Figures \ref{fig:NoGV_TtD} to \ref{fig:022_TtD} are presented in order of increasingly negative coupling strength and corresponding larger (negative) gap variation $\Omv{}$. Additionally, the scenario 6 and Figure~\ref{fig:-02_TtD} show the case of small positive $\Omv{}$. Here and from now on, the negativity and the harvesting estimator are numerically computed using the simplified integrals given in Appendix \ref{apx:decoupled_timeIntegrals}.


First, we observe that the negativity decreases quickly with $T$ for the Gaussian switching, while decreasing much slower with $T$ for the cosine ramps switching and even slower for the isosceles trapezoid switching. This is true regardless of the choice of parameters in the problem. 

Let us now focus on Figure \ref{fig:NoGV_TtD}, which shows the acquired negativity in the weak coupling limit $\lambda\to0$, the case without gap variation. In order to determine the amount of entanglement that is genuinely harvested from the field, we also plot the harvesting estimator from Eq.~\eqref{eq:harvesting_estimator}. Since the negativity is multiplied by $\lambda^2$ at leading order, we plot $\mathcal N/\lambda^2$. For switching functions of compact support (cosine ramps and isosceles trapezoid), the detectors acquire negativity mostly when they have partial or full light contact, which is indicated as being inside the dashed yellow lines in the plots. Nonetheless, there is some negativity outside the dashed lines, which indicates that spacelike and timelike harvesting are both possible. Notice that spacelike harvesting is still possible when the cutoff is taken to be infinitely large $\Omcut\to\infty$, as depicted in Appendix~\ref{apx:largecutoff}. Furthermore, looking at the genuine harvesting indicator on the second row of plots, i.e. the subfigures \ref{fig:NoGV_TtD}(d), \ref{fig:NoGV_TtD}(e), \ref{fig:NoGV_TtD}(f), we observe that a significant amount of entanglement harvesting is possible in causal contact. For Gaussian switching, we observe that when the detectors are in full lightlike contact (yellow solid line at $t_\Delta=1\text{\, ns}$) all the acquired entanglement is harvested (and not acquired through communication), with two peaks of communication at partial lightlike contact. This is not surprising and matches the results of previous literature on derivatively coupled detectors~\cite{adam2024derivative}. For the cosine ramps and isosceles trapezoid switchings, the genuine harvesting estimator displays rapidly oscillatory patterns in the regions of causal contact. Nonetheless, this also means that genuine lightlike entanglement harvesting is possible for these finite duration switching functions.

Next, we analyze how varying the energy gap affects entanglement harvesting by comparing the scenarios of Figures \ref{fig:NoGV_TtD}, \ref{fig:002_TtD}, \ref{fig:007_TtD}, \ref{fig:014_TtD}, \ref{fig:022_TtD}, each with progressively larger (negative) gap variations $\Omv{}$ driven by making the coupling strength $\lambda$ more negative. As $\Omv{}$ becomes more negative, both spacelike and timelike entanglement harvesting decrease. Conversely, in the region inside the dashed lines, where detectors are in causal contact, negativity increases overall. The change in shape of the negativity and genuine harvesting estimator can partially be intuited from the change in detector gap. Since for negative $\Omv{}$ the gap becomes smaller, the $T$ axis in terms of units of $\Omega^{-1}$ is effectively rescaled. Specifically, the shape of the plotted quantities  `stretches' in the horizontal axis around $T=0$ as $\Omv{}$ becomes more negative. We also explored one scenario with small positive $\Omv{}$, and encountered spacelike harvesting, as shown in Figure~\ref{fig:-02_TtD}.

\subsection{Spacelike harvesting for finite duration switchings}
\label{ssec:spacelike_harvesting}
\begin{figure}[ht]
    \centering
    \includegraphics[width=\columnwidth]{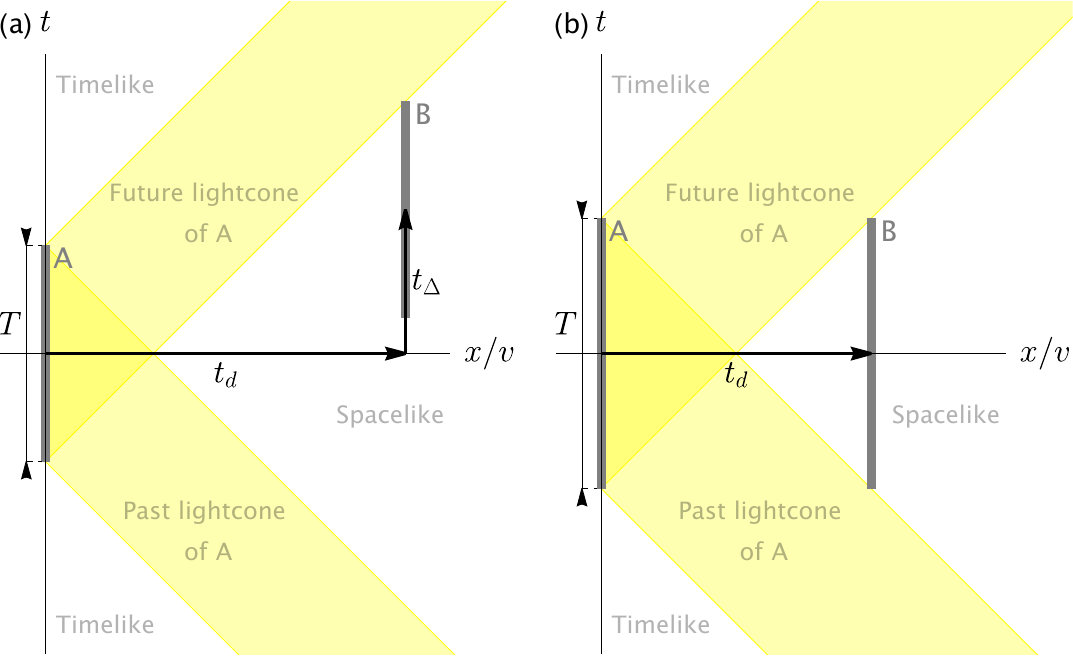}
    \caption{Spacetime diagrams with gray rectangles representing the detectors' interactions. (a) Case $t_\Delta = t_d - T$, with $t_d$ fixed. The entanglement harvested in this case is shown in Figure \ref{fig:N_spacelike_td=1} ($t_d=1\,\unit{\nano\second}$) and Figure \ref{fig:N_spacelike_td=0.5} ($t_d=0.5\,\unit{\nano\second}$). (b) Case $t_d = T$, $t_\Delta = 0$. The entanglement harvested in this case is shown in Figure \ref{fig:N_spacelike_tDelta=0}.
    }
    \label{fig:HarvestingDiagramCombined}
\end{figure}

\begin{figure*}[p]
    \centering
    \includegraphics[height=.85\textheight]{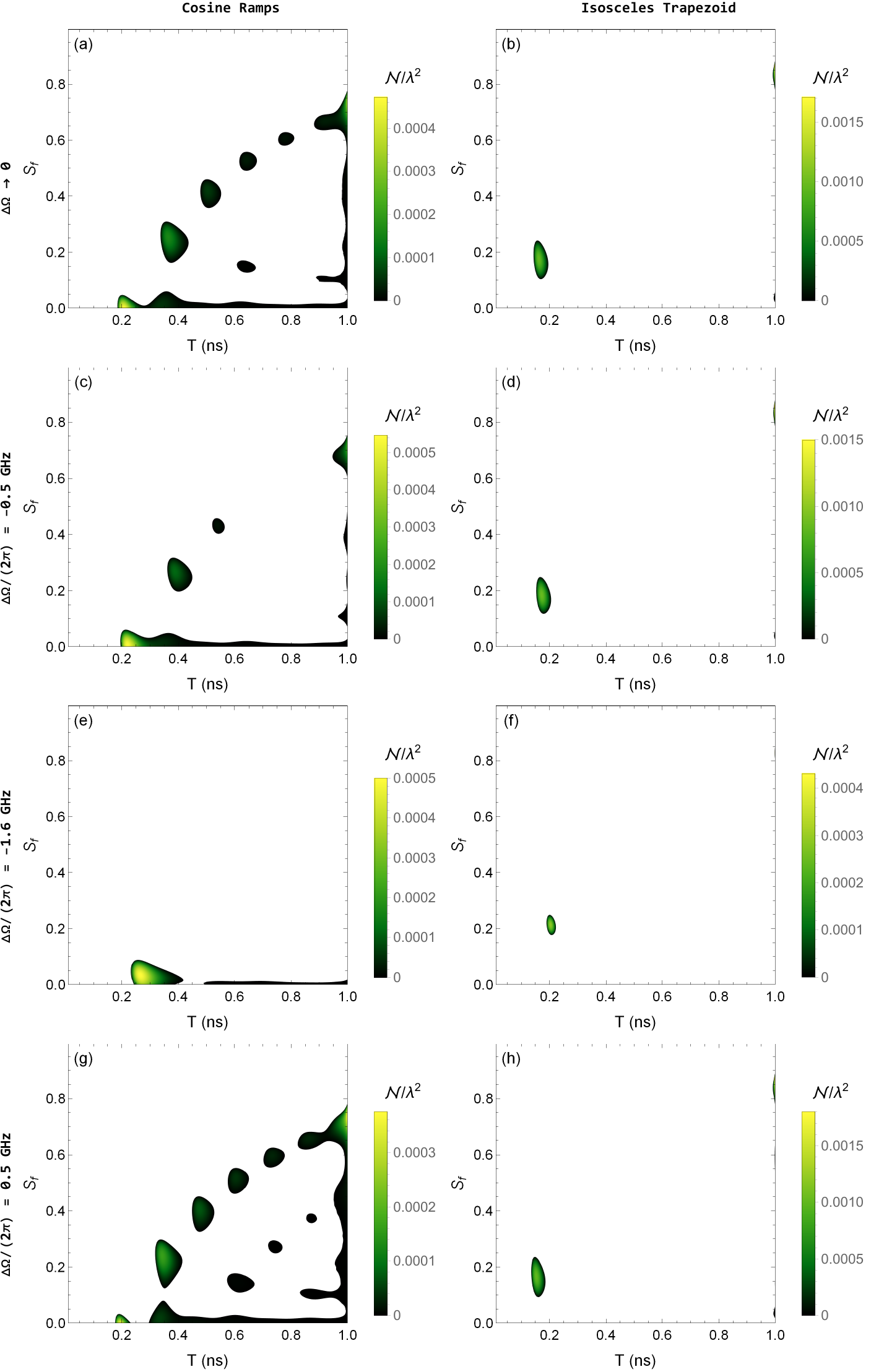}
    \caption{
    Negativity acquired by spacelike detectors. To ensure the negativity is the largest, we explore the boundary where interactions become spacelike, by choosing $t_\Delta=t_d-\Ti$, with $t_d=1\text{\,ns}$. This arrangement is depicted in the spacetime diagram of Figure~\ref{fig:HarvestingDiagramCombined}(a). \textbf{The following applies to Figures \ref{fig:N_spacelike_td=1}, \ref{fig:N_spacelike_td=0.5}, and \ref{fig:N_spacelike_tDelta=0}:} The plots' horizontal axes are the switching durations $\Ti$ and the vertical axes are the portion of the switching function that is flat $S_f$ and at full coupling. The switching function for the left column is the cosine ramps and for the right column the isosceles trapezoid. The rows respectively correspond to the scenarios 1, 2, 3 and 6 of Table~\ref{tb:coupling_strength_scenarios}.}  
    \label{fig:N_spacelike_td=1}
\end{figure*}

\begin{figure*}[p]
    \centering
    \includegraphics[height=.85\textheight]{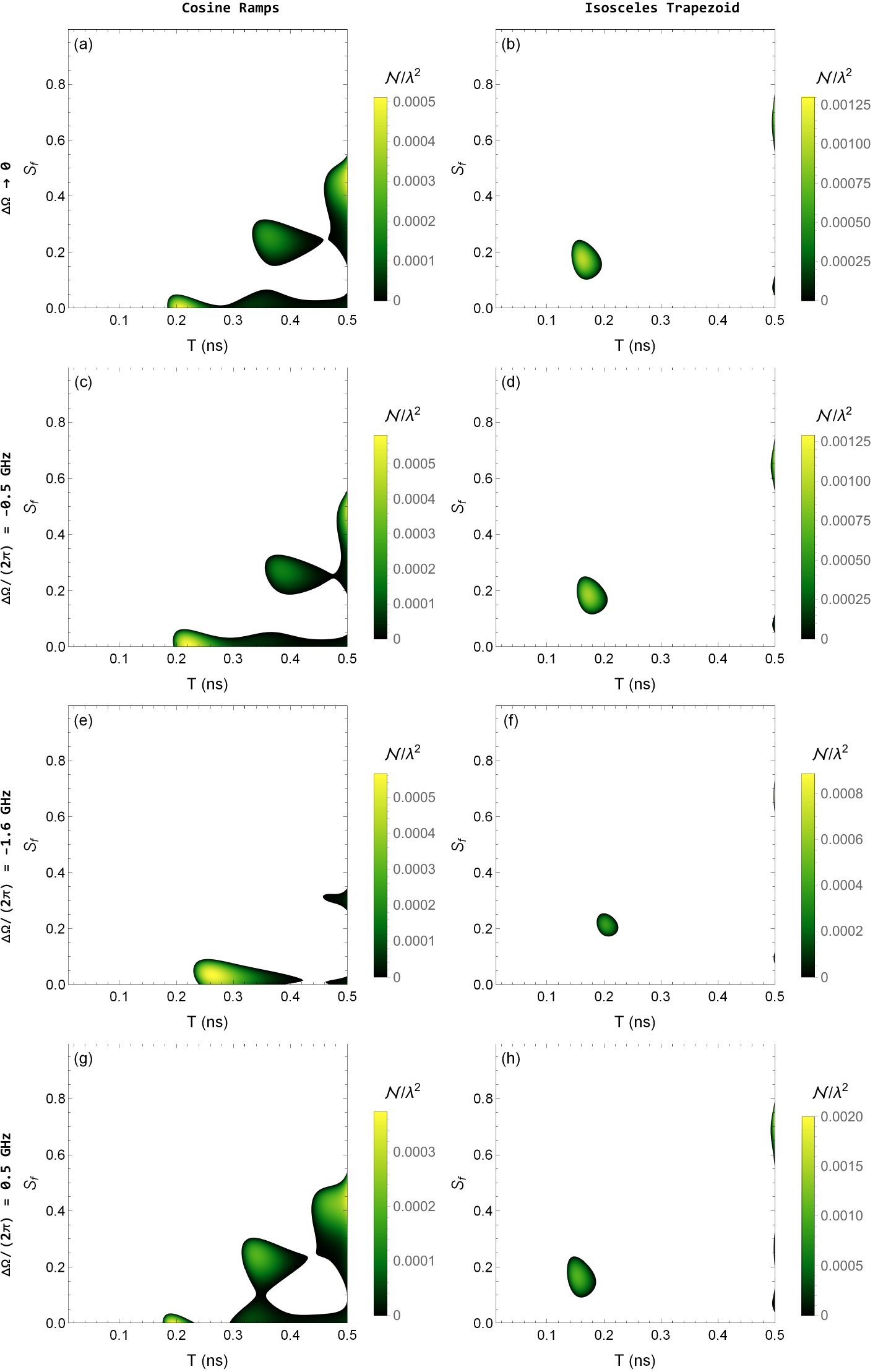}
    \caption{Negativity acquired by spacelike detectors. The plots are analogous to Figure~\ref{fig:N_spacelike_td=1}, with the only difference being that the distance between detectors is reduced so that $t_d=0.5\text{\,ns}$. }
    \label{fig:N_spacelike_td=0.5}
\end{figure*}

\begin{figure*}[p]
    \centering
    \includegraphics[height=.92\textheight]{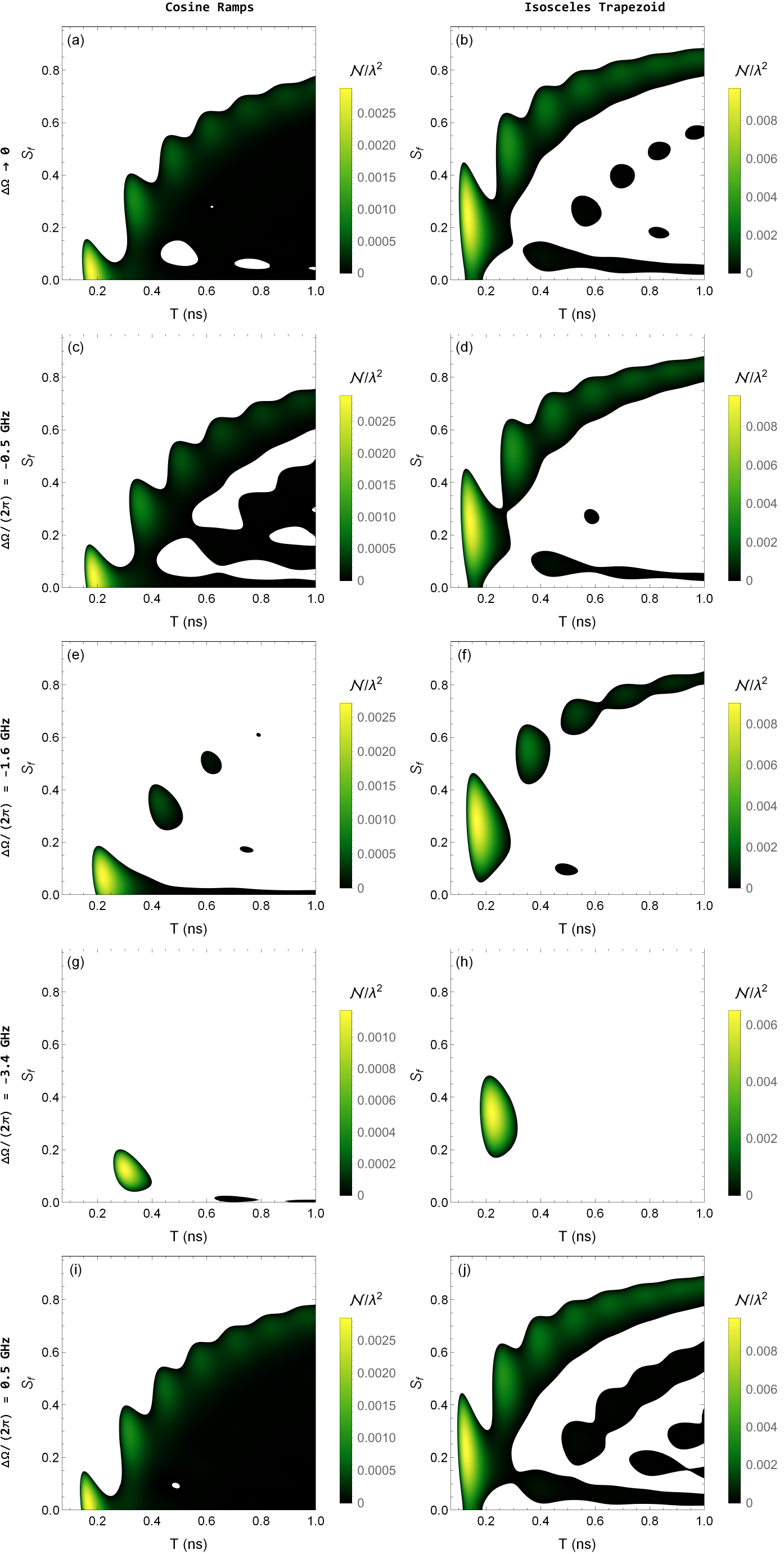}
    \caption{Negativity acquired by spacelike detectors. The plots are analogous to Figure~\ref{fig:N_spacelike_td=1}, with the exceptions of fixing $t_\Delta=0$ and allowing the distance between detectors to vary as $t_d=\Ti$, as depicted in the spacetime diagram of Figure~\ref{fig:HarvestingDiagramCombined}(b). Moreover, the rows show the scenarios 1, 2, 3, 4 and 6 of Table~\ref{tb:coupling_strength_scenarios}. 
    }
    \label{fig:N_spacelike_tDelta=0}
\end{figure*}

We now study harvesting in the regime of strict spacelike separation and, hence, only for the cases of the compactly supported switching functions: the cosine ramps and the isosceles trapezoid. Usually, spacelike detectors acquire more entanglement the closer they are to being in causal contact.  Taking this into account, we consider two ways to place the detectors and to switch on the interactions, depicted in Figure~\ref{fig:HarvestingDiagramCombined}, which maximize the harvested entanglement. These two ways are as follows:
\begin{enumerate}
    \item Place the detectors at a fixed distance, i.e. with a constant $t_d$, and pick their switching delay to be $t_\Delta = t_d - T$. The reason for this choice stems from detectors being spacelike as long as $|t_\Delta| \leq t_d - T$. For larger $|t_\Delta|$, detectors would be in lightlike contact, and for even larger $|t_\Delta|$, timelike separated. Therefore, $t_\Delta = t_d - T$ makes the detectors as close as possible to being in causal contact while keeping them spacelike, maximizing the acquired entanglement. This choice is illustrated in a spacetime diagram in subfigure~\ref{fig:HarvestingDiagramCombined}(a). The acquired negativity is shown in Figure \ref{fig:N_spacelike_td=1} for \mbox{$t_d=1\,\unit{\nano\second}$} and in Figure \ref{fig:N_spacelike_td=0.5} for \mbox{$t_d=0.5\,\unit{\nano\second}$}.
    \item Place two simultaneously switched detectors, \mbox{$t_\Delta=0$}, as close as possible while keeping their interactions spacelike. This is achieved by allowing $t_d$ to take the value $t_d=T$. This choice is illustrated in a spacetime diagram in subfigure~\ref{fig:HarvestingDiagramCombined}(b) and the acquired negativity is shown in Figure \ref{fig:N_spacelike_tDelta=0}.
\end{enumerate}
Importantly, the entanglement that can be harvested from these spacelike configurations does not only originate from the finite cutoff $\Omcut$, and can persist even when $\Omcut\to\infty$, as showcased in Appendix~\ref{apx:largecutoff}.

Figures~\ref{fig:N_spacelike_td=1}, \ref{fig:N_spacelike_td=0.5} and \ref{fig:N_spacelike_tDelta=0} show the final negativity of the detectors as a function of: 1) the interaction duration $T$ and 2) the switching function parameter $S_f$, as defined in Eqs.~\eqref{eq:cosine_switching} and~\eqref{eq:trapezoid_switching}. Each figure provides plots for the scenarios of Table~\ref{tb:coupling_strength_scenarios} that have some non-zero negativity. 
We observe that large negative values of $\Omv{}$ make spacelike harvesting harder by reducing the range of parameters ($T$ and $S_f$) where negativity is non-zero. This is consistent with the findings in subsection \ref{ssec:exploration_duration_delay}. On the other hand, increasing $\lambda$ improves negativity, which has a $\lambda^2$ prefactor. However, a stronger coupling is accompanied by larger gap variation, due to $\Omv{}/(2\pi)\approx 5.2\lambda\,\unit{\giga\hertz}$. Eventually, the larger negative $\Omv{}$ has a stronger negative effect, making spacelike harvesting impossible for strong couplings. For small positive $\Omv{}$ (scenario 6), the regions with non-zero negativity are moderately larger compared to when there is no gap variation (scenario 1), but this comes at the cost of a slight decrease in the maximum negativity that can be harvested.

Looking at the differences caused by the positioning of the detectors, one finds that the most entanglement is found when $t_d=T$ and $t_\Delta=0$, as shown in Figure~\ref{fig:N_spacelike_tDelta=0}. However, this choice is impractical in an actual setup, because the distance between detectors is fixed after the device is built, and cannot be tuned to match the interaction duration $T$. This leaves the tuning of $t_\Delta$ for a fixed $t_d$ as the way to maximize harvested entanglement for a given $T$ in a superconducting implementation. This method allows to reach the negativities shown in Figures~\ref{fig:N_spacelike_td=1} and~\ref{fig:N_spacelike_td=0.5}. Furthermore, by comparing Figure~\ref{fig:N_spacelike_td=1} (fixed $t_d=1\text{\,ns}$) and Figure~\ref{fig:N_spacelike_td=0.5} (fixed $t_d=0.5\text{\,ns}$), we observe that reducing $t_d$ has two main effects: 1) reducing the range of values of $T$ for which the interaction can be spacelike, 2) increasing spacelike negativity, but only for $T\approx t_d$. Specifically, for the explored cases, if $T < 0.45\text{\,ns}$, the negativity is the same regardless of whether $t_d=0.5\text{\,ns}$ or $t_d=1\text{\,ns}$. The difference appears for $T \in[0.45\text{\,ns}, 0.5\text{\,ns}]$, for which negativity is larger in the case $t_d=0.5\text{\,ns}$. For $T \in[0.5\text{\,ns}, 1\text{\,ns}]$, only $t_d=1\text{\,ns}$ allows for the detectors to be spacelike.

Finally, let us focus on the effect of the shape of the switching function. We observe that smoother switching functions provide a wider range of parameters for which spacelike separated detectors can harvest entanglement. In Figures~\ref{fig:N_spacelike_td=1}, \ref{fig:N_spacelike_td=0.5} and \ref{fig:N_spacelike_tDelta=0}, this phenomenon can be seen when comparing the left and right columns of plots, which respectively correspond to smoother cosine ramps switchings and rougher isosceles trapezoid switchings. 
Moreover, for all plots, harvesting entanglement becomes impossible as $S_f\to 1$, that is, as switching functions tend towards rectangular switching functions, which are sudden and rough. These observations match previous literature, where smoother switching functions could harvest entanglement for detectors that are placed further away from each other. 

Sudden switching is known to be particularly detrimental in most scenarios because it increases local noise more than it generates correlations between the detectors~\cite{Pozas2015}. Conversely, smooth switching facilitates harvesting in a wider range of parameters, but at the cost of decreasing the amount of entanglement harvested per run of the protocol. Therefore, the optimal switching profile is expected to lie somewhere between a highly adiabatic regime and a very sudden one. Understanding the effect of the smoothness of the switching function is crucial for future experiments, where it is important to optimize the amount of harvested entanglement to facilitate its detection.


\subsection{Effect of the distance between detectors}


\begin{figure*}[p]
    \centering
    \includegraphics[width=.96\textwidth]{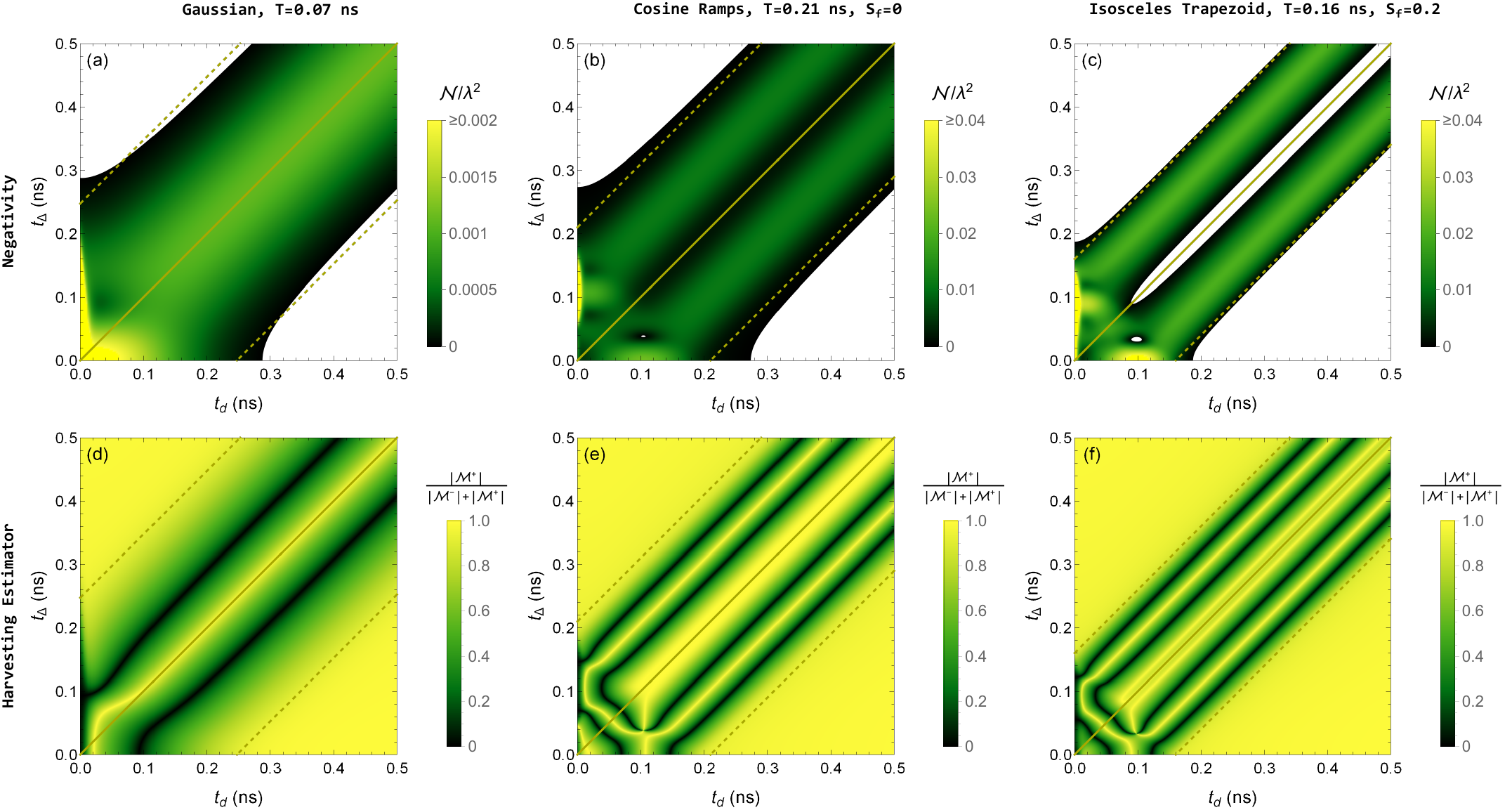}
    \caption{Scenario 1 of Table~\ref{tb:coupling_strength_scenarios}, $\Omv{}\to0$. \textbf{The following applies to Figures \ref{fig:NoGV_tdtD}, \ref{fig:002_tdtD}, \ref{fig:007_tdtD}, \ref{fig:014_tdtD}, \ref{fig:022_tdtD} and \ref{fig:-02_tdtD}:} The first row shows $\mathcal N/\lambda^2$, with $\mathcal N = 0$ in white. The second row shows $|\mathcal M^+|/(|\mathcal M^-|+|\mathcal M^+|)$, which goes from 0 (all entanglement from communication) to 1 (all entanglement from harvesting). $t_d$ is the signaling time and $t_\Delta$ the delay between switchings. The columns use, from left to right, the switching functions: 1) Gaussian, $\Ti=0.07\,\unit{\nano\second}$, 2) cosine ramps, $\Ti=0.21\,\unit{\nano\second}$, $S_f=0$, 3) isosceles trapezoid, $\Ti=0.16\,\unit{\nano\second}$, $S_f=0.2$. The diagonal solid lines indicate full lightlike contact. Outside the dashed lines, either there is no light contact (compact switchings) or only the $5\sigma$ tails of the Gaussian can interact. To aid visualization, the values of negativity larger than the thresholds indicated in the legend are plotted yellow. 
    }
    \label{fig:NoGV_tdtD}
\end{figure*}

\begin{figure*}[p]
    \centering
    \includegraphics[width=.96\textwidth]{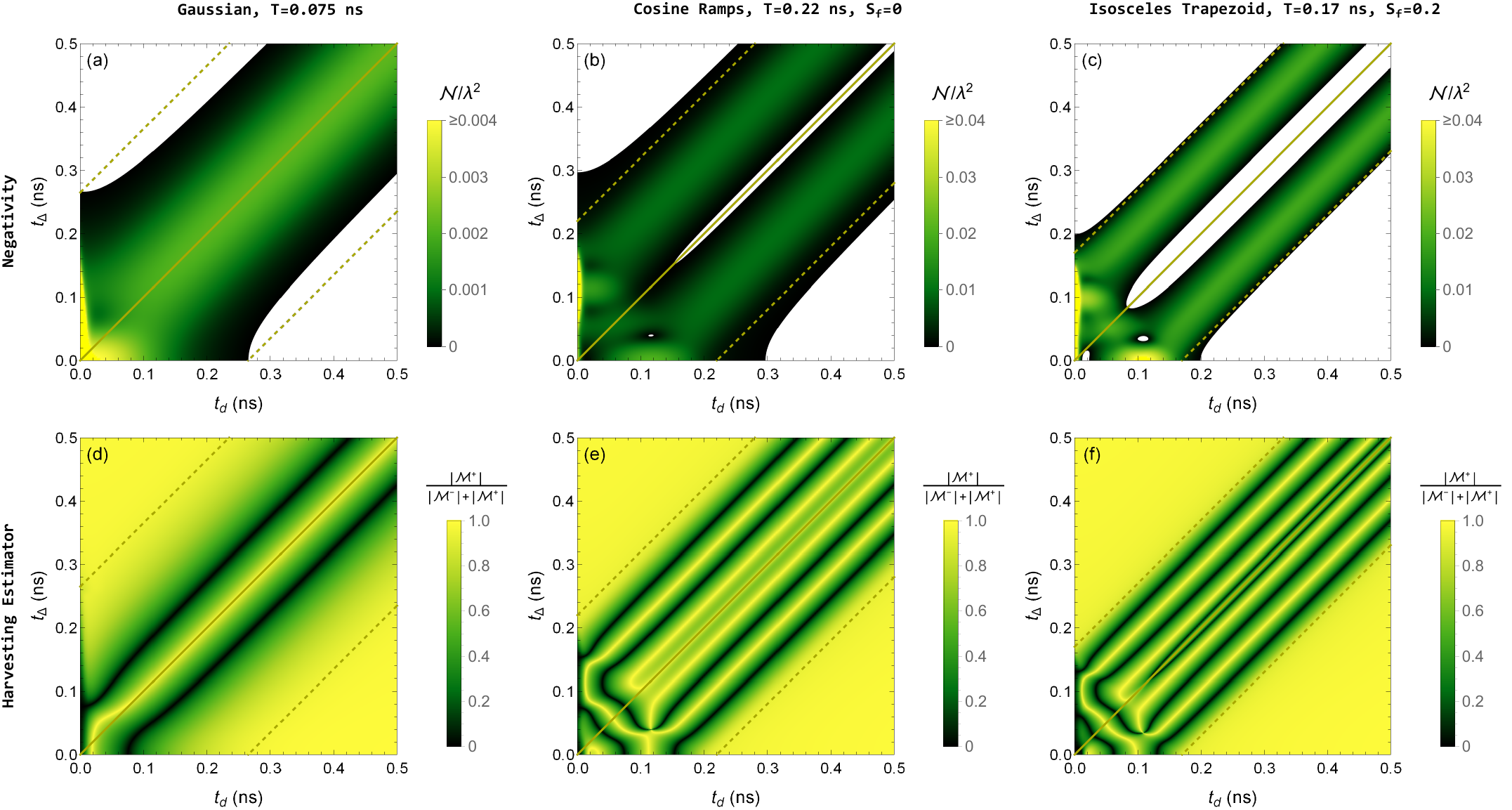}
    \caption{Analogous to Figure \ref{fig:NoGV_tdtD}, but for scenario 2 of Table~\ref{tb:coupling_strength_scenarios}, \mbox{$\Omv{}/(2\pi)=-0.5\text{\,GHz}$}. Moreover, each column of plots uses, from left to right, Gaussian switching with $\Ti=0.075\,\unit{\nano\second}$, cosine ramps switching with $\Ti=0.22\,\unit{\nano\second}$, $S_f=0$, and isosceles trapezoid switching with $\Ti=0.17\,\unit{\nano\second}$, $S_f=0.2$.}
    \label{fig:002_tdtD}
\end{figure*}

\begin{figure*}[p]
    \centering
    \includegraphics[width=\textwidth]{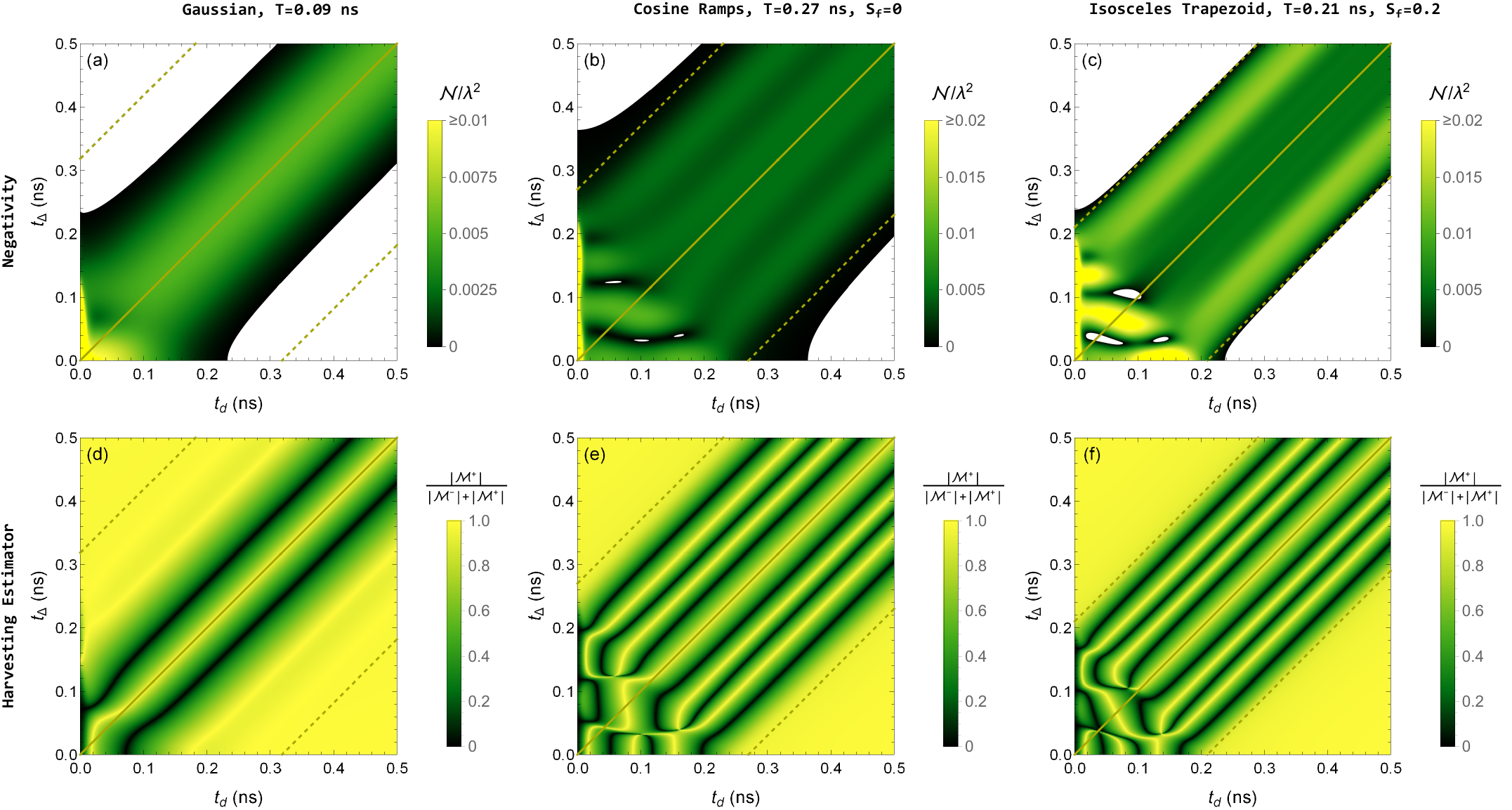}
    \caption{Analogous to Figure \ref{fig:NoGV_tdtD}, but for scenario 3 of Table~\ref{tb:coupling_strength_scenarios}, \mbox{$\Omv{}/(2\pi)=-1.6\text{\,GHz}$}. Moreover, each column of plots uses, from left to right, Gaussian switching with $\Ti=0.09\,\unit{\nano\second}$, cosine ramps switching with $\Ti=0.27\,\unit{\nano\second}$, $S_f=0$, and isosceles trapezoid switching with $\Ti=0.21\,\unit{\nano\second}$, $S_f=0.2$.}
    \label{fig:007_tdtD}
\end{figure*}

\begin{figure*}[p]
    \centering
    \includegraphics[width=\textwidth]{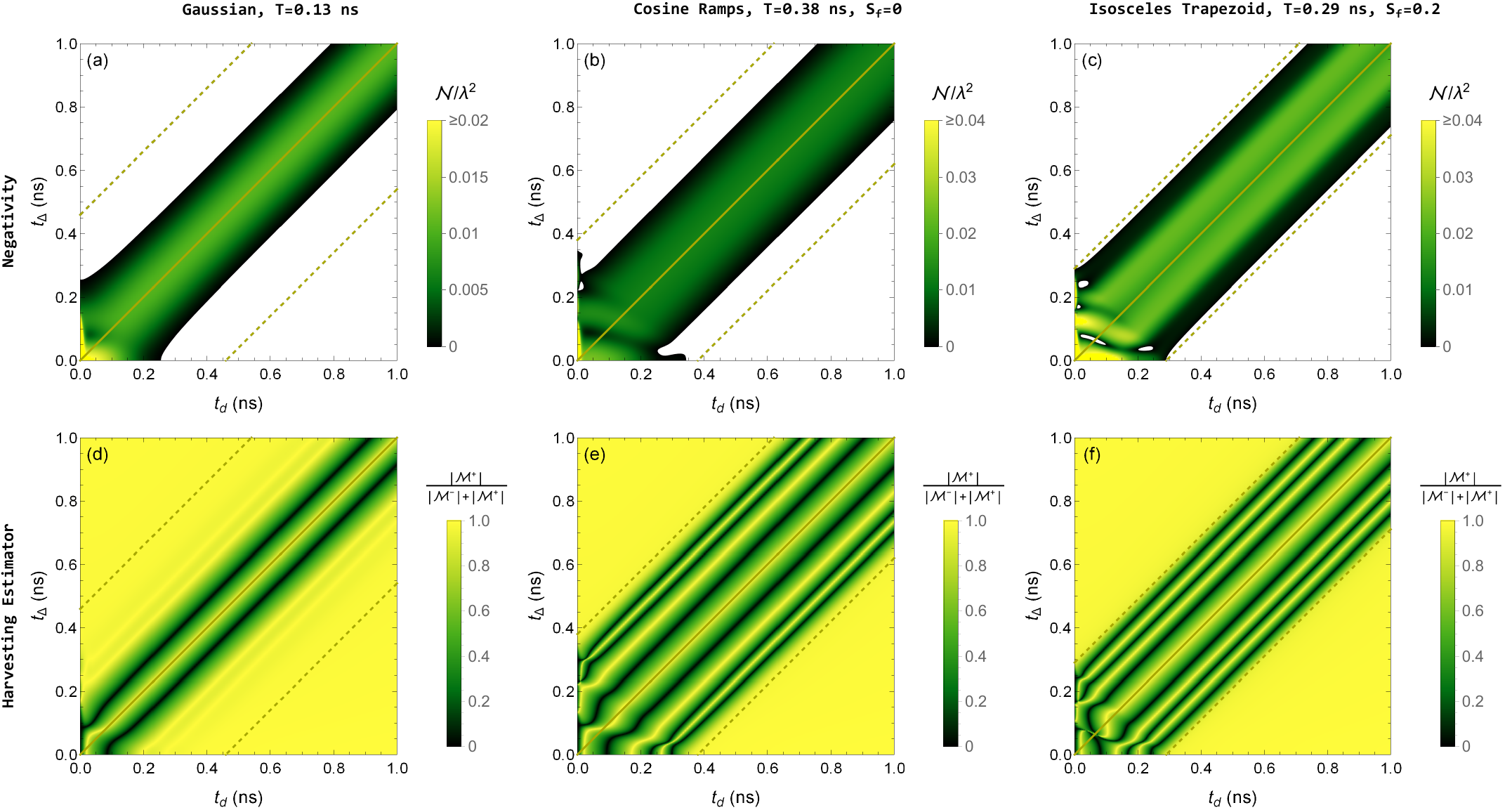}
    \caption{Analogous to Figure \ref{fig:NoGV_tdtD}, but for scenario 4 of Table~\ref{tb:coupling_strength_scenarios}, \mbox{$\Omv{}/(2\pi)=-3.4\text{\,GHz}$}. Moreover, each column of plots uses, from left to right, Gaussian switching with $\Ti=0.13\,\unit{\nano\second}$, cosine ramps switching with $\Ti=0.38\,\unit{\nano\second}$, $S_f=0$, and isosceles trapezoid switching with $\Ti=0.29\,\unit{\nano\second}$, $S_f=0.2$.}
    \label{fig:014_tdtD}
\end{figure*}

\begin{figure*}[p]
    \centering
    \includegraphics[width=\textwidth]{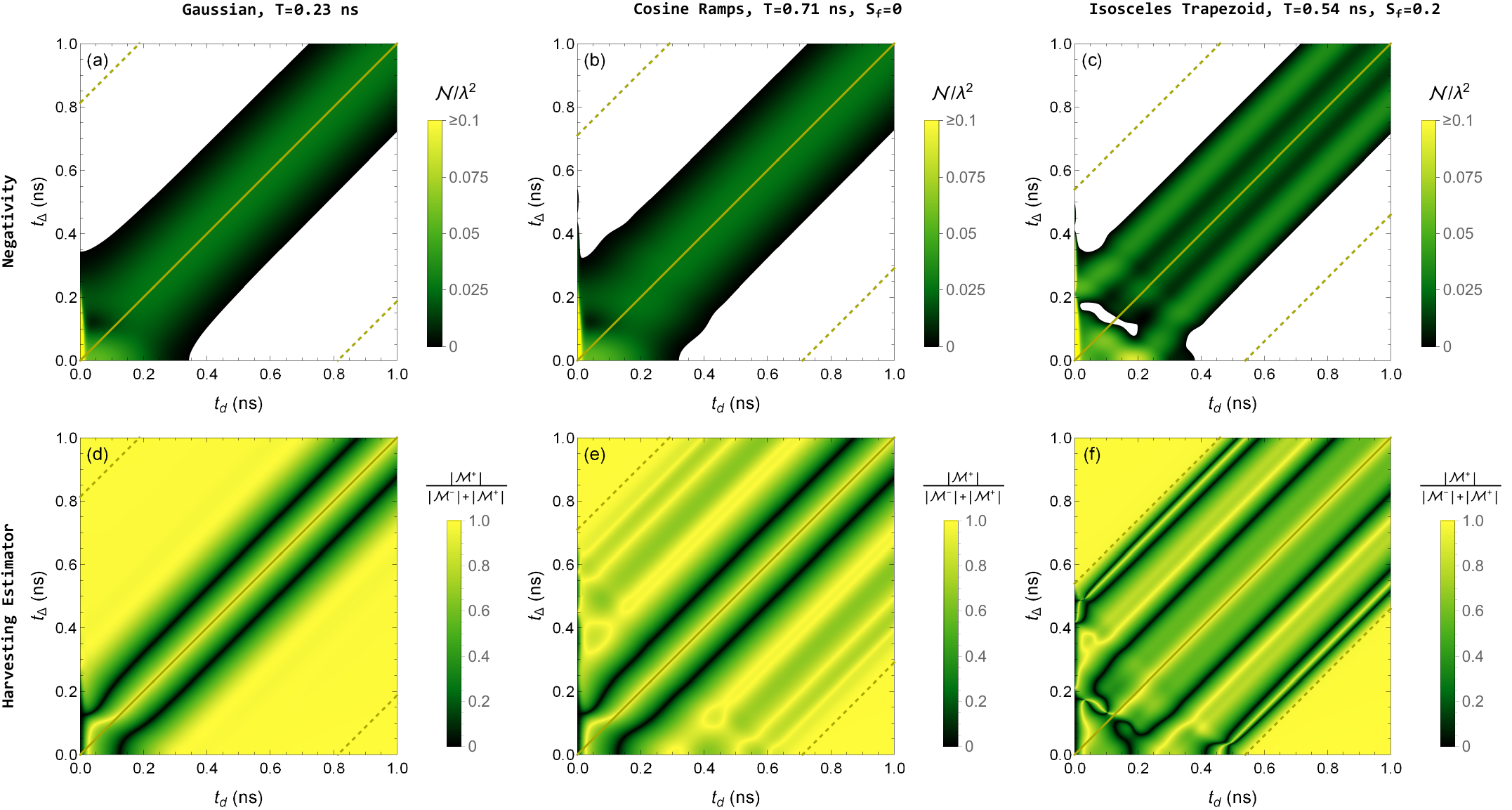}
     \caption{Analogous to Figure \ref{fig:NoGV_tdtD}, but for scenario 5 of Table~\ref{tb:coupling_strength_scenarios}, \mbox{$\Omv{}/(2\pi)=-5.2\text{\,GHz}$}. Moreover, each column of plots uses, from left to right, Gaussian switching with $\Ti=0.23\,\unit{\nano\second}$, cosine ramps switching with $\Ti=0.71\,\unit{\nano\second}$, $S_f=0$, and isosceles trapezoid switching with $\Ti=0.54\,\unit{\nano\second}$, $S_f=0.2$.}
    \label{fig:022_tdtD}
\end{figure*}

\begin{figure*}[p]
    \centering
    \includegraphics[width=\textwidth]{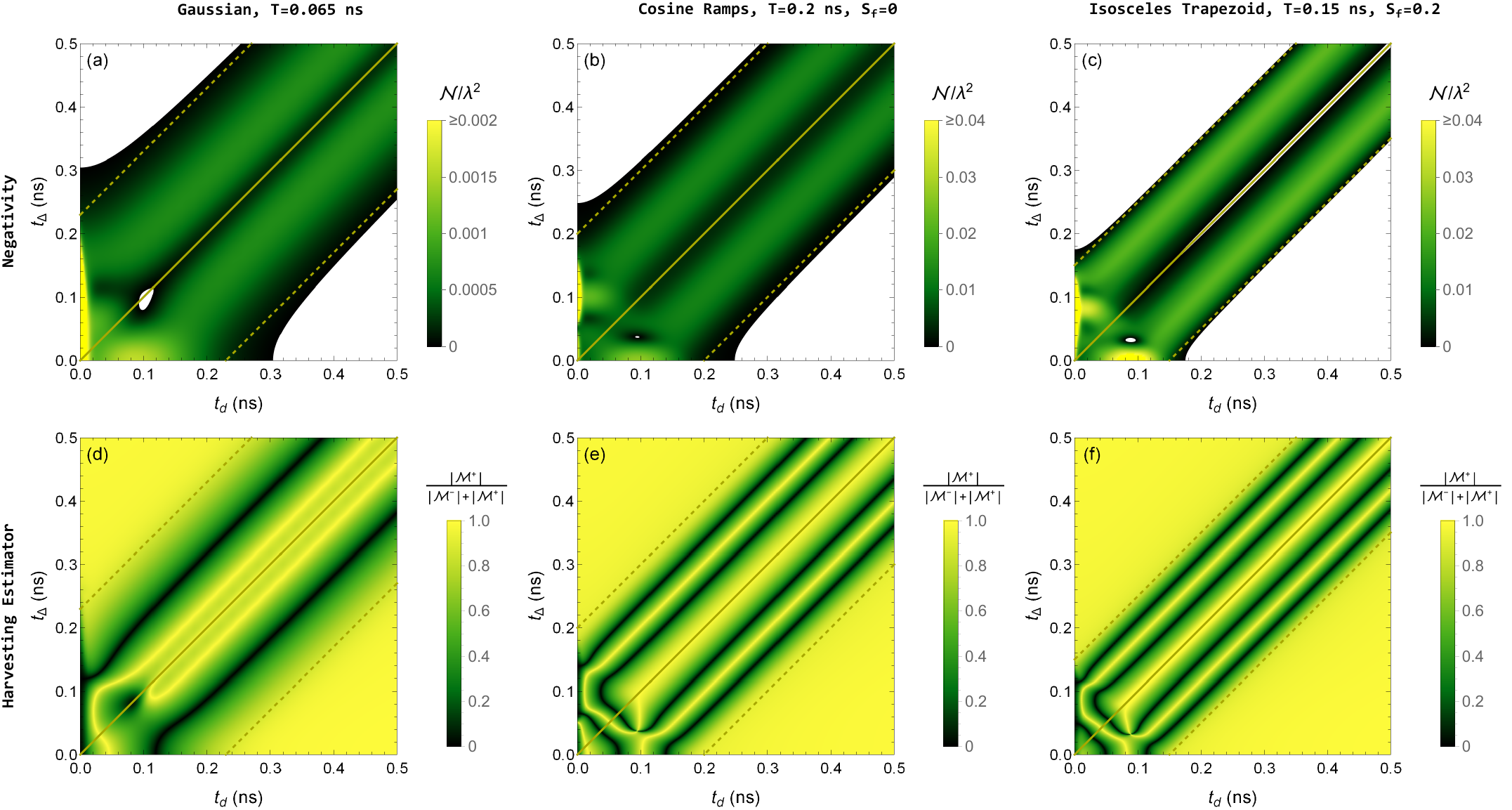}
    \caption{Analogous to Figure \ref{fig:NoGV_tdtD}, but for scenario 6 of Table~\ref{tb:coupling_strength_scenarios}, \mbox{$\Omv{}/(2\pi)=0.5\text{\,GHz}$}. Moreover, each column of plots uses, from left to right, Gaussian switching with $\Ti=0.065\,\unit{\nano\second}$, cosine ramps switching with $\Ti=0.2\,\unit{\nano\second}$, $S_f=0$, and isosceles trapezoid switching with $\Ti=0.15\,\unit{\nano\second}$, $S_f=0.2$.}
    \label{fig:-02_tdtD}
\end{figure*}


Here we show how the entanglement acquired by the detectors changes with the distance between them. Finding an optimal distance is important, since it cannot be changed after building the superconducting device.

Now, when plotting the negativity and the genuine harvesting estimator we will use as independent variables $t_d$ and $t_\Delta$. The variable $t_d$ determines the distance between the probes, and the delay $t_\Delta$ can compensate the reduction in negativity that occurs when separating the detectors, since the amount of harvested entanglement is ruled by the distance of the detectors' interactions to each other's lightcones. Moreover, $t_\Delta$ is easily tunable in the superconducting implementation. We explore the scenarios in Table~\ref{tb:coupling_strength_scenarios}. Results for scenarios 1 to 5 are respectively shown in Figures \ref{fig:NoGV_tdtD}, \ref{fig:002_tdtD}, \ref{fig:007_tdtD}, \ref{fig:014_tdtD}, \ref{fig:022_tdtD}, in order of increasingly negative coupling strength and gap variation. Additionally, results for scenario 6 (small positive gap variation) are shown in Figure \ref{fig:-02_tdtD}. We explore the three switching function shapes given in subsection~\ref{ssec:switching_shapes}. We pick values of $T$ and $S_f$ for scenario 1 ($\Omv{}\to 0$) that have considerable spacelike harvesting. For the rest of scenarios, which have different $\Omv{}$, we pick $T$ so that the adimensional quantity $T(\Omu{}+\Omv{})$ stays constant, and we keep a constant $S_f$.

We observe, consistently with subsections~\ref{ssec:exploration_duration_delay} and \ref{ssec:spacelike_harvesting}, that larger negative $\Omv{}$ concentrates the negativity towards the lightcone, reducing the range of parameters for which spacelike entanglement harvesting occurs. For the explored scenario with largest negative $\Omv{}$ (scenario 5 of Table~\ref{tb:coupling_strength_scenarios}), there is no spacelike harvesting, as in Figure~\ref{fig:022_tdtD}. However, negativity in causal contact grows considerably. Moreover, entanglement can still be harvested for all explored $\Omv{}$ for detectors in causal contact, according to the genuine harvesting estimator.

The figures show that negativity is higher when the detectors are in causal contact, i.e. $|t_\Delta| \approx t_d$. At the same time, the negativity around $|t_\Delta| \approx t_d$ decreases with increasing $t_d$. However, this decay slows down and eventually halts. After such point, the negativity only depends on $t_\Delta - t_d$ and is unaffected by further separating the detectors. This parameter $t_\Delta - t_d$ quantifies the distance of the detector B from the lightcone of the detector A.

This behavior is particular to 1+1D fields, and contrasts with the usual entanglement harvesting results in 3+1D, where the negativity along the lightcone always keeps decaying with distance. In 1+1D the decay along the lightcone stops because the two point correlations of the vacuum of the field tend to non-zero values as the distance along the lightcone increases. This can be seen from Eq.~\eqref{eq:defJ},
\begin{align}
    \mathcal W^{\text{vac}}_{xx'} (t_-,x_-) = \frac{\Jmodes \Big(t_- +\frac{|x_-|}{v}\Big)+\Jmodes\Big(t_- -\frac{|x_-|}{v}\Big)}{4\pi v^2},
\end{align}
and defining $u_\pm = |t_-| \pm \frac{|x_-|}{v}$,
\begin{align}
    \lim_{u_+\to\infty}\mathcal W^{\text{vac}}_{xx'} (t_-,x_-)\big|_{u_-=\text{ctt.}} = \frac{\Jmodes \Big(|t_-| -\frac{|x_-|}{v}\Big)}{4\pi v^2},
\end{align}
which will be non-zero and even large if $|t_-| \approx \frac{|x_-|}{v}$. The negativity exhibits the same limit behavior, but replacing $t_-$ and $|x_-|/v$ by the corresponding $t_\Delta$ and $t_d$. Moreover, if $|t_\Delta| \approx t_d$, the negativity along the lightcone plateaus for $t_d \gg \Omcut^{-1}, T$. This can be deduced from the expressions for $\mathcal L$ and $\mathcal M$ of Eq.~\eqref{eq:ModesFirstLM}, by taking a change of variables $t\to s+t_\textsc{a}$, $t'\to s'+t_\textsc{b}$.

In 3+1D, there would be a $1/|x_-|$ prefactor in the two point correlator of the field amplitudes in the vacuum, $\mathcal W^{\text{vac}}$, which would cause correlations and communication to decay with distance even along the lightcone. However, as already discussed, this decay does not occur in the 1+1D case, which we are interested in due to the field implementation being superconducting transmission lines. This suggests that detectors in causal contact or close to it can be placed far apart without losing most negativity.

\section{Conclusion}
In this study, we strengthened the connection of entanglement harvesting in theoretical Relativistic Quantum Information (RQI) to practical experiments in superconducting circuits. We did so by modeling superconducting circuits as Unruh-DeWitt (UDW) detectors with features such as variable energy gaps and derivative coupling to a 1+1D field. While these features had been previously explored separately, here we explore them together and include an explicit mapping of the parameters of the theoretical models to the experimental parameters in superconducting circuit implementations.

A major contribution is the analysis of the effects on entanglement harvesting of the variable gap detector models that mimic superconducting implementations such as the one in~\cite{Tunable_2023}. Specifically, the variable gap detectors that we consider have a linear reduction in the energy gap due to the coupling (and thus the switching function). For these models, we numerically explored a wide range of parameters, including strength of the gap variation, detector positions, switching functions, and interaction durations. In the scenarios explored, we observe that increasing the gap variation reduces the entanglement acquired by spacelike detectors. Furthermore, this reduction does not occur for detectors in lightlike contact, for which genuine harvesting (subtracting the contributions from communication between the detectors) can even see an enhancement. Moreover, we also see that increasing the distance between detectors does not significantly impair their ability to become entangled, which is a feature of using 1+1D quantum fields. 

Notice that we employed 1+1D quantum fields in order to match the superconducting implementation setting. However, the variable gap detector model we presented can be straightforwardly generalized to fields with any number of spatial dimensions. Analogously to other particle detector models, the major change would be in the expression of the relevant field two point correlator probed by the detectors, which includes the effect of the detectors' smearing functions or cutoff. Such higher dimensional quantum field theories can arise in other experimental platforms such as 2+1-dimensional Bose–Einstein condensates or atoms in 3+1-dimensional spacetime.

Furthermore, for detectors in causal contact, we considered that the entanglement acquired by the detectors can have two contributions: entanglement due to communicating through the field and entanglement that is actually extracted from pre-existing field correlations. We used the tools from \cite{Erickson2021_When} to split these two contributions, and show that having a variable gap does not preclude entanglement harvesting, but rather can even enhance it for detectors in causal contact. This endorses the possibility that this entanglement could be detected in future implementations in those superconducting devices where gap variations cannot be avoided in the ultra-strong coupling regime. Conversely, this work motivates improved implementation designs where the gap variation is reduced or even completely avoided in order to explore spacelike entanglement harvesting.

\acknowledgments

We are grateful to Noah Janzen and Elena Cornick for interesting discussions about the tunable coupler device. This research was undertaken thanks in part to funding from the Canada First Research Excellence Fund. A. T.-B. received the support of a fellowship from ``la Caixa” Foundation (ID 100010434, with fellowship code LCF/BQ/EU21/11890119). Research at Perimeter Institute is supported in part by the Government of Canada through the Department of Innovation, Science and Industry Canada and by the Province of Ontario through the Ministry of Colleges and Universities. AL, EMM acknowledge support through the Discovery Grant Program of the Natural Sciences and Engineering Research Council of Canada (NSERC).

\appendix

\section{Relating the variable gap detector and the spin-boson model}
\label{apx:spin-boson}
The spin-boson model \cite{Legget1987}, commonly used to model superconducting circuits, has the Hamiltonian
\begin{align}
    &\hat H_\textsc{sb} = \hat H_\textsc{s}+\hat H_\textsc{b}+\hat H_\textsc{int-sb},\nonumber\\
    &\hat H_\textsc{s} = - \frac{\hbar \Omega}{2}\hat \sigma_z,\nonumber\\
    &\hat H_\textsc{b} = \sum_k \hbar \omega_k \ad{k}\a{k},\nonumber\\
    &\hat H_\textsc{int-sb} = \sum_k (g_k^x\hat\sigma_x + g_k^z\hat{\sigma}_z)\ad{k}+\text{H.c.},
\end{align}
where the bosonic bath with $\hat H_\textsc{b}$ corresponds to a transmission line of finite length $L$, with $k\in\frac{2\pi}{L}\mathbb{Z}$.

Now, consider the variable gap detector model of Eq.~\eqref{eq:HintApprox}. Then, in the Schrödinger picture, picking the detector position $x_\textsc{qb}=0$, assuming the interaction is fully switched on, and thus $\chi(t)=1$, and ignoring additive constants in $\hat H_\textsc{tl}$,
\begin{align}
    &\hat H_\textsc{vgap} = \hat H_\textsc{qb}+\hat H_\textsc{tl}+\hat H_\textsc{int},\nonumber\\
    &\hat H_\textsc{qb} = - \frac{\hbar (\Omu{} +\Omv{})}{2}\hat \sigma_z,\nonumber\\
    &\hat H_\textsc{tl} = \int \diff k \hbar \omega_k \ad{k}\a{k},\nonumber\\
    &\hat H_\textsc{int} = \varphi_0\gamma \hat\sigma_x\int\diff {k} \cutoff(\omega_k)\sqrt{\frac{\hbar\omega_k}{4\pi\ell_0}} \sign(k) \ii \ad{k}+\text{H.c.}
\end{align}

Therefore, comparing with the spin-boson model,
\begin{equation}
     g_k^x=\varphi_0\gamma\sqrt{\frac{\hbar\omega_k}{4\pi\ell_0}} \cutoff(\omega_k)\sign(k),\  g_k^z=0,
\end{equation}
where the longitudinal coupling is zero because we only kept the transversal coupling in the variable gap detector model.
The spectral density of the bosonic bath (the infinite transmission line in our case) is 
\begin{align}
    J(\omega)&= \frac{2\pi}{\hbar^2} \int\diff k |g_k|^2\delta(\omega - \omega_k)\nonumber\\
    &= 2\int_0^\infty \diff{\omega'} \frac{\varphi_0^2\gamma^2}{2\hbar\ell_0 v}\omega' \cutoff^2(\omega')\delta(\omega - \omega')\nonumber\\
    &= \frac{1}{8\pi}\frac{R_\textsc{k}}{Z_0}\gamma^2\omega e^{-\frac{\omega}{\Omcut}},
\end{align}
where we used the exponential cutoff from Eq.~\eqref{eq:defCutoff}. 

In the limit $\omega \ll \Omcut$,
\begin{equation}
    J(\omega) = \pi \alpha \omega,
\end{equation}
which is the spectral density of an Ohmic bath, with the dimensionless coupling constant $\alpha$ being
\begin{equation}
     \alpha=\frac{R_\textsc{k}}{8\pi^2 Z_0}\gamma^2\approx 6.54 \gamma^2.\label{eq:alpha}
\end{equation}
Since the numerical simulations predict \mbox{$\gamma \in [-0.02,0.22]$} (see subsection \ref{sssec:transversal_approximation}), then $\alpha \in [0,0.32]$.

As the frequencies approach $\Omcut$, the exponential cutoff becomes noticeable, modifying the spectral density,
\begin{align}
    J(\omega)&=  \pi \alpha \omega e^{-\frac{\omega}{\Omcut}}.
\end{align}

\section{Computing the outcome of the harvesting protocol}
\label{apx:final_state}
Here we follow the perturbative procedure outlined in section \ref{sec:time_evolution} to obtain the final state of the probes after they interact with the field. The result is computed to the leading perturbative order, $\mathcal{O}(\lambda^2)$, assuming that $\lambda_\textsc{a}=\lambda_\textsc{b}=\lambda$. 

To simplify the calculations, we define the following operators,
\begin{align}
    \hat\chi_\nu(t) &= \chi_\nu(t) \hat\mu_\nu(t),\nonumber\\
    \hat\phi'_\nu(t) &= \partial_x\hat\phi_{\cutoff}(t,x_\nu)\label{eq:simp_operators}
\end{align}
This notation simplifies the interaction Hamiltonian of 
Eq.~\eqref{eq:Hint} to
\begin{equation}
    \hat H_I = \sum_{\nu}c\lambda\hat\chi_{\nu}\hat\phi'_\nu\,.
\end{equation}
Substituting into Eq.~\eqref{eq:DysonSeries},
\begin{align}
    \hat U^{(1)}=&-\ii c\lambda\int\diff{t} \sum_{\nu}\hat\chi_\nu(t)\hat\phi'_\nu(t)\,,\nonumber\\
    \hat U^{(2)}=&- c^2 \lambda^2\int_{t>t'} \diff {t} \diff {t'}\sum_{\mu, \nu}\hat\chi_{\mu}(t)\hat\chi_{\nu}(t')\hat\phi'_{\mu}(t)\hat\phi'_{\nu}(t').\label{eq:unitaries_substituted}
\end{align}

Now, we can compute the leading order terms of the detectors' final state according to Eq.~\eqref{eq:GenericFinalStateDetector},
\begin{equation}
    \hat \rho_{\textsc{ab}} = \hat\rho_{\textsc{ab},0}+\hat \rho^{(2,0)}_{\textsc{ab}}+\hat \rho^{(1,1)}_{\textsc{ab}}+\hat \rho^{(0,2)}_{\textsc{ab}}+\mathcal{O}(\lambda^4)\,.
\end{equation}
First,
\begin{align}
    \hat \rho^{(1,1)}_{\textsc{ab}} &= \Tr_\phi \bigl( \hat U^{(1)}\hat \rho_{\textsc{ab},0}\otimes\hat\rho_{\phi,0}\hat U^{(1)\dagger}\bigr)\nonumber
    \\
    &=  c^2 \lambda^2\int \diff {t} \diff {t'} \sum_{\mu, \nu}\mathcal W_{xx'}(t',x_{\nu},t,x_{\mu}) \nonumber
    \\
    &\qquad\times\hat\chi_{\mu}(t)\hat \rho_{\textsc{ab},0}\hat\chi_{\nu}(t'),
\end{align}
where
\begin{equation}
    \mathcal W_{xx'}(t,x_{\mu},t',x_{\nu})=\Tr\bigl(\hat\phi'_{\mu}(t)\hat\phi'_{\nu}({t'})\hat\rho_{\phi,0}\bigr),
\end{equation}
which can be seen to match the definition of $\mathcal W_{xx'}$ given in Eq.~\eqref{eq:WGdef}, by using that $\hat\phi'_\nu(t) = \partial_x\hat\phi_{\cutoff}(t,x_\nu)$. 

Second,
\begin{align}
    \hat \rho^{(2,0)}_{\textsc{ab}} &= \Tr_\phi \bigl( \hat U^{(2)}\hat \rho_{\textsc{ab},0}\otimes\hat\rho_{\phi,0}\bigr)\nonumber
    \\
    &= -  c^2 \lambda^2\int \diff {t} \diff {t'} \Heaviside(t-t')\sum_{\mu, \nu}\mathcal W_{xx'}(t,x_{\mu},t',x_{\nu})
    \nonumber\\
    &\qquad\qquad\times \hat\chi_{\mu}(t)\hat\chi_{\nu}({t'})\hat\rho_{\textsc{ab},0}\,,
\end{align}
and finally, $\hat \rho^{(0,2)}_{\textsc{ab}} = \hat \rho^{(2,0)\dagger}_{\textsc{ab}}$.
Notice that the equation for $ \hat \rho^{(2,0)}_{\textsc{ab}}$ can be rewritten as
\begin{align}
    \hat \rho^{(2,0)}_{\textsc{ab}}
    &= -  c^2 \lambda^2\int \diff {t} \diff {t'} \Big( G_{xx'}(t,x_{\textsc{a}},t',x_{\textsc{b}}) \hat\chi_\textsc{a}(t)\hat\chi_\textsc{b}({t'})
    \nonumber\\
    &\ +\sum_{\nu}\Heaviside(t-t') \mathcal W_{xx'}(t,x_{\nu},t',x_{\nu}) \hat\chi_{\nu}(t)\hat\chi_{\nu}({t'})\Big)\hat\rho_{\textsc{ab},0},
\end{align}
where we used the definition of $G_{xx'}$ given in Eq.~\eqref{eq:WGdef}, which states that 
\begin{align}
    G_{xx'}(t,x_{\mu},t',x_{\nu}) =& \Heaviside(t-t')\mathcal W_{xx'}(t,x_{\mu},t',x_{\nu}) \nonumber\\
    &+  \Heaviside(t'-t)\mathcal W_{xx'}(t',x_{\nu},t,x_{\mu}).
\end{align}

This completes the calculation of $\hat \rho_{\textsc{ab}}$ for an arbitrary $\hat\rho_\textsc{ab,0}$. Next, we particularize to the probes starting from the ground state,
\begin{equation}
    \hat\rho_\textsc{ab,0}=\ketbra{0_\textsc{a}0_\textsc{b}}{0_\textsc{a}0_\textsc{b}}.
\end{equation}
Under this assumption, we compute the components of $\hat \rho_{\textsc{ab}}$ in the basis $\ket{0_\textsc{a}0_\textsc{b}}$, $\ket{1_\textsc{a}0_\textsc{b}}$, $\ket{0_\textsc{a}1_\textsc{b}}$, $\ket{1_\textsc{a}1_\textsc{b}}$. To ease the calculation, we denote 
\begin{align}
    \hat A_{\mu\nu}(t,t') =& \hat\chi_{\mu}(t)\hat \rho_{\textsc{ab},0}\hat\chi_{\nu}(t'),
\end{align}
Then,
\begin{align}
    \braket{k_\textsc{a}l_\textsc{b}|\hat A_\textsc{aa}(t,t')|k'_\textsc{a}l'_\textsc{b}}
    &= \delta_{0l}\delta_{0l'}\braket{k_\textsc{a}|\hat\chi_\textsc{a}(t)|0_\textsc{a}}\braket{0_\textsc{a}|\hat\chi_\textsc{a}(t')|k'_\textsc{a}},\nonumber\\
    \braket{k_\textsc{a}l_\textsc{b}|\hat A_\textsc{bb}(t,t')|k'_\textsc{a}l'_\textsc{b}}
    &= \delta_{0k}\delta_{0k'}\braket{l_\textsc{b}|\hat\chi_\textsc{b}(t)|0_\textsc{b}}\braket{0_\textsc{b}|\hat\chi_\textsc{b}(t')|l'_\textsc{b}},\nonumber\\
    \braket{k_\textsc{a}l_\textsc{b}|\hat A_\textsc{ab}(t,t')|k'_\textsc{a}l'_\textsc{b}}
    &=\delta_{0k'}\delta_{0l} \braket{k_\textsc{a}|\hat\chi_\textsc{a}(t)|0_\textsc{a}}\braket{0_\textsc{b}|\hat\chi_\textsc{b}(t')|l'_\textsc{b}},\nonumber\\
    \braket{k_\textsc{a}l_\textsc{b}|\hat A_\textsc{ba}(t,t')|k'_\textsc{a}l'_\textsc{b}}
    &= \delta_{0k}\delta_{0l'}\braket{l_\textsc{b}|\hat\chi_\textsc{b}(t)|0_\textsc{b}}\braket{0_\textsc{a}|\hat\chi_\textsc{a}(t')|k'_\textsc{a}},\nonumber
\end{align}
where, combining Eqs.~\eqref{eq:monopole_vargap_nu} and \eqref{eq:simp_operators} with the definition of $\chi_{\cpx_\nu}(t)$ given in Eq.~\eqref{eq:LMGeneral_complexchi},
\begin{align}
    &\braket{i|\hat\chi_\nu(t)|0} = \delta_{1i} \chi_{\cpx_\nu}(t),\label{eq:components_chi}
\end{align}
and $\braket{0|\hat\chi_\nu(t)|i} = \braket{i|\hat\chi_\nu(t)|0}^*$. Therefore, only one component of each $\hat A_{\mu\nu}$ survives. Precisely, these components correspond to
\begin{align}
    \mathcal L_{\mu\nu} &=  c^2 \lambda^2\int\diff{t} \diff{t'} \mathcal W_{xx'}(t',x_{\nu},t,x_{\mu})\chi_{\cpx_\mu}(t)\chi^*_{\cpx_\nu}(t').\label{eq:calculatedL}
\end{align}
Then,
\begin{align}
    \hat \rho^{(1,1)}_{\textsc{ab}} =&\  \mathcal L_\textsc{aa}\ketbra{1_\textsc{a}0_\textsc{b}}{1_\textsc{a}0_\textsc{b}} + \mathcal L_\textsc{bb}\ketbra{0_\textsc{a}1_\textsc{b}}{0_\textsc{a}1_\textsc{b}}\nonumber\\
    & + \mathcal{L}_\textsc{ab}\ketbra{1_\textsc{a}0_\textsc{b}}{0_\textsc{a}1_\textsc{b}}+ \mathcal{L}_\textsc{ba}\ketbra{0_\textsc{a}1_\textsc{b}}{1_\textsc{a}0_\textsc{b}}.
\end{align}

Moving onto $\hat \rho^{(2,0)}_{\textsc{ab}}$, define for convenience
\begin{align}
    \hat B_{\mu\nu}(t,t') =& \hat\chi_{\mu}(t)\hat\chi_{\nu}(t')\hat \rho_{\textsc{ab},0}.
\end{align}
Then,
\begin{align}
    \braket{k_\textsc{a}l_\textsc{b}|\hat B_\textsc{aa}(t,t')|k'_\textsc{a}l'_\textsc{b}}
    &= \delta_{0l}\delta_{0l'}\delta_{0k'}\braket{k_\textsc{a}|\hat\chi_\textsc{a}(t)\hat\chi_\textsc{a}(t')|0_\textsc{a}},\nonumber\\
    \braket{k_\textsc{a}l_\textsc{b}|\hat B_\textsc{bb}(t,t')|k'_\textsc{a}l'_\textsc{b}}
    &= \delta_{0k}\delta_{0k'}\delta_{0l'}\braket{l_\textsc{b}|\hat\chi_\textsc{b}(t)\hat\chi_\textsc{b}(t')|0_\textsc{b}},\nonumber\\
    \braket{k_\textsc{a}l_\textsc{b}|\hat B_\textsc{ab}(t,t')|k'_\textsc{a}l'_\textsc{b}}
    &= \delta_{0k'}\delta_{0l'}\braket{k_\textsc{a}|\hat\chi_\textsc{a}(t)|0_\textsc{a}}\braket{l_\textsc{b}|\hat\chi_\textsc{b}(t')|0_\textsc{b}}.
\end{align}
Then, using Eqs.~\eqref{eq:monopole_vargap_nu}, \eqref{eq:LMGeneral_complexchi} and \eqref{eq:simp_operators},
\begin{align}
    &\braket{i|\hat\chi_{\nu}(t)\hat\chi_{\nu}(t')|0} = \delta_{0i} \chi_{\cpx_\nu}^*(t)\chi_{\cpx_\nu}(t').
\end{align}
Together with the former Eq.~\eqref{eq:components_chi}, we see that only one component of each $\hat B_{\mu\nu}$ survives, resulting in
\begin{align}
    \hat \rho^{(2,0)}_{\textsc{ab}}
    &= \mathcal M\ketbra{1_\textsc{a}1_\textsc{b}}{0_\textsc{a}0_\textsc{b}} + \sum_\nu \mathcal K_\nu\ketbra{0_\textsc{a}0_\textsc{b}}{0_\textsc{a}0_\textsc{b}},
\end{align}
where 
\begin{align}
    \mathcal M =  &-  c^2 \lambda^2\int \diff {t} \diff {t'}  G_{xx'}(t,x_{\textsc{a}},t',x_{\textsc{b}}) \chi_{\cpx_\textsc{a}}(t)\chi_{\cpx_\textsc{b}}(t')\nonumber,\\
    \mathcal K_\nu = &-  c^2 \lambda^2\int \diff {t} \diff {t'}\Heaviside(t-t') \mathcal W_{xx'}(t,x_{\nu},t',x_{\nu}) \nonumber\\
    &\quad\times\chi_{\cpx_\nu}^*(t)\chi_{\cpx_\nu}(t').\label{eq:calculatedM}
\end{align}
Notice that comparing this expression for $\mathcal K_\nu$ to the one for $\mathcal L_{\nu\nu}$ in Eq.~\eqref{eq:calculatedL} shows that
\begin{equation}
    \mathcal K_\nu + \mathcal K_\nu^* = -\mathcal L_{\nu\nu},
\end{equation}
by using that
\begin{equation}
    \mathcal W_{xx'}^*(t,x_{\nu},t',x_{\nu}) = \mathcal W_{xx'}(t',x_{\nu},t,x_{\nu}).
\end{equation}
Then, when adding to $\hat\rho^{(2,0)}_{\textsc{ab}}$ its Hermitian conjugate $\hat\rho^{(0,2)}_{\textsc{ab}}$,
\begin{align}
    \hat \rho^{(2,0)}_{\textsc{ab}}+\hat \rho^{(0,2)}_{\textsc{ab}}
    =& \mathcal M\ketbra{1_\textsc{a}1_\textsc{b}}{0_\textsc{a}0_\textsc{b}} +\mathcal M^*\ketbra{0_\textsc{a}0_\textsc{b}}{1_\textsc{a}1_\textsc{b}} \nonumber\\
    &- \sum_\nu\mathcal L_{\nu\nu}\ketbra{0_\textsc{a}0_\textsc{b}}{0_\textsc{a}0_\textsc{b}},
\end{align}
Therefore, putting together all the results in this appendix and taking $\hat\rho_\textsc{ab,0}=\ketbra{0_\textsc{a}0_\textsc{b}}{0_\textsc{a}0_\textsc{b}}$ causes $\hat \rho_{\textsc{ab}}$ to become, in the basis $\ket{0_\textsc{a}0_\textsc{b}}$, $\ket{1_\textsc{a}0_\textsc{b}}$, $\ket{0_\textsc{a}1_\textsc{b}}$, $\ket{1_\textsc{a}1_\textsc{b}}$,
\begin{equation}
    \hat\rho_\textsc{ab} = \begin{pmatrix}
    1-\mathcal L_{\textsc{a}\textsc{a}}-\mathcal L_{\textsc{b}\textsc{b}} & 0 & 0 & \mathcal{M}^*\\
    0 & \mathcal L_{\textsc{a}\textsc{a}} & \mathcal L_{\textsc{a}\textsc{b}} & 0 \\
    0 & \mathcal L_{\textsc{b}\textsc{a}} & \mathcal L_{\textsc{b}\textsc{b}} & 0 \\
    \mathcal{M} & 0 & 0 & 0 \\
    \end{pmatrix} + \mathcal{O}(\lambda^4),\label{eq:rho_qubit_apx}
\end{equation}
where from Eqs.~\eqref{eq:calculatedL} and~\eqref{eq:calculatedM},
\begin{align}
    \mathcal L_{\mu\nu} &=  c^2 \lambda^2\int\diff{t} \diff{t'} \mathcal W_{xx'}(t',x_{\nu},t,x_{\mu})\chi_{\cpx_\mu}(t)\chi^*_{\cpx_\nu}(t'),\nonumber\\
    \mathcal M &=-  c^2 \lambda^2\int \diff {t} \diff {t'}  G_{xx'}(t,x_{\textsc{a}},t',x_{\textsc{b}}) \chi_{\cpx_\textsc{a}}(t)\chi_{\cpx_\textsc{b}}(t'),
\end{align}
finishing the derivation of $\hat\rho_\textsc{ab}$.

\section{Simplifying the integrals needed to compute negativity}
\label{apx:simplifyLM}
\subsection{Symmetric switching functions equal up to a time shift and equal detectors}
\label{apx:equal_symmetric_switchings}
We can further simplify the equations $\mathcal L_{\mu\nu}$ and $\mathcal M$ for the type of $\chi_\nu(t)$ explored in this article, which satisfy 
\begin{equation}
    \chi_\nu(t)=\chi(t-t_\nu),\label{eq:shifted_chi}
\end{equation}
with $\chi(t)=\chi(-t)$, and $t_\nu$ controlling the time at which the detector $\nu$ is switched on. Moreover, we assume that the energy gap of both detectors depends linearly on their switching functions, as in Eq.~\eqref{eq:Omega_linear}, as follows,
\begin{equation}
    \Omega_\nu(t) =  \Omu{\nu} + \Omv{\nu}\chi_\nu(t).
\end{equation}
For simplicity, we pick $\Omu{} = \Omu{\nu}$, $\Omv{} = \Omv{\nu}$. 

Under the assumptions above, the expression for $\chi_{\cpx_\nu}$ becomes
\begin{equation}
    \chi_{\cpx_\nu}(t)=
    e^{-\ii\varphi(-t_\nu)}\chi_\cpx (t-t_\nu),\label{eq:SingleChi}
\end{equation}
where we defined
\begin{align}
    &\chi_\cpx(t)=e^{\ii \varphi(t)}\chi(t),\nonumber\\
    &\varphi(t)=\Omu{}t+\Omv{}\int_{0}^t\diff {t'}\chi(t').\label{eq:cpxChi_apx}
\end{align}
Moreover, $\chi(t)=\chi(-t)$ implies 
\begin{equation}
    \varphi(-t) = -\varphi(t),\quad \chi_\cpx(-t)=\chi_\cpx(t)^*.\label{eq:timeSymm}
\end{equation}

\subsubsection{Integrating field modes last}
Substituting Eq.~\eqref{eq:SingleChi} into Eq.~\eqref{eq:LMfinal},
\begin{align}
    &\mathcal L= \frac{\lambda^2}{2\pi} \int_{0}^\infty\!\diff \omega \omega \cutoff(\omega)^2 |\widetilde{\chi_{\cpx}}(\omega)|^2,\nonumber\\
    &\mathcal L_\textsc{ab}= \frac{\lambda^2}{2\pi}e^{-\ii(\varphi(-t_\textsc{a})-\varphi(-t_\textsc{b}))}
    \nonumber\\
    &\qquad\quad\cross\int_{0}^\infty\!\diff \omega \omega \cutoff(\omega)^2 \cos(\omega t_d)e^{-\ii\omega t_\Delta} |\widetilde{\chi_{\cpx}}(\omega)|^2,\label{eq:L_LAB_from_Fourier}
\end{align}
where we defined $\mathcal L = \mathcal L_\textsc{aa}= \mathcal L_\textsc{bb}$. Moreover,
\begin{align}
    &\mathcal M=-\frac{\lambda^2}{2\pi}e^{-\ii(\varphi(-t_\textsc{a})+\varphi(-t_\textsc{b}))}\Mclean,\nonumber\\
    &\Mclean= \int_{0}^\infty\!\diff \omega\omega \cutoff(\omega)^2\cos(\omega t_d)\nonumber\\
    &\qquad\cross\int\diff{t} \diff{t'}e^{-\ii\omega|t-t'-t_\Delta|}\chi_{\cpx}(t) \chi_{\cpx}(t').\label{eq:defMclean}
\end{align}
Here, we also defined the delay between the times of switching $t_\Delta = t_\textsc{b}-t_\textsc{a}$ and used that
\begin{equation}
    \widetilde{\chi_{\cpx_\nu}}(\omega)= e^{\ii(\omega t_\nu-\varphi(-t_\nu))}\widetilde{\chi_{\cpx}}(\omega).
\end{equation}
Moreover due to $\chi(t)=\chi(-t)$,
\begin{align}
    \Re(\Mclean)&= \int_{0}^\infty\!\!\diff \omega\omega \cutoff(\omega)^2\cos(\omega t_d)\cos(\omega t_\Delta)\widetilde{\chi_{\cpx}}(\omega) \widetilde{\chi_{\cpx}}(-\omega),\nonumber\\
    \Im(\Mclean)&=-\int_{0}^\infty\!\diff \omega \diff{t} \diff{t'} \omega \cutoff(\omega)^2\cos(\omega t_d) \nonumber\\
    &\qquad\quad\cross\sin(\omega|t-t'-t_\Delta|)\chi_{\cpx}(t) \chi_{\cpx}(t').\label{eq:Mclean_ReIm_modesfirst}
\end{align}
These expressions for $\Re(\Mclean)$ and $\Im(\Mclean)$ follow from using that
\begin{align}
    \Mclean^* &= \int_{0}^\infty\!\diff \omega\omega \cutoff(\omega)^2\cos(\omega t_d)\nonumber\\
    &\qquad\cross\int\diff{t} \diff{t'}e^{\ii\omega|t-t'-t_\Delta|}\chi_{\cpx}(t) \chi_{\cpx}(t'),\label{eq:Mstar}
\end{align}
which in turn was obtained by performing the change of variables $t \to -t'$, $t' \to -t$ in Eq.~\eqref{eq:defMclean} and then applying Eq.~\eqref{eq:timeSymm}. 

When comparing Eq.~\eqref{eq:defMclean} to Eq.~\eqref{eq:Mstar}, we see that complex conjugating $\Mclean$ is equivalent to complex conjugating $\mathcal W^{\text{vac}}$, since we only need to conjugate the complex phase $e^{-\ii\omega|t-t'-t_\Delta|}$, which comes from $\mathcal W^{\text{vac}}$, and not $\chi_{\cpx}$. Consequently, for the switching functions considered in this appendix (of the kind defined in Eq.~\eqref{eq:shifted_chi}, with $\chi(t)=\chi(-t)$), we have a shortcut to compute $\mathcal M^\pm$,
\begin{align}
    \mathcal M^+ &= -\frac{\lambda^2}{2\pi}e^{-\ii(\varphi(-t_\textsc{a})+\varphi(-t_\textsc{b}))}\Re(\Mclean),\nonumber\\
    \mathcal M^- &= -\frac{\lambda^2}{2\pi}e^{-\ii(\varphi(-t_\textsc{a})+\varphi(-t_\textsc{b}))}\Im(\Mclean),\label{eq:M+M-_expressions}
\end{align}
with $\Mclean$ as defined in Eq.~\eqref{eq:defMclean} and $\varphi(t)$ from Eq.~\eqref{eq:cpxChi_apx}. Additional expressions for $\mathcal L$, $\mathcal L_\textsc{ab}$ and $\Mclean$ are provided next.

\subsubsection{Integrating field modes first}
Substituting Eq.~\eqref{eq:SingleChi} into Eq.~\eqref{eq:ModesFirstLM},
\begin{align}
    &\mathcal L=\frac{\lambda^2}{2\pi}\int\diff{t} \diff{t'} \Jmodes(t'-t)\chi_{\cpx}(t)\chi^*_{\cpx}(t'),\nonumber\\
    &\mathcal L_{\textsc{ab}}=\frac{\lambda^2}{2\pi}e^{-\ii(\varphi(-t_\textsc{a})-\varphi(-t_\textsc{b}))}\nonumber\\
    &\qquad\ \ \cross\int\diff{t} \diff{t'}\Imodes(t'-t+t_\Delta) \chi_{\cpx}(t)\chi_{\cpx}^*(t'),\nonumber\\
    &\mathcal M=-\frac{\lambda^2}{2\pi}e^{-\ii(\varphi(-t_\textsc{a})+\varphi(-t_\textsc{b}))}\Mclean,\nonumber\\
    &\Mclean=\int\diff{t} \diff{t'}\Imodes(|t-t'-t_\Delta|)\chi_{\cpx}(t) \chi_{\cpx}(t'),\label{eq:defMclean2}
\end{align}
where $\mathcal L = \mathcal L_\textsc{aa}= \mathcal L_\textsc{bb}$, and $\Jmodes(t)$ and $\Imodes(t)$ are defined as in Eq.~\eqref{eq:defJ} and Eq.~\eqref{eq:ModesFirstLM} respectively.

Moreover, due to $\chi(t)=\chi(-t)$,
\begin{align}
    \Re(\Mclean)&= \int\diff{t} \diff{t'}\Re\bigl(\Imodes(t-t'-t_\Delta)\bigr)\chi_{\cpx}(t) \chi_{\cpx}(t'),\nonumber\\
    \Im(\Mclean)&=\int\diff{t} \diff{t'}\Im\bigl(\Imodes(|t-t'-t_\Delta|)\bigr)\chi_{\cpx}(t) \chi_{\cpx}(t').
    \label{eq:Mclean_ReIm_modeslast}
\end{align}
These expressions for $\Re(\Mclean)$ and $\Im(\Mclean)$ follow from using that $\Imodes(-t)=\Imodes^*(t)$ together with
\begin{align}
    \Mclean^* &= \int\diff{t} \diff{t'}\Imodes^*(|t-t'-t_\Delta|)\chi_{\cpx}(t) \chi_{\cpx}(t').
\end{align}
In turn, this expression was obtained by performing the change of variables $t \to -t'$, $t' \to -t$ in Eq.~\eqref{eq:defMclean2} and then applying Eq.~\eqref{eq:timeSymm}.

\subsection{Changing variables to partially decouple time integrals}
\label{apx:decoupled_timeIntegrals}
The following change of variables facilitates numerical integration by uncoupling one of the time integrals from  $\Jmodes$ and $\Imodes$ in the expressions for $\mathcal L_{\mu\nu}$ and $\mathcal M$,
\begin{align}
    t_\pm=t\pm t',
\end{align}
with absolute value of the Jacobian 
\begin{equation}
    \bigg|\frac{\partial(t_+,t_-)}{\partial(t,t')}\bigg| = 2.
\end{equation}
Then, performing the change of variables in Eq.~\eqref{eq:defMclean2}, 
\begin{align}
    &\mathcal L=\frac{\lambda^2}{4\pi}\int\diff{t_-} \Jmodes^*(t_-)I_L(t_-),\nonumber\\ 
    &\mathcal L_{\textsc{ab}}=\frac{\lambda^2}{4\pi}e^{-\ii(\varphi(-t_\textsc{a})-\varphi(-t_\textsc{b}))}\int\diff{t_-}\Imodes^*(t_- -t_\Delta)I_L(t_-),
    \nonumber\\
    &\mathcal M=-\frac{\lambda^2}{2\pi}e^{-\ii(\varphi(-t_\textsc{a})+\varphi(-t_\textsc{b}))}\Mclean,\nonumber\\
    &\Mclean=\frac{1}{2}\int\diff{t_-}\Imodes(|t_- -t_\Delta|)I_M(t_-),\label{eq:IsinLM}
\end{align}
where we used $\Jmodes(-t)=\Jmodes^*(t)$, $\Imodes(-t)=\Imodes^*(t)$, and defined
\begin{align}
    &I_L(t_-)=\int\diff{t_+}\chi_{\cpx}\bigg(\frac{t_+ +t_-}{2}\bigg)\chi_{\cpx}^*\bigg(\frac{t_+ -t_-}{2}\bigg),\nonumber\\
    &I_M(t_-)=\int\diff{t_+}\chi_{\cpx}\bigg(\frac{t_+ +t_-}{2}\bigg)\chi_{\cpx}\bigg(\frac{t_+ -t_-}{2}\bigg).
\end{align}
For finite time switching functions supported on $[-\frac{\Ti}{2},\frac{\Ti}{2}]$, the integration limits for $t_\pm$ are
\begin{align}
    -\Ti\leq t_- \leq \Ti,\quad -\Ti+|t_-|\leq t_+ \leq \Ti-|t_-|.
\end{align}
Moreover, when $\chi(t)=\chi(-t)$ and thus $\chi_{\cpx}(-t)=\chi_\cpx^*(t)$,
\begin{align}
    &I_L(t_-)=2\int_0^\infty\diff{t_+}\chi_{\cpx}\bigg(\frac{t_+ +t_-}{2}\bigg)\chi_{\cpx}^*\bigg(\frac{t_+ -t_-}{2}\bigg),\nonumber\\
    &I_M(t_-)=2\int_0^\infty\diff{t_+}\Re\biggl(\chi_{\cpx}\bigg(\frac{t_+ +t_-}{2}\bigg)\chi_{\cpx}\bigg(\frac{t_+ -t_-}{2}\bigg)\biggr).
\end{align}

Furthermore, because of Eq.~\eqref{eq:M+M-_expressions}, $\mathcal M^\pm$ can be computed in terms of $\Re(\Mclean)$ and $\Im(\Mclean)$.

The numerical evaluation of the integrals above provided the results for Figures~\ref{fig:NoGV_TtD}-\ref{fig:-02_tdtD} and~\ref{fig:LargeCutoff}. For the numerical integration algorithm we used Mathematica's \texttt{NIntegrate}. We used the default parameters except for: i) \texttt{PrecisionGoal}, which sets the intended digits of precision for the final result and ii) \texttt{WorkingPrecision}, which sets the digits of precision kept in internal computations. We adjusted these parameters differently depending on the integral:
\begin{enumerate}
    \item $I_L$: \texttt{PrecisionGoal}$\to 8$, \texttt{WorkingPrecision}$\to80$
    \item $\mathcal L$: \texttt{PrecisionGoal}$\to 8$, \texttt{WorkingPrecision}$\to40$
    \item $I_M$ for Gaussian switching:  \texttt{PrecisionGoal}$\to 8$, \texttt{WorkingPrecision}$\to16$
    \item $I_M$ for cosine ramps and isosceles trapezoid switchings:  \texttt{PrecisionGoal}$\to 6$, \texttt{WorkingPrecision}$\to60$
    \item $\Mclean$ for Gaussian switching:  \texttt{PrecisionGoal}$\to 6$, \texttt{WorkingPrecision}$\to16$
    \item $\Mclean$ for cosine ramps and isosceles trapezoid switchings:  \texttt{PrecisionGoal}$\to 6$, \texttt{WorkingPrecision}$\to30$
\end{enumerate}

\section{Validity of the second order perturbative approach}
\label{apx:fourth_order}
The entanglement acquired by the detectors was computed via a perturbative expansion to second order, as described in Section~\ref{sec:time_evolution} and further detailed in Appendix~\ref{apx:final_state}. Such perturbative expansion can fail for large values of the coupling constant $\lambda$. The present appendix intends to estimate for which $\lambda$ the second order calculation becomes unreliable. To do so, we compute fourth order corrections and check for which values of $\lambda$ they become larger than the second order. Specifically, for simplicity, we consider single detectors, and compute their excitation probability up to fourth order.

\vspace{0.5cm}
\subsection{Computing the detector excitation probability to fourth order}

The calculation proceeds analogously to Appendix~\ref{apx:final_state}. A first difference is that, since we are considering a single detector, we drop the sum and the subindices in the interaction Hamiltonian of Eq.~\eqref{eq:Hint}, leaving
\begin{align}
    &\hat H_I =  c \lambda\hat\chi\hat\phi'\,,\nonumber\\
    &\hat\chi(t) = \chi(t) \hat\mu(t),\nonumber\\
    &\hat\phi'(t) = \partial_x\hat\phi_{\cutoff}(t,x_\textsc{det}). \label{eq:HI_single_detector}
\end{align}

Next, we perform a Dyson expansion of the time evolution of Eq.~\eqref{eq:time_ordered_evolution} up to fourth order,
\begin{equation}
    \hat U = \hat U^{(0)}+\hat U^{(1)}+\hat U^{(2)}+\hat U^{(3)}+\hat U^{(4)}+\mathcal{O}(\lambda^5), 
\end{equation}
\begin{align}
    &\hat U^{(1)}=-\ii c \lambda\int\diff{t_1} \hat\chi(t_1)\hat\phi'(t_1)\,,\nonumber\\
    &\hat U^{(2)}=- c^2 \lambda^2\int_{t_1>t_2} \diff {t_1} \diff {t_2}\hat\chi(t_1)\hat\chi(t_2)\hat\phi'(t_1)\hat\phi'(t_2)\,,\nonumber\\
    &\hat U^{(3)}=\ii c^3 \lambda^3\int_{t_1>t_2>t_3} \diff {t_1} \diff {t_2} \diff {t_3} \hat\chi(t_1)\hat\chi(t_2)\hat\chi(t_3)
    \nonumber\\
    &\qquad\qquad\qquad\qquad\cross\hat\phi'(t_1)\hat\phi'(t_2)\hat\phi'(t_3)\,,\nonumber\\
    &\hat U^{(4)}= c^4 \lambda^4\int_{t_1>t_2>t_3>t_4} \!\diff {t_1} \diff {t_2} \diff {t_3} \diff {t_4} \hat\chi(t_1)\hat\chi(t_2)\hat\chi(t_3)
    \nonumber\\
    &\qquad\qquad\qquad\qquad\cross\hat\chi(t_4)\hat\phi'(t_1)\hat\phi'(t_2)\hat\phi'(t_3)\hat\phi'(t_4)\,.
\end{align}

As usual, let us initialize the detector and the field in an uncorrelated state,
\begin{equation}
    \hat\rho_0 = \hat\rho_{\textsc{det},0}\otimes\hat\rho_{\phi,0}\,,
\end{equation}
and assume that the initial field state $\hat\rho_{\phi,0}$ has vanishing odd-point functions. That is, \mbox{$\forall n\in \{0,1,2,\ldots\}$},
\begin{align}
&\Tr(\hat\rho_{\phi,0}
\hat \phi(t_0,x_0)\ldots \hat \phi(t_{2n+1},x_{2n+1}))=0,
\end{align}
which is the case for the vacuum, thermal states and Fock states. Tracing over the field to obtain the final detector state,
\begin{align}
    \hat \rho_{\textsc{det}} &= \Tr_\phi \hat U \hat \rho_0 \hat U^\dagger \nonumber\\
    &=\hat\rho_{\textsc{det},0}
    +\hat \rho^{(2,0)}_{\textsc{det}}+\hat \rho^{(1,1)}_{\textsc{det}}+\hat \rho^{(0,2)}_{\textsc{det}}\nonumber\\
    &\quad+\hat \rho^{(4,0)}_{\textsc{det}}+\hat \rho^{(3,1)}_{\textsc{det}}+\hat \rho^{(2,2)}_{\textsc{det}}+\hat \rho^{(1,3)}_{\textsc{det}}+\hat \rho^{(0,4)}_{\textsc{det}}
    +\mathcal{O}(\lambda^6)\,,
\end{align}
with 
\begin{equation}
    \hat \rho^{(i,j)}_\textsc{det}=\Tr_\phi  \hat U^{(i)}\hat\rho_0\hat U^{(j)\dagger}\,.
\end{equation}

From now on, take the detector to be initialized in the ground state,
\begin{align}
    \hat\rho_{\textsc{det},0} = \ketbra{0}{0}.
\end{align}
Then, the final detector state, in the basis $\ket{0}$, $\ket{1}$, is of the form
\begin{equation}
    \hat\rho_\textsc{det} = \begin{pmatrix}
    1-\mathcal L^{(2)}-\mathcal L^{(4)} & 0 \\
    0 & \mathcal L^{(2)} + \mathcal L^{(4)}  \\
    \end{pmatrix} + \mathcal{O}(\lambda^6), \label{eq:finalstate_fourth_order}
\end{equation}
where $\mathcal L^{(n)}$ is a term of order $\lambda^n$. This is a consequence of only corrections $\hat\rho_\textsc{det}^{(i,j)}$ with even values $i+j$ participating in $\hat\rho_\textsc{det}$, together with
\begin{align}
    \hat\chi(t)\ket{0} = \chi_{\varphi}(t) \ket{1}, \quad \hat\chi(t)\ket{1} = \chi_{\varphi}^* (t) \ket{0}.
\end{align}
Here, $\chi_{\varphi}(t) = e^{\ii \varphi(t)} \chi(t)$, $\varphi(t) = \int_0^t \diff{t'} \Omega(t')$, matching Eq.~\eqref{eq:monopole_vargap_nu}. For example, $\hat \rho^{(1,1)}_{\textsc{det}}$ is a double integral over terms proportional to $\hat\chi(t_1)\ketbra{0}{0}\hat\chi(t_2) \propto \ketbra{1}{1}$, and thus
\begin{align}
    \hat \rho^{(1,1)}_{\textsc{det}} = \mathcal L^{(2)}\ketbra{1}{1}.
\end{align}
Analogously, $\hat \rho^{(2,0)}_{\textsc{det}}$ is a double integral over terms proportional to \mbox{$\hat\chi(t_1)\hat\chi(t_2)\ketbra{0}{0} \propto \ketbra{0}{0}$}. The same holds for $\hat \rho^{(0,2)}_{\textsc{det}} = \hat \rho^{(2,0)\dagger}_{\textsc{det}}$, and, since the Dyson expansion conserves the trace at each order,
\begin{align}
    \hat \rho^{(2,0)}_{\textsc{det}} + \hat \rho^{(0,2)}_{\textsc{det}}= -\mathcal L^{(2)}\ketbra{0}{0}.
\end{align}

The same arguments can be extended to the fourth order, leaving us with
\begin{align}
    &\hat \rho^{(3,1)}_{\textsc{det}} + \hat \rho^{(1,3)}_{\textsc{det}} = \mathcal L^{(4)}\ketbra{1}{1},\nonumber\\
    &\hat \rho^{(4,0)}_{\textsc{det}} + \hat \rho^{(2,2)}_{\textsc{det}}+ \hat \rho^{(0,4)}_{\textsc{det}} = -\mathcal L^{(4)}\ketbra{0}{0},
\end{align}
and finishing the justification of Eq.~\eqref{eq:finalstate_fourth_order}. 

Now, let us move onto the expression for $\mathcal L^{(2)}$,
\begin{align}
    \mathcal L^{(2)} &= \mel{1}{\hat \rho^{(1,1)}_{\textsc{det}}}{1} \nonumber\\
    &=  c^2 \lambda^2 \int \diff{t_1} \diff{t_2} \chi_\varphi(t_1)\chi_\varphi^*(t_2) \Tr\bigl(\hat\phi'(t_2)\hat\phi'(t_1)\hat\rho_{\phi,0}\bigr)\nonumber\\
    &=  c^2 \lambda^2\! \int \diff{t_1} \diff{t_2} \chi_\varphi^*(t_1)\chi_\varphi(t_2) \mathcal{W}_{xx'}(t_1,x_\textsc{det},t_2,x_\textsc{det}), 
\end{align}
where we performed a change of variables in the last line and used 
\begin{equation}
    \Tr\bigl(\hat\phi'(t_2)\hat\phi'(t_1)\hat\rho_{\phi,0}\bigr) = \mathcal W_{xx'}(t_2,x_\textsc{det},t_1,x_\textsc{det}). \label{eq:phi'phi'_wightman}
\end{equation}
Here, the two point correlator $\mathcal W_{xx'}$ is as defined in Eq.~\eqref{eq:WGdef}. 

Next, let us find $\mathcal L^{(4)}$ using
\begin{align}
    \mathcal L^{(4)} &= \mel{1}{\hat \rho^{(3,1)}_{\textsc{det}}+\hat \rho^{(1,3)}_{\textsc{det}}}{1} \nonumber\\
    &= 2\Re\big(\mel{1}{\hat \rho^{(3,1)}_{\textsc{det}}}{1}\big) \nonumber\\
    &= 2c^4 \lambda^4 \Re \int_{t_1>t_2>t_3} \diff{t_1} \diff{t_2} \diff{t_3} \diff{t_4} \chi_\varphi(t_1)\chi_\varphi^*(t_2)\nonumber\\
    &\qquad\cross \chi_\varphi(t_3)\chi_\varphi^*(t_4) \Tr\bigl(\hat\phi'(t_4)\hat\phi'(t_1)\hat\phi'(t_2)\hat\phi'(t_3)\hat\rho_{\phi,0}\bigr)\nonumber\\
    &= 2c^4 \lambda^4 \Re \int_{t_2>t_3>t_4} \diff{t_1} \diff{t_2} \diff{t_3} \diff{t_4} \chi_\varphi^*(t_1)\chi_\varphi(t_2)\nonumber\\
    &\qquad\cross \chi_\varphi^*(t_3)\chi_\varphi(t_4) \Tr\bigl(\hat\phi'(t_1)\hat\phi'(t_2)\hat\phi'(t_3)\hat\phi'(t_4)\hat\rho_{\phi,0}\bigr),
\end{align}
where $\hat\phi'(t) = \partial_x\hat\phi_{\cutoff}(t,x_\textsc{det})$, as previously defined in Eq.~\eqref{eq:HI_single_detector}.

Furthermore, let us assume that the field is prepared in its vacuum state,
\begin{align}
    \hat\rho_{\phi,0}=\ketbra{0_\phi}{0_\phi}.
\end{align}
For the vacuum, the correlator $\mathcal{W}_{xx'}(t,x,t',x')$ only depends on $t_- = t - t'$ and $x_- = x - x'$. Specifically,
\begin{align}
    \mel{0_\phi}{\hat\phi'(t_1)\hat\phi'(t_2)}{0_\phi} &= \mathcal{W}_{xx'}^\text{vac}(t_1-t_2, 0)\nonumber\\
    &= \frac{1}{2\pi c^2}\Jmodes(t_1-t_2). \label{eq:phi'phi'_vac}
\end{align}
The expression for $\mathcal{W}_{xx'}^\text{vac}$ is provided in Eq.~\eqref{eq:WG_vac}, and for $\Jmodes$ in Eq.~\eqref{eq:defJ}.

Furthermore,
\begin{align}
    &\Tr\bigl(\hat\phi'(t_1)\hat\phi'(t_2)\hat\phi'(t_3)\hat\phi'(t_4)\ketbra{0_\phi}{0_\phi}\bigr)\nonumber\\
    &=\frac{1}{4\pi^2 c^4}\Big(\Jmodes(t_1-t_2)\Jmodes(t_3-t_4)\nonumber\\
    &\qquad \qquad \ \  +\Jmodes(t_1-t_3)\Jmodes(t_2-t_4) \nonumber\\
    &\qquad \qquad \ \  +\Jmodes(t_1-t_4)\Jmodes(t_2-t_3)\Big).
\end{align}
This follows from Eq.~\eqref{eq:phi'phi'_vac} together with
\begin{align}
    &\mel{0_\phi}{\hat\phi'(t_1)\hat\phi'(t_2)\hat\phi'(t_3)\hat\phi'(t_4)}{0_\phi}
    \nonumber\\
    &=
    \mel{0_\phi}{\hat\phi'(t_1)\hat\phi'(t_2)}{0_\phi}\mel{0_\phi}{\hat\phi'(t_3)\hat\phi'(t_4)}{0_\phi}\nonumber\\
    &\quad +\mel{0_\phi}{\hat\phi'(t_1)\hat\phi'(t_3)}{0_\phi}\mel{0_\phi}{\hat\phi'(t_2)\hat\phi'(t_4)}{0_\phi}\nonumber\\
    &\quad +\mel{0_\phi}{\hat\phi'(t_1)\hat\phi'(t_4)}{0_\phi}\mel{0_\phi}{\hat\phi'(t_2)\hat\phi'(t_3)}{0_\phi}. \label{eq:4_to_2_point_correlator}
\end{align}
This equality can be derived, for example, using the canonical commutation relations \mbox{$[\a{k},\ad{k'}]=\delta(k-k')$}, the fact that all $\a{k}$ annihilate the vacuum $\a{k}\ket{0_\phi} = 0$, the definition \mbox{$\hat\phi'(t) = \partial_x\hat\phi_{\cutoff}(t,x_\textsc{det})$} and the expression for $\hat\phi_{\cutoff}$ provided in Eq.~\eqref{eq:field_cutoff}. Alternatively, the equality in Eq.~\eqref{eq:4_to_2_point_correlator} also follows from the vacuum being a zero mean Gaussian state.

Summarizing, a detector initialized in the ground state, after interacting with the vacuum, ends up in the state
\begin{equation}
    \hat\rho_\textsc{det} = \begin{pmatrix}
    1-\mathcal L^{(2)}-\mathcal L^{(4)} & 0 \\
    0 & \mathcal L^{(2)} + \mathcal L^{(4)}  \\
    \end{pmatrix} + \mathcal{O}(\lambda^6). 
\end{equation}
From here we can read that the detector gets excited with a probability
\begin{equation}
    \mel{1}{\hat\rho_\textsc{det}}{1} = \mathcal{L}^{(2)} + \mathcal{L}^{(4)} + \mathcal{O}(\lambda^6).
\end{equation}
Moreover, we found that
\begin{align}
    \mathcal L^{(2)} 
    &=  \frac{\lambda^2}{2\pi} \int \diff{t_1} \diff{t_2} \chi_\varphi^*(t_1)\chi_\varphi(t_2)  \Jmodes(t_1-t_2), \nonumber\\
    \mathcal L^{(4)} 
    &= \frac{\lambda^4}{2\pi^2} \Re \int_{-\infty}^{\infty} \diff{t_4} \int_{t_4}^{\infty} \diff{t_3} \int_{t_3}^{\infty} \diff{t_2} \int_{-\infty}^{\infty}\diff{t_1} \nonumber\\
    &\ \ \cross \chi_\varphi^*(t_1)\chi_\varphi(t_2)\chi_\varphi^*(t_3)\chi_\varphi(t_4) \Big(\Jmodes(t_1-t_2)\Jmodes(t_3-t_4)\nonumber\\
    &\quad\   +\Jmodes(t_1-t_3)\Jmodes(t_2-t_4)  +\Jmodes(t_1-t_4)\Jmodes(t_2-t_3)\Big). \label{eq:apx_L2_L4_integralform}
\end{align}
Here, $\chi_{\varphi}(t) = e^{\ii \varphi(t)} \chi(t)$, $\varphi(t) = \int_0^t \diff{t'} \Omega(t')$,
\begin{align}
    \Jmodes(t)=\int_{0}^\infty\!\diff \omega\omega \cutoff(\omega)^2 e^{-\ii\omega t}.
\end{align}
Notice that, combining the above equations,
\begin{align}
    \mathcal L^{(2)} 
    &=  \frac{\lambda^2}{2\pi} \int_0^\infty \diff{\omega} \omega \cutoff(\omega)^2 |\widetilde{\chi_\varphi}(\omega)|^2,\label{eq:L2_FT}
\end{align}
with the Fourier transform convention of Eq.~\eqref{eq:fourier_convention}.

Finally, assuming the exponential cutoff of Eq.~\eqref{eq:defCutoff},
\begin{equation}
    \Jmodes(t) = \frac{\Omcut^2}{(1+\ii \Omcut t)^{2}}.
\end{equation}

\subsection{Evaluating the excitation probability for the relevant switching functions}
Ultimately, we intend to compute the excitation probability, and thus $\mathcal L^{(2)}$ and $\mathcal L^{(4)}$, for scenarios that match the exploration of entanglement harvesting in Section \ref{sec:harvesting_results}.

Aiming to provide a simple and concise estimation for the failure of the second order approximation, we remain in the case of detectors with a constant gap $\Omega$. This implies
\begin{equation}
    \chi_{\varphi}(t) = e^{\ii \Omega t} \chi(t), \label{eq:chiphi_cttgap}
\end{equation}
which facilitates providing analytical expressions for some of the nested integrals, as shown next for each choice of switching function.

\subsubsection{Gaussian switching}
The Gaussian switching function is given by
\begin{equation}
    \chi(t) = e^{-\frac{t^2}{T^2}}. \label{eq:gaussian_switching_apx}
\end{equation}
Notice that the switching does not include a delay (time translation). We do not need to include it because the time at which the switching is turned on does not affect the excitation probability for single detectors.

We move onto finding an analytical expression for the following integral, which appears in $\mathcal L^{(2)}$ and $\mathcal L^{(4)}$,
\begin{align}
    \Jchi(t')&=\int_{-\infty}^{\infty} \diff{t} e^{-\ii \Omega t} \chi(t) \Jmodes(t-t')
    \nonumber\\
    &=\int_{-\infty}^{\infty} \diff{t} e^{-\ii \Omega t} \chi(t) \int_{0}^\infty\!\diff \omega \omega \cutoff(\omega)^2 e^{-\ii\omega (t-t')} 
    \nonumber\\
    &= \int_{0}^\infty\!\diff \omega \omega \cutoff(\omega)^2 \widetilde{\chi}^*(\omega+ \Omega) e^{\ii\omega t'},\label{eq:Jchi}
\end{align}
where we expressed $\Jmodes(t-t')$ as an integral using Eq.~\eqref{eq:defJ} and used $\widetilde{\chi}(\omega)$ to denote the Fourier transform of $\chi(t)$, with the convention specified in Eq.~\eqref{eq:fourier_convention}. 

For the current Gaussian switching,
\begin{align}
    \widetilde{\chi}(\omega) = \sqrt{\pi} T e^{-\frac{1}{4}\omega^2T^2}, \label{eq:FT_Gaussian}
\end{align}
and using $\cutoff(\omega) = e^{-\frac{|\omega|}{2\Omcut}}$, as in Eq.~\eqref{eq:defCutoff},
\begin{align}
    \Jchi(t')&= \sqrt{\pi} T \int_{0}^\infty\!\diff \omega \omega e^{-\frac{\omega}{\Omcut}} e^{-\frac{1}{4}(\omega+\Omega)^2T^2}e^{\ii\omega t'} \nonumber\\
    &= \frac{2\sqrt{\pi}}{T}\JGaussian\bigg(\frac{\Omega T}{2}, \frac{t'}{T} +  \frac{\ii}{T \Omcut} \bigg).
\end{align}
Here,
\begin{align}
    \JGaussian(\upsilon, \kappa) &= e^{-\upsilon^2}\big(1-2(\kappa+\ii\upsilon)\DawsonF(\kappa+\ii\upsilon)\big) \nonumber\\
    &\quad + \ii \sqrt{\pi} \kappa e^{-\kappa^2 - 2\ii\kappa\upsilon},
\end{align}
$\DawsonF(z)=\frac{\sqrt{\pi}}{2} e^{-z^2}\erfi(z)$ denoting the Dawson function, $\erfi(z) = -\ii \erf(\ii z)$ and $\erf$ denoting the error function. 

Using the above result, $\mathcal L^{(4)}$ becomes a triple integral, facilitating its numerical evaluation,
\begin{align}
    \mathcal L^{(4)} &= \frac{\lambda^4}{\pi^{\frac{3}{2}} T} \Re \int_{-\infty}^\infty \diff{t_4}\! \int_{t_4}^\infty\! \diff{t_3} \! \int_{t_3}^\infty\! \diff{t_2} 
     e^{\ii\Omega(t_2-t_3+t_4)} \nonumber\\
    &\ \ \cross  e^{-\frac{t_2^2+t_3^2+t_4^2}{T^2}} \Bigg(\JGaussian\bigg(\!\frac{\Omega T}{2}, \frac{t_2}{T} + \frac{\ii}{T \Omcut} \bigg)\Jmodes(t_3-t_4) \nonumber\\
    &\qquad\qquad \qquad\! +\JGaussian\bigg(\frac{\Omega T}{2}, \frac{t_3}{T} + \frac{\ii}{T \Omcut} \bigg)\Jmodes(t_2-t_4)   \nonumber\\
    &\qquad\qquad \qquad\! +\JGaussian\bigg(\frac{\Omega T}{2}, \frac{t_4}{T} + \frac{\ii}{T \Omcut} \bigg)\Jmodes(t_2-t_3)\Bigg). 
\end{align}
In the following subsection, the integral is evaluated in Python, using the numerical method \texttt{nquad} from SciPy. We use the default parameters, except for setting the absolute error tolerance to $10^{-8}$ and the relative error tolerance to $10^{-4}$. Furthermore we use finite integration limits, replacing the  $\pm\infty$ limits by $\pm 8 T/\sqrt{2}$. For these and larger values of time, the value of the Gaussian switching function is negligible. Specifically, replacing these limits corresponds to setting the Gaussian switching to zero when the time is larger than 8 standard deviations. Increasing the limits beyond 8 standard deviations did not significantly change the numerical results. Under these choices, the \texttt{nquad} integrator uses the ``\texttt{qagse}"\cite{qagse} method from the Fortran library QUADPACK.

Similarly to the last expression for $\mathcal L^{(4)}$, $\mathcal L^{(2)}$ can be written as
\begin{align}
    \mathcal L^{(2)} 
    &=  \frac{\lambda^2}{\sqrt{\pi} T} \int \diff{t_2} e^{-\frac{t_2^2}{T^2}}e^{\ii\Omega t_2} \JGaussian\bigg(\frac{\Omega T}{2}, \frac{t_2}{T} + \frac{\ii}{T \Omcut} \bigg).
\end{align}
However, $\mathcal L^{(2)}$ also admits an expression solely in terms of special functions. Using Eqs.~\eqref{eq:L2_FT}, \eqref{eq:FT_Gaussian} and \eqref{eq:defCutoff},
\begin{align}
    \mathcal L^{(2)} 
    &= \frac{\lambda^2 T^2}{2} \int_0^\infty \diff{\omega} \omega e^{-\frac{\omega}{\Omcut}} e^{-\frac{1}{2}\omega^2T^2}\nonumber\\
    &= \frac{\lambda^2}{2}\Bigg(e^{-\frac{1}{2}\Omega^2 T^2} - \sqrt{\frac{\pi}{2}}\frac{1 + \Omega\Omcut T^2}{\Omcut T} \nonumber\\
    & \qquad\quad  \cross\erfc\bigg(\frac{1 + \Omega\Omcut T^2}{\sqrt{2}\Omcut T}\bigg)\Bigg),
\end{align}
where $\erfc(z)=1-\erf(z)$ is the complementary error function.

\subsubsection{Cosine ramps switching}
Consider now the cosine ramps switching function, which is given by
\begin{align}
    &\chi(t)=\aswitch\bigg(\frac{t}{T}\bigg),\nonumber\\
    &\aswitch(\atc)=\begin{cases}
    1 & |s|\leq \frac{\aTf}{2}, \\
    \left(\cos \left(\frac{\pi}{2}\frac{2 |s|-\aTf}{1-\aTf}\right)\right)^2 &
   \frac{\aTf}{2}<|s|<\frac{1}{2}, \\
     0 & \frac{1}{2}\leq |s|. \label{eq:apx_cosineramps}
\end{cases}
\end{align}

We will follow a different approach to reducing the number of integrals in the expression of $\mathcal L^{(4)}$ from Eq.~\eqref{eq:apx_L2_L4_integralform}.

First, substituting into Eq.~\eqref{eq:apx_L2_L4_integralform} the definition of $\Jchi$ from Eq.~\eqref{eq:Jchi},
\begin{align}
    \mathcal L^{(4)} &= \frac{\lambda^4}{2\pi^2} \Re \int_{t_2>t_3>t_4} \diff{t_2}\diff{t_3}\diff{t_4}  \chi_\varphi(t_2)\chi_\varphi^*(t_3) \chi_\varphi(t_4)\nonumber\\
    &\qquad\qquad \cross\Big(\Jchi(t_2)\Jmodes(t_3-t_4) +\Jchi(t_3)\Jmodes(t_2-t_4)\nonumber\\
    &\qquad \qquad \qquad  +\Jchi(t_4)\Jmodes(t_2-t_3)\Big).
\end{align}
Then, notice that
\begin{align}
    &\int_{t_2>t_3>t_4} \diff{t_2}\diff{t_3}\diff{t_4} \chi_\varphi(t_2)\chi_\varphi^*(t_3)\chi_\varphi(t_4) \Jchi(t_3)\Jmodes(t_2-t_4)\nonumber\\
    &=\int_{-\infty}^\infty \diff{t_4} \int_{t_4}^\infty \diff{t_3}  \int_{t_4}^{t_3} \diff{t_2} \chi_\varphi(t_3)\chi_\varphi^*(t_2)\chi_\varphi(t_4) \nonumber\\
    &\qquad\qquad\qquad\qquad\qquad\cross\Jchi(t_2)\Jmodes(t_3-t_4),
\end{align}
by performing the change of variables $t_2 \to t_3$, $t_3 \to t_2$. Similarly,
\begin{align}
    &\int_{t_2>t_3>t_4} \diff{t_2}\diff{t_3}\diff{t_4} \chi_\varphi(t_2)\chi_\varphi^*(t_3)\chi_\varphi(t_4) \Jchi(t_4)\Jmodes(t_2-t_3)\nonumber\\
    &=\int_{-\infty}^\infty \diff{t_4} \int_{t_4}^\infty \diff{t_3}  \int_{-\infty}^{t_4} \diff{t_2} \chi_\varphi(t_3)\chi_\varphi^*(t_4)\chi_\varphi(t_2) \nonumber\\
    &\qquad\qquad\qquad\qquad\qquad\cross\Jchi(t_2)\Jmodes(t_3-t_4),
\end{align}
by performing the change of variables $t_2 \to t_3$, $t_3 \to t_4$, $t_4 \to t_2$.

Therefore, substituting back into $\mathcal L^{(4)}$,
\begin{align}
    \mathcal L^{(4)} &= \frac{\lambda^4}{2\pi^2} \Re \int_{-\infty}^\infty \diff{t_4} \int_{t_4}^\infty \diff{t_3} \Jmodes(t_3-t_4)\nonumber\\
    &\quad \cross\bigg(
    \chi_\varphi^*(t_3) \chi_\varphi(t_4)\int^{\infty}_{t_3} \diff{t_2} \chi_\varphi(t_2)\Jchi(t_2)\nonumber\\
    &\qquad\quad +\chi_\varphi(t_3)\chi_\varphi(t_4)\int_{t_4}^{t_3} \diff{t_2} \chi_\varphi^*(t_2)\Jchi(t_2) \nonumber\\
    &\qquad\quad +\chi_\varphi(t_3)\chi_\varphi^*(t_4)\int_{-\infty}^{t_4} \diff{t_2} \chi_\varphi(t_2)\Jchi(t_2)\bigg).
\end{align}

Now, let us focus on the innermost integrals, starting with
\begin{align}
    &\int^{\infty}_{t_3} \diff{t_2} \chi_\varphi(t_2)\Jchi(t_2)\nonumber\\
    &=\int_{0}^\infty\!\diff \omega \omega \cutoff(\omega)^2\widetilde{\chi}^*(\omega+ \Omega)\int \diff{t_2}  \Heaviside_{t_3}(t_2)\chi(t_2)e^{\ii(\omega+\Omega) t_2}
    \nonumber\\
    &=\int_{0}^\infty\!\diff \omega \omega \cutoff(\omega)^2\widetilde{\chi}^*(\omega+ \Omega) \widetilde{\Heaviside_{t_3}\chi}(\omega+\Omega),
\end{align}
where we used Eqs.~\eqref{eq:chiphi_cttgap} and~\eqref{eq:Jchi}, and denoted $\Heaviside_{t_0}(t)= \Heaviside(t-t_0)$, where $\Heaviside$ is the Heaviside step function. Similarly,
\begin{align}
    &\int_{t_4}^{t_3} \diff{t_2} \chi_\varphi^*(t_2)\Jchi(t_2) \nonumber\\
    &=\int_{0}^\infty\!\diff \omega \omega \cutoff(\omega)^2\widetilde{\chi}^*(\omega+ \Omega)\nonumber\\
    &\qquad\cross\int \diff{t_2}  \big(\Heaviside_{t_4}(t_2)-\Heaviside_{t_3}(t_2)\big)\chi(t_2)e^{\ii(\omega-\Omega) t_2}
    \nonumber\\
    &=\int_{0}^\infty\!\diff \omega \omega \cutoff(\omega)^2\widetilde{\chi}^*(\omega+ \Omega) \nonumber\\
    &\qquad\cross\Big(\widetilde{\Heaviside_{t_4}\chi}(\omega-\Omega)-\widetilde{\Heaviside_{t_3}\chi}(\omega-\Omega)\Big).
\end{align}
The remaining innermost integral is
\begin{align}
    &\int_{-\infty}^{t_4} \diff{t_2} \chi_\varphi(t_2)\Jchi(t_2)\nonumber\\
    &=\int^{\infty}_{-t_4} \diff{t_2} \chi_\varphi(-t_2)\Jchi(-t_2)\nonumber\\
    &=\int_{0}^\infty\!\diff \omega \omega \cutoff(\omega)^2\widetilde{\chi}^*(\omega+ \Omega)\nonumber\\
    &\qquad\cross\int \diff{t_2}  \Heaviside_{-t_4}(t_2)\chi(t_2)e^{-\ii(\omega+\Omega) t_2}
    \nonumber\\
    &=\int_{0}^\infty\!\diff \omega \omega \cutoff(\omega)^2\widetilde{\chi}^*(\omega+ \Omega) \widetilde{\Heaviside_{-t_4}\chi}(-\omega-\Omega).
\end{align}

The last step is to provide analytical expressions for $\widetilde \chi$ and $\widetilde{\Heaviside_t\chi}$ for the cosine ramps switching (Eq.~\eqref{eq:apx_cosineramps}). First, 
\begin{align}
    &\widetilde \chi(\omega) = \frac{ \pi^2}{4}\frac{(1+S_f)T}{4} \sinc\!\left(\frac{(1+S_f)T}{4}\omega\right)\nonumber\\
    &\qquad \cross\sinc\!\bigg(\frac{\pi}{4}\big(1+\middlebar{T}_r\omega\big)\bigg)\sinc\!\bigg(\frac{\pi}{4}\big(1-\middlebar{T}_r\omega\big)\bigg),\nonumber\\
    &\middlebar{T}_r= \frac{T_r}{2\pi},
    \nonumber\\
    &T_r= (1-S_f)T.\label{eq:FTCosineRamps}
\end{align}
Second, since the expression for $\widetilde{\Heaviside_t\chi}$ is lengthy, let us split it into three parts, 
\begin{align}
    &\widetilde{\Heaviside_t\chi} = \widetilde{\Heaviside_t\chi_\textsc{l}}(\omega)+\widetilde{\Heaviside_t\chi_\textsc{m}}(\omega)+\widetilde{\Heaviside_t\chi_\textsc{r}}(\omega), \label{eq:HeavisideFT_CosineRamps}
\end{align}
each corresponding to the left ramp ($\chi_\textsc{l}$), flat middle ($\chi_\textsc{m}$) and right ramp ($\chi_\textsc{r}$) of the cosine ramps switching function, so that $\chi(t) = \chi_\textsc{l}(t)+\chi_\textsc{m}(t)+\chi_\textsc{r}(t)$. 

The Fourier transform for each part is:
\begin{align}
    & \widetilde{\Heaviside_t\chi_\textsc{l}}(\omega) \nonumber\\
    &= 
    \begin{cases}
      0 
      & 2 t \leq -T ,
      \\[12pt]
      \frac{
        e^{-\frac{\ii S_f T \omega}{2}}
      }{
        1 - \middlebar{T}_r^2 \omega^2
      }\! \left(\! 
          \frac{T_r}{4} 
          e^{-\ii \frac{T_r}{4}\omega}\sinc\!\left(
          \frac{T_r}{4}\omega
          \right)
          +
          \ii \middlebar{T}_r^2 \omega
        \!\right)
      & 2 t \geq -S_f T,
      \\[12pt]
      \frac{
        e^{\ii \omega t} 
      }{
        1 - \middlebar{T}_r^2 \omega^2
      }
      \bigg[ 
        \frac{T + 2 t}{4} 
        e^{-\ii \frac{T + 2 t}{4}\omega}\sinc\!\left(
          \frac{T + 2 t}{4}\omega
        \right)
        \\[6pt]
        \ + \ii \middlebar{T}_r^2
          \omega
          \sin^2\!\left( \frac{T + 2 t}{4 \middlebar{T}_r} \right)
        - \frac{\middlebar{T}_r}{2} \sin\!\left( \frac{T + 2 t}{2 \middlebar{T}_r} \right)
      \bigg]
      & \text{otherwise.}
    \end{cases}
\end{align}
\begin{align}
    & \widetilde{\Heaviside_t\chi_\textsc{m}}(\omega) \nonumber\\
    &= 
    \begin{cases}
      0 
      & 2 t \leq -S_f T,
      \\[6pt]
      S_f T \, \sinc\!\left( \frac{S_f T \omega}{2} \right) 
      & 2 t \geq S_f T,
      \\[6pt]
      e^{\ii \omega t} \frac{S_f T + 2t}{2}e^{-\ii \frac{S_f T + 2t}{4}\omega}\sinc\!\left(\frac{S_f T + 2t}{4} \omega \right)
      & \text{otherwise.}
    \end{cases}
\end{align}
\begin{align}
    & \widetilde{\Heaviside_t\chi_\textsc{r}}(\omega)\nonumber\\
    &= 
    \begin{cases}
      0 
      & 2 t \leq S_f T ,
      \\[12pt]
      \frac{
        e^{\frac{\ii S_f T \omega}{2}}
      }{
        1 - \middlebar{T}_r^2 \omega^2
      } \left( 
          \frac{T_r}{4} 
          e^{\ii \frac{T_r}{4}\omega}\sinc\!\left(
          \frac{T_r}{4}\omega
          \right)
          -
          \ii \middlebar{T}_r^2 \omega
        \right)
      & 2 t \geq T,
      \\[12pt]
      \frac{
        e^{\ii \omega t} 
      }{
        1 - \middlebar{T}_r^2 \omega^2
      }
      \bigg[ 
        \frac{\middlebar{T}_r}{2} \sin\!\left( \frac{T - 2 t}{2 \middlebar{T}_r} \right)
        \\[6pt]
        \ - \ii \middlebar{T}_r^2
          \omega e^{\frac{\ii (S_f T-2t) \omega}{2}} + \ii \middlebar{T}_r^2
          \omega
          \sin^2\!\left( \frac{T - 2 t}{4 \middlebar{T}_r} \right) 
          \\[6pt]
          \ + 
          \frac{2 t - S_f T}{4} 
        e^{-\ii \frac{2 t - S_f T}{4}\omega}\sinc\!\left(-
          \frac{2 t - S_f T}{4}\omega
        \right)
      \bigg]
        & \text{otherwise.}
    \end{cases}\label{eq:HeavisidedFT_RightCosineRamp}
\end{align}

Finally, $\mathcal L^{(4)}$ becomes
\begin{align}
    \mathcal L^{(4)} &= \frac{\lambda^4}{2\pi^2} \Re \int_0^\infty\diff{\omega} \omega\cutoff(\omega)^2\widetilde{\chi}^*(\omega+ \Omega)  \nonumber\\
    &\quad \cross \int_{-\frac{T}{2}}^\frac{T}{2} \diff{t_4} \int_{t_4}^\frac{T}{2} \diff{t_3} \Jmodes(t_3-t_4)\chi(t_3)\chi(t_4) \nonumber\\
    &\quad \cross \bigg(
    e^{\ii\Omega(-t_3+t_4)}\widetilde{\Heaviside_{t_3}\chi}(\omega+\Omega)\nonumber\\
    &\qquad\quad +e^{\ii\Omega(t_3+t_4)}\Big(\widetilde{\Heaviside_{t_4}\chi}(\omega-\Omega)-\widetilde{\Heaviside_{t_3}\chi}(\omega-\Omega)\Big) \nonumber\\
    &\qquad\quad +e^{\ii\Omega(t_3-t_4)}\widetilde{\Heaviside_{-t_4}\chi}(-\omega-\Omega)\bigg).
\end{align}
The analytic expressions for $\widetilde{\chi}(\omega)$ and $\widetilde{\Heaviside_t\chi}(\omega)$ are provided in Eqs.~\eqref{eq:FTCosineRamps} to~\eqref{eq:HeavisidedFT_RightCosineRamp}.

The numerical integration is carried out using \texttt{nquad} from SciPy, with default parameters, except for setting the absolute error tolerance to $10^{-8}$ and the relative error tolerance to $10^{-4}$. Furthermore, we approximate the infinite integration bounds over $\omega\in[0,\infty)$ by finite bounds: $\omega\in[0,6\Omcut)$. For $\omega>6\Omcut$, the exponentially decaying cutoff $\cutoff(\omega)$ significantly reduces the integrand, multiplying it by a factor smaller than $\cutoff(6\Omcut)^2 =e^{-6}\approx 10^{-3}$. Increasing the approximate upper bound on $\omega$ did not significantly affect the numerical results. Due to the finite integration bounds, the \texttt{nquad} integrator uses the ``\texttt{qagse}"\cite{qagse} method from the Fortran library QUADPACK.

\subsubsection{Isosceles trapezoid switching}
Consider now the isosceles trapezoid switching function, which is given by
\begin{align}
    \chi(t)=\aswitch\bigg(\frac{t}{T}\bigg),\quad\aswitch(\atc)=\begin{cases}
     1 & |s|\leq \frac{\aTf}{2},\\
     \frac{1-2 |s|}{1-\aTf} & \frac{\aTf}{2}<|s|<\frac{1}{2},\\
     0 & \frac{1}{2}\leq |s|.
    \end{cases}
\end{align}

Analogously to the Gaussian switching case, we will provide an expression for the integral $\Jchi$ (see Eq.~\eqref{eq:Jchi}) in terms of special functions,
\begin{align}
    &\Jchi(t')
    = \frac{2}{(1-S_f)T}\JIsoTrap\bigg(S_f,\,\frac{\Omega T}{2},\, -\ii \Omega t'  +  \frac{\Omega}{\Omcut} \bigg), \nonumber\\
    &\JIsoTrap(S_f,\upsilon, z) = \JIsoTrapAux(\upsilon,z) - \JIsoTrapAux(S_f\upsilon,z), 
\end{align}
\begin{align}
    \JIsoTrapAux(\upsilon,z) = 2\cos(\upsilon) - e^z \big(&(1+z+\ii\upsilon)E_1(z+\ii\upsilon)\nonumber\\
    &+(1+z-\ii\upsilon)E_1(z-\ii\upsilon) \big).
\end{align}
Here, $\expint(z)$ is the exponential integral function, defined as
\begin{align}
    \expint(z) = \int_z^\infty \diff{t} \frac{e^{-t}}{t},
\end{align}
with a branch cut discontinuity in the real negative axis of the complex plane.\footnote{The branch cut is not crossed when integrating $\Jchi(t')$ over any interval of $t'\in\mathbb{R}$.}

The above result can be obtained from the fact that, for the current isosceles trapezoid switching,
\begin{align}
    \widetilde{\chi}(\omega) =\frac{T}{2(1-S_f)} \bigg(\sinc^2\bigg(\frac{\omega T}{4}\bigg)-S_f^2\sinc^2\bigg(\frac{S_f \omega T}{4}\bigg) \bigg).
\end{align}
Then, choosing $\cutoff(\omega) = e^{-\frac{|\omega|}{2\Omcut}}$, the last line of Eq.~\eqref{eq:Jchi} can be integrated by using symbolic integration software, such as Mathematica.

Finally, substituting into $\mathcal L^{(4)}$ from Eq.~\eqref{eq:apx_L2_L4_integralform},
\begin{align}
    \mathcal L^{(4)} &= \frac{\lambda^4}{\pi^2 (1-S_f)T} \Re \int_{-\frac{T}{2}}^\frac{T}{2} \diff{t_4}\! \int_{t_4}^\frac{T}{2}\! \diff{t_3} \! \int_{t_3}^\frac{T}{2}\! \diff{t_2} 
      \nonumber\\
    &\quad   \cross  e^{\ii\Omega(t_2-t_3+t_4)}\chi(t_2)\chi(t_3)\chi(t_4)\nonumber\\
    &\quad  \cross\Bigg(\JIsoTrap\bigg(S_f,\,\frac{\Omega T}{2},\, -\ii \Omega t_2 +  \frac{\Omega}{\Omcut} \bigg)\Jmodes(t_3-t_4) \nonumber\\
    &\qquad  +\JIsoTrap\bigg(S_f,\,\frac{\Omega T}{2},\, -\ii \Omega t_3  +  \frac{\Omega}{\Omcut} \bigg)\Jmodes(t_2-t_4)   \nonumber\\
    &\qquad +\JIsoTrap\bigg(S_f,\,\frac{\Omega T}{2},\, -\ii \Omega t_4  +  \frac{\Omega}{\Omcut} \bigg)\Jmodes(t_2-t_3)\Bigg).
\end{align}
Notice that this expression is indeterminate for the case $S_f = 1$, which corresponds to the limit case of a square switching function. However, this is not an issue, since we explore the case $S_f = 0.15$ in this appendix.

For the numerical integration using \texttt{nquad} from SciPy, we use the default parameters, except for setting the absolute error tolerance to $10^{-6}$ and the relative error tolerance to $10^{-3}$. Under these choices, the \texttt{nquad} integrator uses the ``\texttt{qagse}"\cite{qagse} method from the Fortran library QUADPACK.

\begin{figure*}[t]
    \centering
    \includegraphics[width=0.95\textwidth]{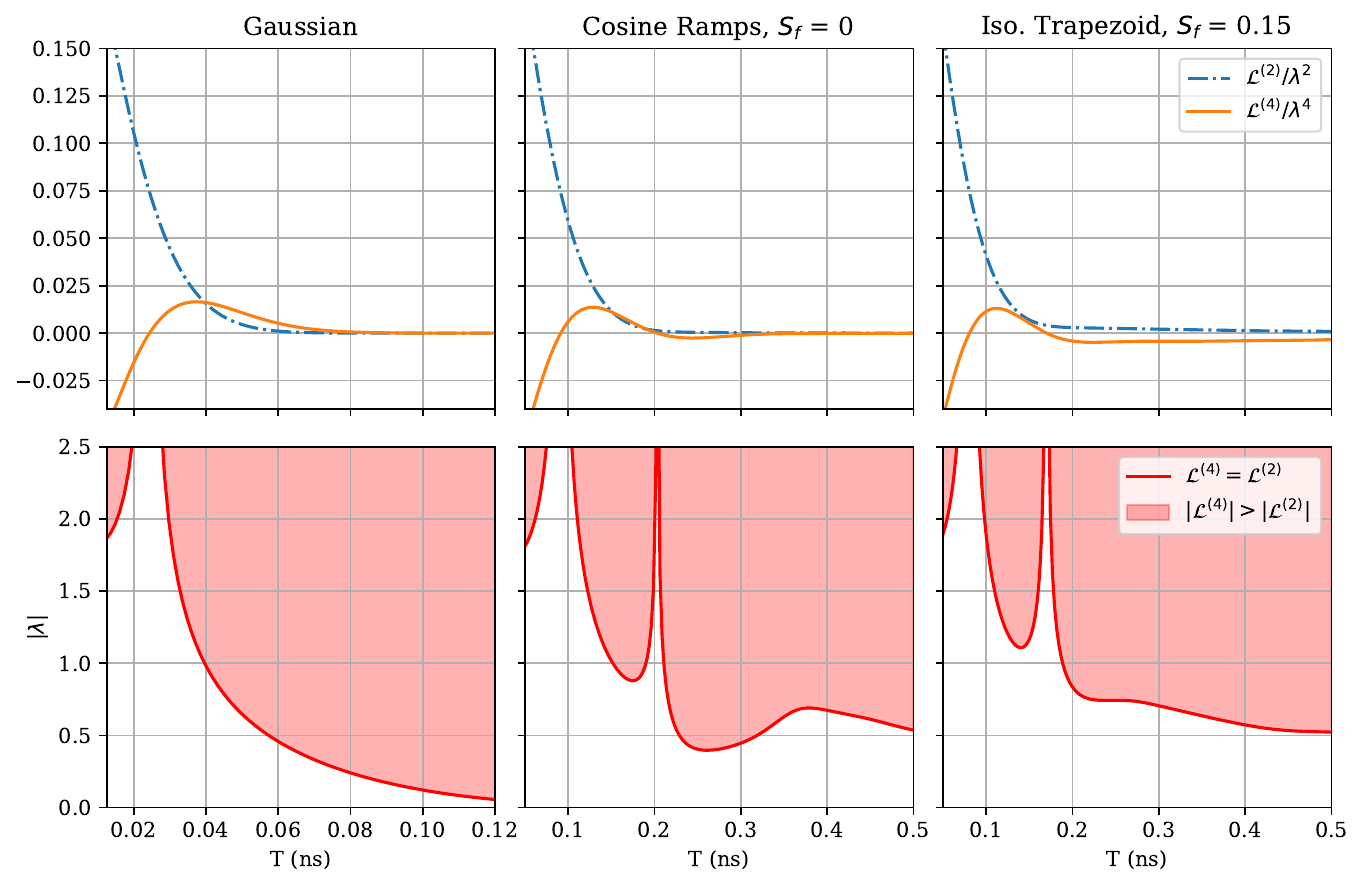}
    \caption{Comparison between second and fourth order corrections to the excitation probability of single detectors. All plots show dependencies on the switching duration $T$ (horizontal axis). The first row shows $\mathcal L^{(2)}/\lambda^2$ and $\mathcal L^{(4)}/\lambda^4$, that is, the second and fourth order corrections, normalized by their corresponding power of the coupling strength. The second row shows the values of $\lambda$ for which $\mathcal L^{(4)} = \mathcal L^{(2)}$ (red line) and $\mathcal L^{(4)} > \mathcal L^{(2)}$ (red region). 
    Each column of plots corresponds to a choice of switching function, from left to right being: Gaussian, cosine ramps with $S_f=0$, isosceles trapezoid with $S_f=0.15$. The detector model is as in Eq.~\eqref{eq:Hint}, but for a single detector with constant gap $\Omega/(2\pi) = 7.3 \,\unit{\giga\hertz}$ instead. The cutoff is taken to be exponential (Eq.~\eqref{eq:defCutoff}) with $\Omcut/(2\pi) = 50 \,\unit{\giga\hertz}$.}
    \label{fig:comparison_fourth_order}
\end{figure*}

\subsection{Estimating the breakdown of the second order calculations}

Finally, we provide a best case estimate for when the second order is not enough for single detectors. We do so by comparing the fourth order correction $\mathcal L^{(4)}$ to the second order correction $\mathcal L^{(2)}$. Furthermore, this best case estimate should also apply to the entanglement harvesting setup, since if the second order is already not enough for single detectors, we also expect it to not be enough for a pair of detectors.

We intend to estimate the failure of the second order approximation for scenarios as in the entanglement harvesting exploration of Section \ref{sec:harvesting_results}. Thus we use the parameters that match the experimental implementation described in Section \ref{ssec:parameter_choices}, except for the detector gap $\hbar \Omega$, which we chose to be constant to provide a simpler estimation. Specifically we are choosing \mbox{$\Omega = \Omu{} = 2\pi\times 7.3 \,\unit{\giga\hertz}$}, i.e. the gap remaining equal to the uncoupled gap. 

The first row of plots in Figure~\ref{fig:comparison_fourth_order} provides the numerical results for $\mathcal L^{(2)}$ and $\mathcal L^{(4)}$, as a function of the switching duration $T$. The figure includes plots for the Gaussian switching and for a choice of cosine ramps switching and isosceles trapezoid switching which can harvest entanglement. The second row of plots in Figure~\ref{fig:comparison_fourth_order} depicts the values of the coupling strength $\lambda$ for which we know that the second order computation is not reliable. Specifically, the region painted in red indicates the $|\lambda|$ (vertical axis) and $T$ (horizontal axis) for which the fourth order contribution is larger than the second order contribution, $|\mathcal L^{(4)}|>|\mathcal L^{(2)}|$. Thus, we know that the second order perturbative approach is not valid in the red region. Notice that for larger $T$, smaller values of $\lambda$ are already not acceptable. Conversely, for small but non-zero $T$, around $0.02\,\unit{\nano\second}$ for Gaussian switching and $0.1\,\unit{\nano\second}$ for cosine ramps and isosceles trapezoid switching, the $\mathcal L^{(4)}$ contribution gets close to zero. For these small $T$, our estimation cannot say with certainty that the second order approach is invalid until for much larger $\lambda$.

Finally, let us compare the second row of plots in Figure~\ref{fig:comparison_fourth_order} to the values of $\lambda$ that can be achieved by the detector implementation, depicted in Table~\ref{tb:coupling_strength_scenarios}. From this table we see that the implementation can achieve $|\lambda|$ as large as 1. Then, whether such values of $|\lambda|$ invalidate the second order approximation strongly depends on the choice of switching function and its duration $T$, according to the estimator presented in the second row of Figure~\ref{fig:comparison_fourth_order}.

\section{Spacelike harvesting in the large cutoff limit}
\label{apx:largecutoff}

The model explored in this article includes an exponential cutoff, which can be reinterpreted as the detector being smeared in space according to Eq.~\eqref{eq:effSmearing_ExpCutoff}. This relationship is shown in subsection~\ref{sec:TL_quantization} and further discussion can be found in~\cite{McKay_2017}. This effective smearing can raise the question of whether the harvesting found in section \ref{sec:harvesting_results} for spacelike separated point-like detectors is truly present or is merely a byproduct of their effective size induced by the cutoff. 

\begin{figure*}[t]
    \centering
    \includegraphics[width=0.95\textwidth]{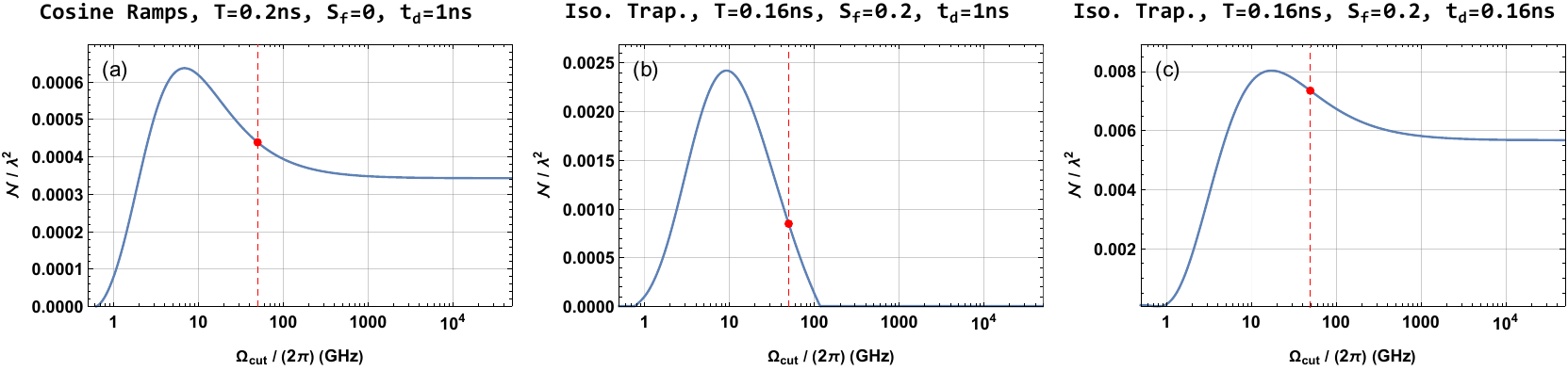}
    \caption{$\mathcal N /\lambda^2$ plotted against the cutoff frequency $\Omcut/(2\pi)$, for $\Omu{}/(2\pi)=7.3\,\unit{\giga\hertz}$, $\Omv{}=0$, $|t_\Delta| = t_d - T$ (spacelike detector interactions). The cutoff frequency $\Omcut/(2\pi) = 50\,\unit{\giga\hertz}$, used throughout this work, is highlighted with a vertical dashed line. The subfigures are for: (a)~Cosine ramps switching, $T=0.2\,\unit{\nano\second}$, $S_f=0$, $t_d=1\,\unit{\nano\second}$. (b)~Isosceles trapezoid switching, $T=0.16\,\unit{\nano\second}$, $S_f=0.2$, $t_d=1\,\unit{\nano\second}$. (c)~Isosceles trapezoid switching, $T=0.16\,\unit{\nano\second}$, $S_f=0.2$, $t_d=0.16\,\unit{\nano\second}$. We emphasize that for all the scenarios analyzed the detectors are in strict spacelike separation.}
    \label{fig:LargeCutoff}
\end{figure*}

In this appendix, we show that spacelike entanglement harvesting can be achieved in the explored scenarios. Specifically, we provide examples of $\mathcal N$ tending to positive values as $\Omcut$ grows large. This corresponds to the limit where the effective detector size becomes negligible. Three spacelike harvesting examples are presented in Figure~\ref{fig:LargeCutoff}. For the cosine ramps switching example in Figure~\ref{fig:LargeCutoff}(a), harvesting remains possible as $\Omcut\to\infty$. There is only a 20\% reduction of $\mathcal N$ in the limit $\Omcut\to\infty$ with respect to $\Omcut/(2\pi) = 50\,\unit{\giga\hertz}$. For the isosceles trapezoid switching, the cutoff seems to play a larger role when $t_d=1\,\unit{\nano\second}$, as shown in Figure~\ref{fig:LargeCutoff}(b), where spacelike harvesting becomes impossible as $\Omcut\to\infty$. However, spacelike harvesting for isosceles trapezoid switching and $\Omcut\to\infty$ is still possible by reducing the distance between the detectors to $t_d = 0.16\,\unit{\nano\second}$, as shown in Figure~\ref{fig:LargeCutoff}(c).

\bibliography{bibliography}

\end{document}